\documentclass[12pt]{article}

\usepackage{verbatim,color,amssymb}
\usepackage{amsmath}					
\usepackage{amsthm}					
\usepackage{natbib}
\usepackage{multirow}
\usepackage{setspace}
\usepackage[mathscr]{euscript}
\usepackage{fancyhdr}
\usepackage{enumitem}
\usepackage{graphicx}
\usepackage{geometry}
\usepackage[bookmarks=false]{hyperref}
\usepackage{subcaption}
\usepackage{adjustbox}
\usepackage{soul}
\usepackage{xcolor}

\usepackage{multibib}
\newcites{latex}{References}

\usepackage{tabularx, booktabs}
\newcolumntype{Y}{>{\centering\arraybackslash}X}
\usepackage{lscape}

\usepackage{float}
\usepackage{lineno}

\newtheorem*{Proof*}{Proof}

\newtheorem{Lem}{\underline{\bf Lemma}}

\def\eE{\mathbb{E}}

\def\X{{\cal X}}

\def\Ind{\hbox{I}}
\def\wh{\widehat}
\def\wt{\widetilde}

\def\data{\hbox{data}}

\def\Beta{\hbox{Beta}}

\def\Ga{\hbox{Ga}}

\def\IG{\hbox{Inv-Ga}}
\def\InvG{\hbox{Inv-Gaussian}}

\def\MVN{\hbox{MVN}}

\def\Normal{\hbox{Normal}}

\def\P_25_ICML{{\it Proceedings of the 25th international conference on Machine learning}}

\def\bse{\begin{eqnarray*}}
\def\ese{\end{eqnarray*}}
\def\bsa#1\esa{\begin{align*}#1\end{align*}}
\def\bal#1\eal{\begin{align}#1\end{align}}
\def\be{\begin{eqnarray}}
\def\ee{\end{eqnarray}}
\def\bq{\begin{equation}}
\def\eq{\end{equation}}

\def\wh{\widehat}

\def\trans{^{\rm T}}

\def\th{^{th}}

\def\cB{{\mathcal B}}
\def\cG{{\mathcal G}}

\def\cX{{\mathcal X}}
\def\cY{{\mathcal Y}}

\def\bA{{\mathbf A}}

\def\bB{{\mathbf B}}

\def\bD{{\mathbf D}}

\def\b1e{{\mathbf e}}
\def\b1f{{\mathbf f}}

\def\bH{{\mathbf H}}
\def\bI{{\mathbf I}}
\def\bJ{{\mathbf J}}

\def\bu{{\mathbf u}}
\def\bU{{\mathbf U}}

\def\bV{{\mathbf V}}

\def\bx{{\mathbf x}}
\def\bX{{\mathbf X}}
\def\by{{\mathbf y}}
\def\bY{{\mathbf Y}}

\def\bzero{{\mathbf 0}}

\newcommand{\bmu}{\mbox{\boldmath $\mu$}}
\newcommand{\bnu}{\mbox{\boldmath $\nu$}}

\newcommand{\bphi}{\mbox{\boldmath $\phi$}}

\newcommand{\bvarepsilon}{\mbox{\boldmath $\varepsilon$}}
\newcommand{\btheta}{\mbox{\boldmath $\theta$}}

\newcommand{\bbeta}{\mbox{\boldmath $\beta$}}

\newcommand{\bSigma}{\mbox{\boldmath $\Sigma$}}
\newcommand{\balpha}{\mbox{\boldmath $\alpha$}}
\newcommand{\bomega}{\mbox{\boldmath $\omega$}}

\newcommand{\bpsi}{\mbox{\boldmath $\psi$}}

\newcommand{\btau}{\mbox{\boldmath $\tau$}}

\newcommand{\abs}[1]{\left\vert#1\right\vert}

\renewcommand\footnoterule{\kern-3pt \hrule \textwidth 2in \kern 2.6pt}

\def\colred#1{\textcolor{red}{#1}}

\def\boxit#1{\vbox{\hrule\hbox{\vrule\kern6pt \vbox{\kern6pt \textcolor{blue}{#1}\kern6pt}\kern6pt\vrule}\hrule}}

\def\authorfootnote#1{{\let\thefootnote\relax\footnotetext{#1}}}

\allowdisplaybreaks

\begin{document}
\thispagestyle{empty}
\baselineskip=28pt

\begin{center}
{\LARGE{\bf Bayesian Tensor Factorized Vector Autoregressive Models 
for Inferring 
Granger Causality Patterns 
from High-Dimensional Multi-subject Panel Neuroimaging Data
}}
\end{center}
\baselineskip=12pt
\vskip 3mm

\begin{center}
Jingjing Fan$^{a}$ (jfan25@utexas.edu)\\
Kevin Sitek$^{b}$(kevin.sitek@pitt.edu)\\
Bharath Chandrasekaran$^{b}$(b.chandra@pitt.edu)\\
Abhra Sarkar$^{a}$ (abhra.sarkar@utexas.edu)\\

\vskip 3mm
$^{a}$Department of Statistics and Data Sciences, \\
University of Texas at Austin,\\ 2317 Speedway (D9800), Austin, TX 78712-1823, USA
\vskip 8pt 
$^{b}$Department of Communication Science and Disorders,\\ 
University of Pittsburgh,\\
4028 Forbes Tower, Pittsburgh, PA 15260, USA
\end{center}

\vskip 3mm
\begin{center}
{\Large{\bf Abstract}} 
\end{center}
Understanding the dynamics of functional brain connectivity patterns 
using noninvasive neuroimaging techniques
is an important focus in human neuroscience. 
Vector autoregressive (VAR) processes and Granger causality analysis (GCA) have been extensively used for such purposes. 
While high-resolution multi-subject neuroimaging data are routinely collected now-a-days, 
the literature on VAR models has remained heavily focused on small-to-moderate dimensional problems and single-subject data. 
Motivated by these issues, 
we develop a novel Bayesian random effects VAR model for high-dimensional multi-subject panel neuroimaging data. 
We begin with a model that 
structures the VAR coefficients as a three-way tensor, 
then reduces the dimensions by applying a Tucker tensor decomposition. 
A novel sparsity-inducing shrinkage prior 
allows data-adaptive rank and lag selection.
We then extend the approach to a novel random model for multi-subject data 
that carefully avoids the dimensions getting exploded with the number of subjects but also flexibly accommodates subject-specific heterogeneity. 
We design a Markov chain Monte Carlo algorithm for posterior computation. 
Finally, GCA with posterior false discovery control is performed on the posterior samples. 
The method shows excellent empirical performance in simulation experiments. 
Applied to our motivating functional magnetic resonance imaging study, 
the approach allows the directional connectivity of human brain networks to be studied in fine detail, 
revealing meaningful but previously unsubstantiated  cortical connectivity patterns.

\vspace{3mm}
\baselineskip=12pt

\vspace{3mm}
\baselineskip=12pt
\noindent\underline{\bf Key Words}: 
Bayesian methods, 
Granger causality,  
Neuroimaging data, 
Tensor decomposition, 
Shrinkage priors, 
Vector autoregressive processes

\par\medskip\noindent
\underline{\bf Short/Running Title}: Vector Autoregressive Models for Granger Causality

\par\medskip\noindent
\underline{\bf Corresponding Author}: Abhra Sarkar (abhra.sarkar@utexas.edu)

\pagenumbering{arabic}
\setcounter{page}{0}
\baselineskip=16pt

\newpage

\section{Introduction}
{\bf Scientific Background:}
The human brain is an incredibly complex organ, both in its structure and its functionality. 
Its constituent regions are functionally specialized and show common patterns of activity even at rest.
It has long been recognized that these connections are often directional in nature \citep{ding2006granger}. 
Assessing these connectivity patterns 
is crucial for understanding how the brain works \citep{menon2010saliency}.

Techniques for evaluating the effects of one brain region on another include stimulating one region and investigating the outcome in the second \citep{bressler2011wiener}.
The utility of such techniques is, however, diminished by 
high levels of convergence (one neuron receiving input signals from many others) and divergence (one neuron sending output signals to many others)  
as well as high levels of feedback in intracranial communication 
\citep{man2013neural}. 
\cite{bressler2011wiener} described the benefits of 
analyzing the relations between time series recordings of neural activity 
for determining causal relationships between different regions of interest (ROIs) instead.
One major benefit of such inference over stimulation based approaches {is} that 
recordings of neural activity can be obtained relatively easily and also importantly noninvasively 
using electroencephalography (EEG), magnetoencephalography (MEG), or, 
as we will use in our study, functional magnetic resonance imaging (fMRI). 

During resting state fMRI data acquisition, participants are presented with no overt stimuli, and functional connectivity is inferred based on the similarity of the time series from the given ROIs \citep{biswal1995functional, fox2007spontaneous}. 
Using whole-brain parcellations, we can identify functional connectomes across hundreds of regions of interest \citep{smith2013functional, craddock2012whole, glasser2016multi, schaefer2018local}.
Estimation of such fine-grained whole-brain connectomes can benefit from datasets with multiple participants \citep{smith2013resting,sudlow2015uk} and hours of data per participant \citep{poldrack2015long,gordon2017precision}.

{\bf Brain Connectivity via Granger Causality Analysis (GCA):} 
One way to determine the directional connectivity of different brain regions 
from fMRI time series data is via Granger Causality Analysis (GCA).
Introduced by \cite{granger1963economic}, the concept of Granger causality (GC) proposes that there is causal influence between two univariate time series if the prediction of one series is improved by knowledge of the other.
GCA for resting state fMRI data is promising given the dominance of fMRI within neuroimaging \citep{roebroeck2005mapping}. 
A potential complication is that
the hemodynamic response function (HRF), which links neural activity with the fMRI signal, characterizes changes in the blood oxygenation level dependent (BOLD) signal over time, and typically peaks around three to five seconds after stimulus presentation \citep{martindale2003hemodynamic,hirano2011spatiotemporal}. 
\cite{bressler2011wiener} 
have, however, demonstrated the efficacy of GCA for fMRI data 
through a thorough review of numerous studies 
demonstrating the robustness of GCA to HRF artifacts 
\citep{deshpande2010effect, 
david2008identifying, vakorin2007inferring}.
Despite its promise, however, GCA has mostly been used 
in small scale brain networks due to the difficulty of obtaining accurate estimates in high-dimensional problems \citep{wang2020large}.

{\bf GCA via Vector Autoregressive (VAR) Models:} 
Classically, GCA has been performed by 
comparing separately fitted reduced and full models via F-tests 
\citep{granger1963economic}. 
{To avoid recovering spurious effects through joint dependencies, 
e.g., region S1 in the brain appearing to influence S2 when in truth 
a third region S3 is the driving causal influence for both S1 and S2, 
it is generally preferable to consider pairwise-conditional causalities, 
which account for the influence of 
all by-standing regions.} 
Estimating pairwise conditional GC can, however, become very expensive for even a moderate number of brain regions.

When high-resolution data with fine parcelizations is available,
simply limiting the scope of the GCA to a portion of the brain network is not a good solution to the aforementioned problem.
\cite{roebroeck2011identification} demonstrated that exploring causal relationships within a partial brain network leads to the generation of pseudo-connections caused by confounding from un-modeled regions that are connected to the ones included in the network.
Partial Granger causality 
helps to remove the effects of unobserved confounders 
but does not scale well for large brain networks \citep{geweke1984measures, ding2006granger}.
Recent works addressing the difficulty of performing GCA for full brain networks focus mainly on pruning the indirect effects between regions \citep{wang2020large} 
but do not reduce the number of VAR model fits necessary to obtain the GC estimates.
\cite{barnett2014mvgc} offer an alternative that avoids fitting separate VARs for each pairwise analysis by using the cross-power spectral density (CPSD) of the full VAR process. 
CPSD, however, 
requires knowing the transition coefficients of the VARs, which itself can become difficult to estimate if the number of network nodes approaches or is larger than the number of time series observations.

{\bf Statistical Approaches to VAR Processes:}
The traditional VAR(L) process models a $K$-dimensional time series $\{\by_{t}\}_{t=1}^{T}$ as a linear combination of $L$ lagged observations $\{\by_{t-1},\dots,\by_{t-L}\}$ plus error.
The regression coefficients associated with the $L$ lagged observations are all $K \times K$ matrices 
and hence the model size ($K^{2}L$) grows quadratically with $K$. 
When $K$ is relatively large compared to $T$, 
conventional estimation methods such as ordinary least squares (OLS) suffer from ill-conditioned design matrices, 
and 
can fail to provide an estimate. 
Frequentist approaches to high-dimensional VARs \citep{nicholson2020high, chen2018inference} have thus used penalized likelihood based methods such as the LASSO \citep{tibshirani1996regression} to address the poor conditioning problem.

Bayesian approaches \citep{litterman1986forecasting}
implemented via iterative Markov chain Monte Carlo (MCMC) schemes can also provide promising alternatives.
Shrinkage priors have been shown to improve performance for large VARs by 
mitigating overfitting  \citep{banbura2010large, giannone2015prior, miranda2018bayesian, billio2019bayesian}.
They are useful for when each variable 
is influenced by a small number of others,  
and for lag selection \citep{ahelegbey2016sparse, kastner2020sparse, song2011large}. 
Shrinkage priors applied directly to the VAR coefficients, however, do not reduce the number of parameters needed to be sampled in each MCMC iteration 
and hence may not be adequate for high-dimensional problems. 
The model we introduce in this paper combines the advantages of Bayesian shrinkage with structural dimension reduction to address these issues.

In addition, we propose a random-effects VAR model for performing GCA on panel VARs with multi-subject data $\{\by_{i,t}\}_{i=1,t=1}^{n,T}$.
In brain connectivity analyses, 
healthy individuals have been found to have common connectivity patterns with low levels of heterogeneity \citep{damoiseaux2006consistent}. 
Random effects models let 
shared parameters capture the common patterns 
while individual specific effects explain the random heterogeneity. 
The literature on random effects VARS remains extremely sparse. 
Existing methods estimate shared and random effects
by restricting their supports to be orthogonal \citep{skripnikov2019regularized}. 
An alternative approach to accommodate individual heterogeneity is to cluster subjects with similar VAR parameters \citep{kundu2021non}. 
Both these strategies could be restrictive in large $K$ problems. 
To our knowledge, there are currently no random effects VARs for panel data in which all series in the VAR have systematic fixed and random effects components. 
The model introduced in this paper also attempts to address these gaps.

{\bf Our Proposed Approach:}
{The Bayesian tensor decomposed VAR (BTDVAR) model we introduce in this paper can provide accurate estimation of very large VARs.
The resultant posterior samples also provide an easy method for performing GCA.}

To begin with, the BTDVAR structures the transition matrices as a three-way tensor, 
then decomposes it \citep{tucker1966some} 
into a smaller core tensor and three factor matrices using a Tucker decomposition \citep{tucker1966some},
obtaining a massive reduction in the number of parameters. 
A novel horseshoe Gamma-Gamma prior, adapted from \cite{polson2012local} and \cite{bhattacharya2011sparse}, 
allows data-adaptive rank selection 
as well as automatic lag selection.

We next extend the BTDVAR to BPTDVAR, a novel random effects panel VAR model.
Our main motivation is to find shared brain connectivity patterns while also accommodating subject heterogeneity \citep{beckmann2003general}. 
The BPTDVAR assigns each individual 
fixed and random intercepts \citep{epskamp2018gaussian} 
as well as fixed and random transition matrices, 
containing shared and subject-specific information. 
Taking a carefully designed structured approach again, 
we avoid a rapid multiplicative increase in the number of model parameters
by allowing only the factor matrices for the `row' dimension 
in Tucker decomposed transition matrices to have 
shared and an individual components  
while still obtaining a highly flexible model 
for capturing subject-specific heterogeneity in effect sizes as well as GC networks.

We design MCMC algorithms to sample from the posteriors of the tensor components, 
and perform GCA on the reconstructed VAR parameters while controlling for false positives and false negatives \citep{muller2004optimal}.  
Simulation studies show that the proposed B(P)TDVAR models provide 
excellent estimates even when $K$ is relatively large compared to $T$.

Applied to human neuroimaging data, our approach recovers meaningful connectivity patterns within and between brain sub-regions, 
some corroborating prior studies, 
while some others, to our knowledge, being novel findings that have not previously been scientifically rigorously validated.
Our models also identify multiple lags as being influential, 
as opposed to the single-lag models that are currently standard for fMRI analysis using VAR models \citep{martinez2004concurrent}.

An innovative approach to (single-subject) VARs that also combines the power of shrinkage priors with structured dimension reduction 
was recently introduced in \cite{yuwen2022low} where they restricted the non-zero coefficients to appear only on the union of a few spanning trees. 
Tensor decomposed {(single-subject)} VARs have previously been considered in the literature. 
\cite{wang2021high} used a Tucker decomposed VAR
for a 40-variate financial time series data set 
but required that the Tucker ranks be fixed prior to the analysis. 
\cite{zhang2021bayesian} analyzed a multi-subject 27-variate fMRI data set, 
modeling the VAR parameters as mixtures of time varying subsets of a collection of parallel factor (PARAFAC) decomposed atoms 
but fitted the model separately to each subject.  
The main focus of this article, however, is on Tucker decomposed random effects VARs 
to estimate shared and individual  
GC brain networks from multi-subject 200-variate longitudinal fMRI data. 
Our carefully designed structured approach flexibly accommodates individual heterogeneity while addressing an order-of-magnitude more massive dimensionality challenges. 
Our novel shrinkage priors and our MCMC based implementation 
also allow data-adaptive on-the-fly selection of the Tucker ranks as well as the selection of the important lags, 
while also allowing for straightforward uncertainty quantification.

{\bf Outline of the Article:}
Section \ref{sec: Data} describes our motivating fMRI data set. 
Section \ref{sec: prelims} presents some background on VARs and tensor factorization techniques. 
Section \ref{sec: method} presents the single-subject BTDVAR, 
its random effects multi-subject panel data extension BPTDVAR, 
our global-local priors for rank and lag selection, 
and an overview of posterior computation. 
Section \ref{sec: GCausality} describes GCA via posterior FPR and FNR control.
Section \ref{sec: Simulations MP} demonstrates the efficacy of our proposed approach on simulated single and multi-subject panel data. 
Section \ref{sec: Application} describes the results obtained by our method applied to the fMRI dataset.
Section \ref{sec: Discussion} concludes with a discussion.

\section{The 7T-HCP-fMRI Data Set} \label{sec: Data}
\begin{figure}[!ht]
 \begin{subfigure}{0.5\textwidth}
      \centering
      \includegraphics[width=\linewidth]{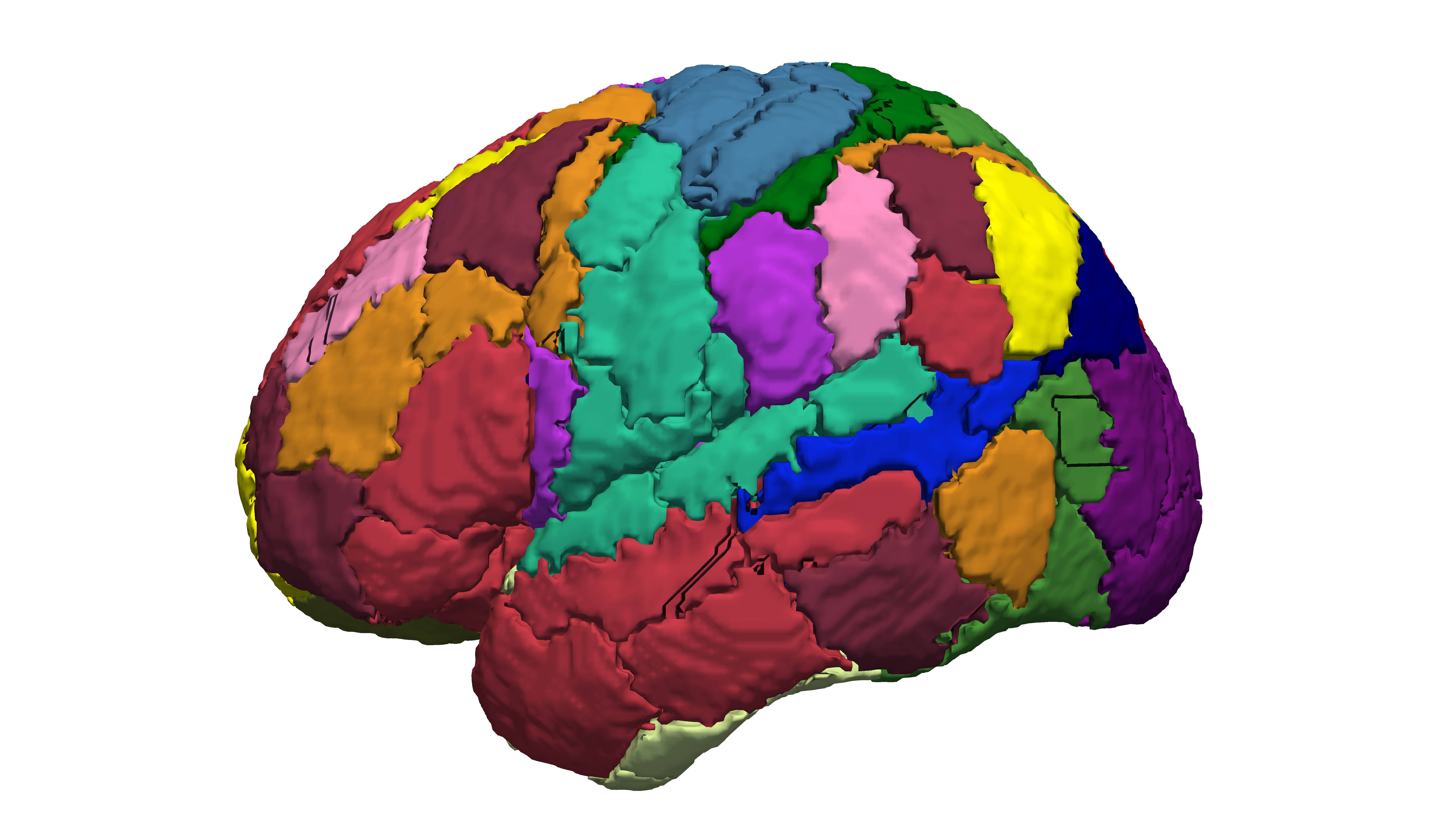}
      \caption{Left lateral surface.}
    \end{subfigure}%
  \begin{subfigure}{0.5\textwidth}
      \centering
      \includegraphics[width=\linewidth]{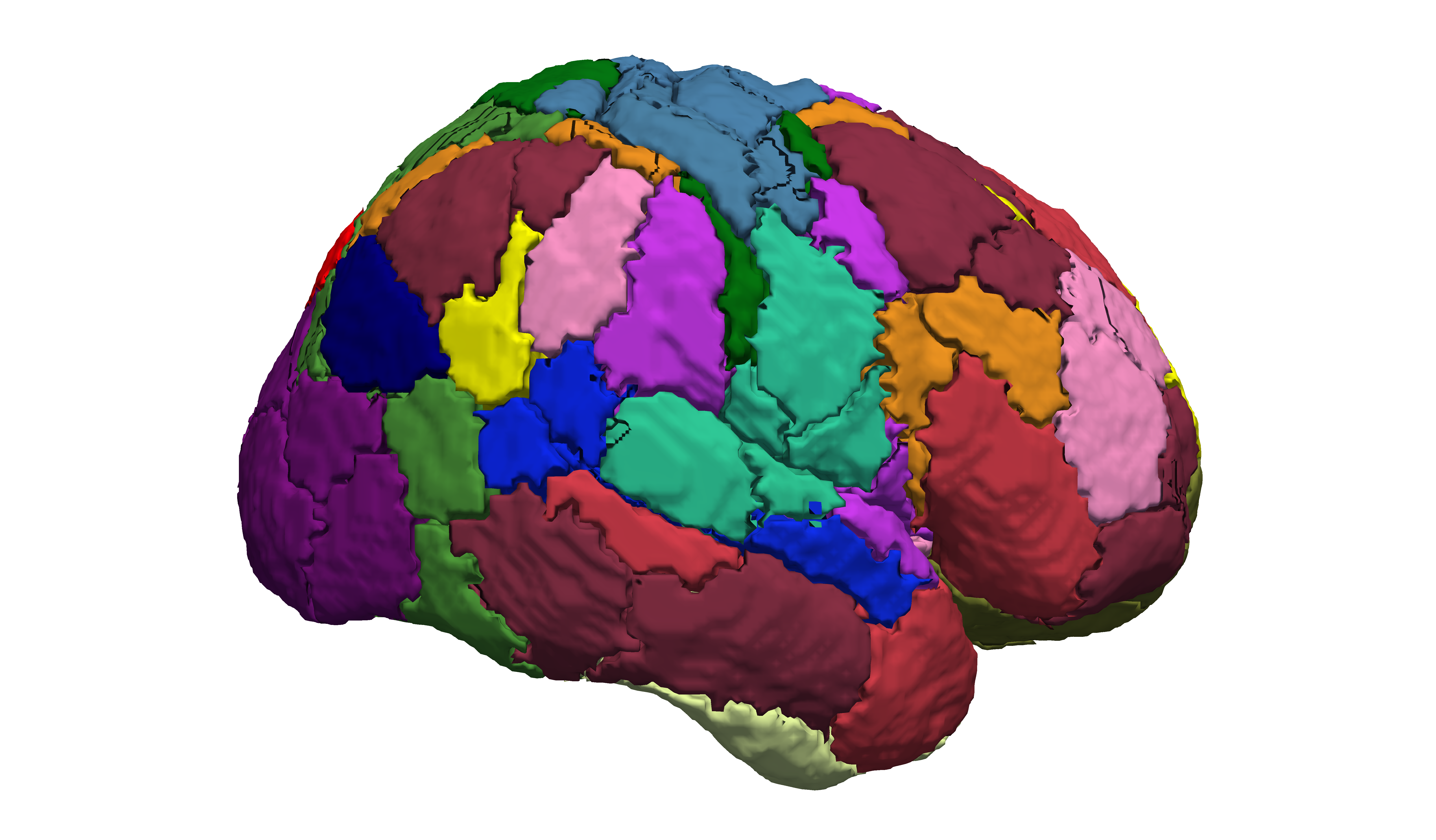}
     \caption{Right lateral surface.}
    \end{subfigure}
  \caption{Cortical regions of interest from the Schaefer 200 brain atlas. 
  Color denotes network membership of each region utilizing the Yeo 2011 17-Network schema.}
  \label{fig: Schaefer_atlas}
\end{figure}

The Human Connectome Project (HCP) is one of the most popular large-scale public neuroimaging datasets, with over 1000 participants having resting state fMRI data collected with 3 Tesla MRI and almost 200 participants with 7 Tesla resting state fMRI data \citep{smith2013resting,uugurbil2013pushing,elam2021human}. 
Both the 3 Tesla and 7 Tesla HCP data include approximately 60 minutes of resting state fMRI data per participant. 
The largest difference between the two is in their spatial resolutions: While the 3 Tesla data were acquired with 2 mm isotropic voxel sizes ($2\times2\times2$ mm), the higher signal-to-noise ratio at the stronger 7 Tesla magnetic field allowed for 1.25 mm isotropic voxels, 
enabling finer-grained investigations of functional connectivity.
In our analyses, we are particularly interested in studying the within and between network connectivity under the Yeo 17-network schema \citep{yeo2011organization} (Figure \ref{fig: Schaefer_atlas}).

\vskip -0pt

Modeling these sub-networks separately may allow us to `divide-and-conquer' the dimensionality challenges but will only allow the study of within region patterns. 
However, as discussed in the Introduction,
estimation of fine-grained whole-brain connectomes benefits from 
joint analysis of sub-networks \citep{roebroeck2011identification}, multi-subject data \citep{smith2013resting,sudlow2015uk}, 
and hours of data per participant \citep{poldrack2015long,gordon2017precision}.
We will thus analyze data for all 17 sub-networks jointly across multiple subjects using our proposed BPTVARs via an MCMC based implementation. 
This is a massive modeling and computational endeavor 
but will allow us to simultaneously study both within and between sub-network connectivity patterns, 
borrowing information across different brain regions and subjects, 
while also allowing straightforward assessment of uncertainty.

\section{Preliminaries} \label{sec: prelims}

\subsection{VAR Basics} \label{sec:VAR}  

Given a multivariate auto-regressive time series with $K$ series and $T$ total observations, a VAR process with $L$ lags, referred to henceforth as VAR(L), is specified as 
\vspace{-5ex}\\
\be
    \by_{t} = \bnu + \bA_{1} \by_{t-1} + \bA_{2} \by_{t-2} + \dots + \bA_{L} \by_{t-L} + \bvarepsilon_{t}, ~~~~~\bvarepsilon_{t} \sim \MVN(\bzero, \sigma_{\varepsilon}^{2}\bI_{K}),
    \label{eq:time series}
\ee
\vspace{-5ex}\\
where $\by_{t}$ and $\bvarepsilon_{t}$ are $K \times 1$ vectors and each $\bA_{\ell}=((a_{\ell,j,k})), \ell=1,\dots,L$, is a $K\times K$ matrix.
With some algebraic manipulations, any VAR(L) process with $L>1$ can be rewritten as a VAR(1) process,
\vspace{-5ex}\\
\be
    \bY_{t} = \bV + \bA \bY_{t-1} + \bU_{t}, ~~~\text{where}
    \label{eq:var1}
\ee
\vspace{-10ex}\\
\bse
\bY_t = \begin{bmatrix}
    \by_{t}\\
    \by_{t-1}\\
    \vdots\\
    \by_{t-L+1}
\end{bmatrix}, 
\quad
\bV = \begin{bmatrix}
    \bnu\\
    0\\
    \vdots\\
    0
\end{bmatrix}, 
\quad
\bA = \begin{bmatrix}
    \bA_{1} & \bA_{2} & \dots & \bA_{L-1} & \bA_{L}\\

    \bI_{K} & \bzero & \dots & \bzero & \bzero\\
    \bzero & \bI_{K} & & \bzero & \bzero\\
    \vdots & \vdots & \ddots & \vdots & \vdots\\
    \bzero & \bzero & \dots & \bI_{K} & \bzero
\end{bmatrix}, 
\quad
\bU_{t} = \begin{bmatrix}
    \bvarepsilon_{t}\\
    0\\
    \vdots\\
    0
\end{bmatrix}.
\ese
\vspace{-5ex}\\

Multiplying both sides of (\ref{eq:var1}) by $\bJ = [\bI_{K}:\bzero:\dots:\bzero]$ gives 
\vspace{-5ex}\\
\be
    & \by_{t} = \bnu +  \bB\bx_{t} + \bvarepsilon_{t},
    \label{eq:likelihood}
\ee
\vspace{-5ex}\\
where $\bx_{t}^{KL \times 1} = (\by_{t-1}\trans,\dots,\by_{t-L}\trans)\trans$, and $\bB^{K\times KL} = \bJ\bA = [\bA_{1}, \bA_{2},\dots,\bA_{L}]$.
The VAR process defined ins \eqref{eq:time series} and \eqref{eq:var1} is stable if all eigenvalues of the companion matrix $\bA$ have modulus less than $1$ \citep{lutkepohl2005new}.

As discussed in the Introduction, one of the main challenges of VAR processes is how quickly the number of parameters $K^2\times L$ grows with the dimension $\by_t$. 
When $K$ is very large, 
accurate estimation of the VAR parameters using traditional methods can become prohibitively costly, 
or, if there are relatively few observations, even impossible.  
This indeed is the case with our motivating fMRI data, 
in which the top cortical layer of the brain is partitioned into $K=200$ regions of interest (ROIs).

\subsection{Tensor Basics}\label{sec: tensor_basics}
In this section, we briefly review some tensor operations and properties relevant for the rest of this paper.
For a more complete review, please refer to \cite{de2000multilinear,kolda2009tensor}, \cite{rabanser2017introduction}, etc. 
We denote vectors by lowercase bold letters $\bx,\by$, etc., matrices by capital bold letters $\bX, \bY$, etc., and tensors by curly script capital letters $\cX, \cY$, etc. 

An $N$-dimensional tensor, also known as an $N\th$-order or an $N$-mode tensor, is defined as $\cX \in \mathbb{R}^{I_{1} \times I_{2} \times \dots \times I_{N}}$. 
Its mode-n fibers are vectors obtained by fixing every index save for the index corresponding to the $n\th$ dimension.  
Its mode-n matricization $\bX_{(n)} \in \mathbb{R}^{I_{n} \times (I_{1}  I_{n-1}   I_{n+1}   I_{N})}$ is the unfolding of $\cX$ into an $I_{n} \times (I_{1}  I_{n-1}   I_{n+1}   I_{N})$ matrix whose columns are the mode-n fibers of $\cX$.
Matricization is an important way to link tensors to generalized properties of matrices.

Some common tensor operations are defined as follows.
The mode-n product of a tensor $\cX \in \mathbb{R}^{I_{1} \times I_{2} \times \dots \times I_{N}}$ by a matrix $\bU \in \mathbb{R}^{J_n \times I_n}$, 
denoted $\cX \times_{n} \bU$, 
is a $(I_{1} \times I_{2} \times \dots \times I_{n-1} \times J_{n} \times I_{n+1} \times \dots \times I_{N})$ tensor whose entries are given by 
\vspace{-5ex}\\
\bse
& (\cX \times_{n} \bU)_{i_1i_2 \dots i_{n-1} j_n i_{n+1} \dots, i_{N}} = \sum_{i_n} x_{i_1i_2 \dots i_{n-1} j_n i_{n+1} \dots, i_{N}} u_{j_n i_n}.
\ese
\vspace{-5ex}\\
The outer product of two 1-mode tensors, or vectors, $\bu^{(1)} \in \mathbb{R}^{I_1}$ and $\bu^{(2)} \in \mathbb{R}^{I_2}$ is a 2-mode tensor $\bA \in \mathbb{R}^{I_1 \times I_2}$ where $a_{i_{1}i_{2}} = u^{(1)}_{i_1} u^{(2)}_{i_2}$. 
The outer product of $N$ vectors is defined analogously, and is written as $\bu^{(1)} \circ \bu^{(2)} \circ \cdots \circ \bu^{(N)}$.
The Kronecker product of two matrices $\bA \in \mathbb{R}^{I_{1} \times I_{2}}$ and $\bB \in \mathbb{R}^{I_{3} \times I_{4}}$ is a $\mathbb{R}^{I_{1}I_{3} \times I_{2}I_{4}}$ block matrix where 
\vspace{-6ex}\\
\bse
\bA \otimes \bB = \begin{bmatrix}
    &a_{11} \bB, &\dots, &a_{1I_{2}} \bB,\\
    &\vdots &\ddots &\vdots\\
     &a_{I_{1}1} \bB, &\dots, &a_{I_{1}I_{2}} \bB
\end{bmatrix}.
\ese
\vspace{-4ex}

The Tucker decomposition is crucial to the models introduced in this paper. 
Given a N-mode tensor $\cX\in \mathbb{R}^{{I_{1} \times I_{2}} \times \dots \times I_{N}}$, the Tucker decomposition of $\cX$ is
\vspace{-5ex}\\
\bse
\textstyle\cX = \sum_{r_{1}=1}^{R_{1}} \sum_{r_{2}=1}^{R_{2}}\cdots \sum_{r_{N}=1}^{R_{N}} g_{r_{1}\cdots r_{N}} \bbeta_{1,\cdot,r_{1}} \circ  \bbeta_{2,\cdot,r_{2}} \circ \cdots \circ  \bbeta_{N,\cdot,r_{N}},
\ese
\vspace{-5ex}\\
where $g_{r_{1}\cdots r_{N}}$ are elements of the core tensor $\cG$, and $\bbeta_{n,\cdot,r_{n}}$ is the $r_{n}$-th column of the factor matrix $\bbeta_{n}$. 
The Tucker ranks $(R_{1},...,R_{N})$ of $\cX$ refers to the number of columns in each factor matrix in the Tucker decomposition of $\cX$.
The $n$-ranks of $\cX$, on the other hand, are the matrix ranks of the mode-n matricization of $\cX$, i.e., 
$\wt{r}_{n} = rank_{n}(\cX) = rank(\cX_{(n)})$.
The $n$-ranks of $\cX$ satisfy a set of inequalities which state that no one $n$-rank can be greater than the product of all others \citep{tucker1966some}. 
For any tensor, its Tucker ranks and $n$-ranks and are not necessarily equal, 
but must satisfy $\wt{r}_{n} = \min\{R_{n},R_{1} \times \cdots \times R_{n-1} \times R_{n+1} \times \cdots \times R_{N}\}$ \citep{de2000multilinear}.

The PARAFAC representation is obtained as a restrictive special case when the ranks $R_{1} = \dots = R_{n}$ are all equal 
and hence $\cG$ is diagonal.

\section{Bayesian Tensor Decomposed VAR Processes} \label{sec: method}

\subsection{Tucker Decomposed VAR} \label{sec: BTDVAR}

\begin{figure}[!h]
    \centering
    \includegraphics[width=0.75\linewidth, trim={1cm 1cm 1cm 1cm},clip=true]{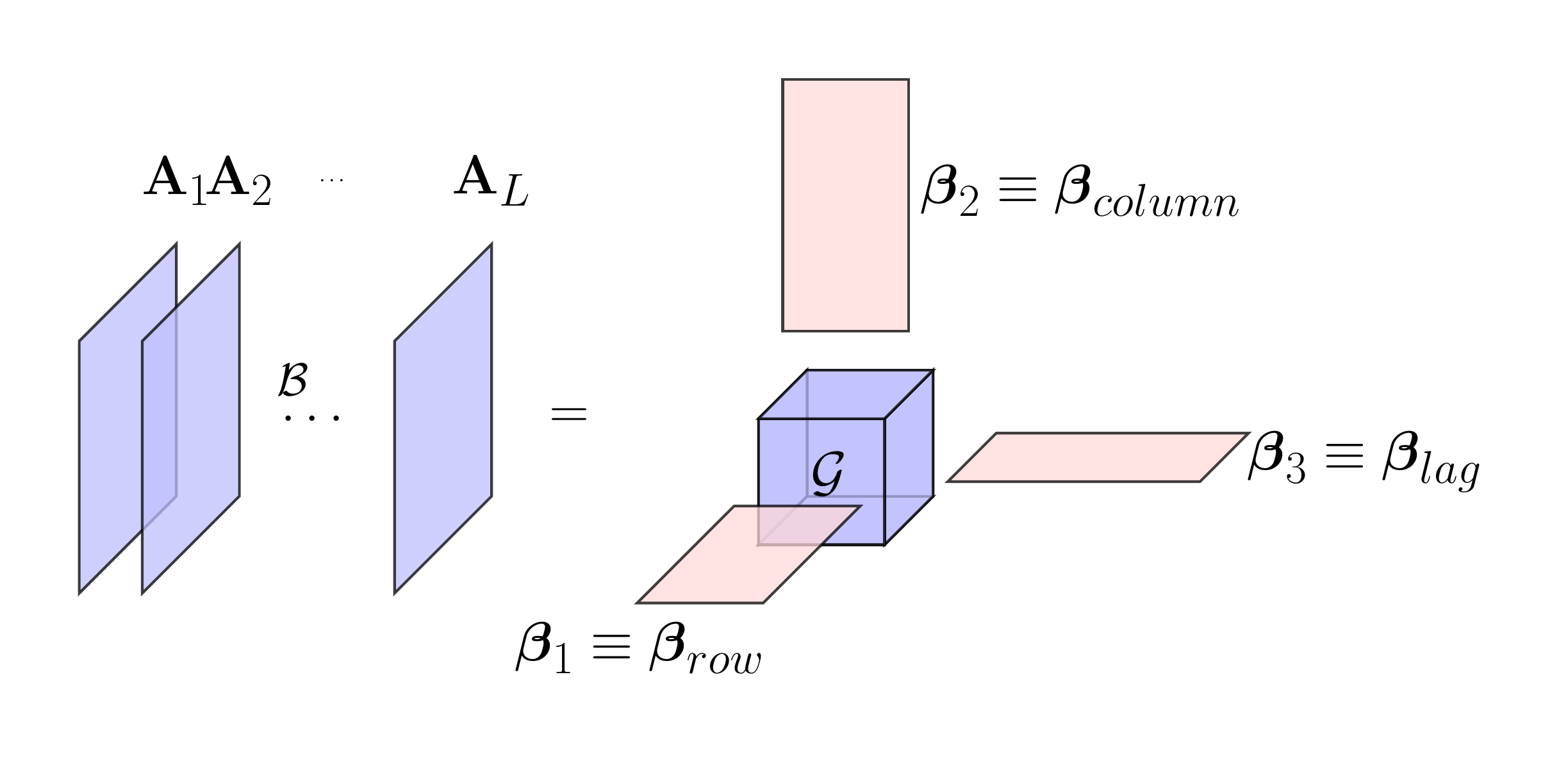}
    \caption{
     In the TDVAR model, the VAR parameter matrices for different lags, $A_{\ell}, \ell=1,\dots,L$, are stacked into a 3-way tensor $\cB^{K \times K \times L}$ such that each $A_{\ell}$ is the $\ell\th$ slice of $\cB$. The tensor $\cB$ is then Tucker decomposed into a 3-way core tensor $\cG^{R_{1} \times R_{2} \times R_{3}}$ and associated factor matrices $\bbeta_{1}^{K \times R_{1}}, \bbeta_{2}^{K \times R_{2}}$ and $\bbeta_{3}^{L \times R_{3}}$. 
     Significant reduction in dimension is achieved when the size of the core tensor $\cG$ is much smaller than the size of the original tensor $\cB$, i.e., $\prod_{j=1}^{3}R_{j} \ll K^{2}L$. 
   }
    \label{fig:tucker}
\end{figure}

To combat the curse of dimensionality presented by high-dimensional VARs, we reduce the number of estimated parameters by applying a Tucker decomposition to a 3-mode tensor of stacked VAR transition matrices, then estimating the components of the decomposition, rather than the original parameters directly (Figure \ref{fig:tucker}).
We refer to this model as the Tucker decomposed VAR (TDVAR).

Using the notation from \cite{rabanser2017introduction}, we define the 3-mode parameter tensor for a VAR$(L)$ process with $K$ component series as $\cB^{K \times K \times L} = [[\bA_{1}, \dots, \bA_{L}]]$, where $\bA_{\ell}$ are the frontal slices of $\cB$. 
In Figure 1, $\cB$ is rotated to show the $\bA_{\ell}$'s laterally for clarity of viewing. 
The Tucker decomposition of $\cB$ can then be expressed as 
\vspace{-5ex}\\
\be
\textstyle\cB = \sum_{r_{1}=1}^{R_{1}} \sum_{r_{2}=1}^{R_{2}}\sum_{r_{3}=1}^{R_{3}} g_{r_{1}r_{2}r_{3}} \bbeta_{1,\cdot,r_{1}} \circ  \bbeta_{2,\cdot,r_{2}} \circ  \bbeta_{3,\cdot,r_{3}},
\label{eq: tucker}
\ee
\vspace{-5ex}\\
where $g_{r_{1}, r_{2}, r_{3}}\in \cG$ is the element of the core tensor indexed by $(r_{1},r_{2},r_{3})$, and $\bbeta_{1,\cdot,r_{1}}$ is the $r_{1}$-th column of the factor matrix $\bbeta_{1}$. Similarly, $\bbeta_{2,\cdot,r_{2}}$ and $\bbeta_{3,\cdot,r_{3}}$ are columns $r_{2}$ and $r_{3}$ of $\bbeta_{2}$ and $\bbeta_{3}$ respectively. 
The $\circ$ operator denotes the outer product.

While (\ref{eq: tucker}) makes clear the role of the Tucker ranks, 
$\{R_{1}, R_{2}, R_{3}\}$, an equivalent and more convenient representations of the Tucker decomposition is 
\vspace{-5ex}\\
\be
\cB = \cG \times_{1} \bbeta_{1} \times_{2} \bbeta_{2} \times_{3} \bbeta_{3},
\label{eq:tuckermodemult}
\ee
\vspace{-5ex}\\
where $\times_{n}$ is the mode-$n$ tensor product. 
The original parameters of the VAR process $\bB$ can be recovered from $\cB$ by taking the mode-1 matricization of $\cB$,
$\bB = \cB_{(1)}  = \bbeta_{1} \cG_{(1)} (\bbeta_{3} \otimes \bbeta_{2})^{\trans}$.
Thus, (\ref{eq:likelihood}) can be rewritten as 
\vspace{-5ex}\\
\be
\by_{t} =\bnu + (\bbeta_{1} \cG_{(1)} (\bbeta_{3} \otimes \bbeta_{2})^{\trans}) \bx_{t} + \bvarepsilon_{t}.
\label{eq:Tucker VAR}
\ee
\vspace{-5ex}

Tucker decomposition techniques to estimate large tensor valued parameters have been proposed before. 
See, e.g., \cite{spencer2022parsimonious, li2018tucker, ahmed2020tensor} in a regression context, 
and \cite{wang2021high} in a VAR context. 
A major advantage of the Tucker decomposition is that, 
unlike the PARAFAC \citep{guhaniyogi2017bayesian, chen2019bayesian,papadogeorgou2021soft}, 
it does not place constraints on the structure of the core tensor, which allows for very flexible linear combinations of the factor matrix components. 
In addition, it allows for different Tucker ranks $R_{1}, R_{2}, R_{3}$ along different directions of the decomposition. 
For a high-dimensional problems with $K \gg L$, 
the ability to have different amounts of compression for the rows and columns of $\bB$ versus the lags is crucial.
In our simulation studies, the TDVAR is shown to have accurate parameter recovery even in extremely challenging settings where $K$ is very large 
and $T$ is comparatively small.

\subsubsection{Identifiability}\label{sec: identifiability}

The core tensor and the factor matrices in a Tucker decomposition are not separately identifiable without additional complex orthogonality restrictions on these components. 
Existing Bayesian literature utilizing the simpler PARAFAC decomposition for tensor-valued linear regressions are able to recover identifiable parameters without requiring the individual components of the decomposition be identifiable \citep{guhaniyogi2017bayesian, chen2019bayesian}.
Since the emphasis in this article is on the accurate estimation and inference of $\bB$, which are identified as is \citep{spencer2022parsimonious}, 
we avoid imposing identifiability restrictions on the tensor components.
Our simulation studies suggest that the non-identifiability of $\bbeta_{j}$ and $\cG$ does not cause convergence issues,  
and greatly simplifies the design of our MCMC algorithm.

\subsubsection{Lag Selection}\label{sec: lagselection}

An advantage of our proposed TDVAR model is that the geometry of the Tucker decomposition naturally lends itself to lag selection. 
Given a VAR model as defined in (\ref{eq:time series}), 
the $\ell\th$ lag $\by_{t-\ell}$ has no influence on $\by_{t}$ if its corresponding parameter matrix $\bA_{\ell}$ contains no non-zero elements. 
Under the BTDVAR model, each $\bA_{\ell}$ is the $\ell\th$ frontal slice of the mode-3 parameter tensor $\cB$. 
In lemma \ref{lem: lag selection}, we outline the requirements for the model to generate zero-valued $\bA_{\ell}$.

\begin{Lem} \label{lem: lag selection}
$\bA_{\ell}$ in (\ref{eq:time series}) has no non-zero elements if the $\ell\th$ row of the factor matrix $\bbeta_{3}$ in (\ref{eq: tucker}) also contains no non-zero elements. 
The converse is also true provided at least one transition matrix $\bA_{p\neq \ell}$ includes one non-zero value, and all columns of each $\bbeta_{j}$ make a non-zero contribution to the reconstruction of $\cB$. 
\end{Lem}

The proof follows from different mathematical representations of the Tucker decomposition. 
Details are provided in section \ref{sec: proof} of the supplementary materials.

\subsubsection{Low Rank Estimation}\label{sec:lowrankVAR}
As discussed in Section \ref{sec: tensor_basics}, 
the $n$-rank of $\cB_{(j)}$ is constrained by the Tucker rank of $\bbeta_{j}$, $\wt{r}_{j}\leq R_{j}$. 
In application, this means that the number of linearly independent rows in the parameter matrix $\bB$ is bounded by the number of columns in $\bbeta_{1}$, 
the number of linearly independent columns in each $\bA_{\ell}$ is bounded by the number of columns in $\bbeta_{2}$,
and the number of linearly independent transition matrices $vec(\bA_{\ell})$ within $\bB$ is bounded by the number of columns in $\bbeta_{3}$. 
Low rank VAR parametrization is thus obtained when $R_{1}, R_{2} < K$ and $R_{3}<L$. 
Such low rank parametrizations have been shown to be useful in cases when 
groups of variables move in a concerted manner
\citep{basu2019low, alquier2020high}. 
This scenario is plausible for brain networks  
since regions belonging to the same network often tend to cross-fire in a collaborative manner \citep{sharaev2016effective}.
Importantly, the novel shrinkage priors we employ in Section \ref{sec: priors} allows data-adaptive rank and lag selection, 
precluding the need to a-priori know the appropriate rank structure of $\bB$, 
and enabling our approach to exploit low rank structures in the data with minimal prior knowledge and supervision.

\subsection{Panel TDVAR with Random Effects} \label{sec: BPTDVAR}

We now introduce an intuitive extension of the TDVAR model which fits random effects VARs in the presence of multi-subject data $\{\by_{i,t}\}_{i=1,t=1}^{n,T}$. 
Our applied motivation is again the accurate recovery of brain connectivity patterns from resting state fMRI data, now from multiple subjects. 
A challenge here is  
that the variations in signal strength between the brains of different participants can obscure the common processing network shared by the participants \citep{ramsey2010six}. 
To solve this problem, we design a random effects VAR model in which
the fixed effects would identify a common structure across the subjects 
while random effects would accommodate variations in signal strength and connectivity patterns across the subjects.
We call this model the Panel TDVAR (PTDVAR).

The literature on VARs for multi-subject data 
is quite sparse. 
Some recent works addressing moderate dimensional problems
have focused on drawing distinctions between a subset of variables for which the connectivity is shared between all subjects, and a subset for which connectivity can vary between subjects \citep{chiang2017bayesian, skripnikov2019regularized}.
\cite{kundu2021non} took a clustering approach instead, 
quantifying shared information by forcing all subjects in the same cluster to share the same VAR parameters. 
Most similar to our proposal, \cite{gorrostieta2013hierarchical} considered each subject's VAR transition matrices to be the sums of a shared fixed and a subject-specific random  matrices.

In addition to fixed and random transition matrices, the random effects VAR we consider also allows for fixed and random intercepts. 
Let $i$ index data and parameters for different subjects.
Our random effects VAR model is defined as 
\vspace{-5ex}\\
\be
& \by_{i,t} = \bnu + \balpha_{i} + (\bB^{fixed} + \bB^{random}_{i}) (\bx_{i,t} - \balpha_{i}) + \bvarepsilon_{i,t}. \label{eq: random-effect}
\ee
\vspace{-5ex}\\
Here, the fixed effects parameters consist of a shared transition matrix $\bB^{fixed}$ 
and a shared intercept $\bnu$. 
The random effects parameters consist of subject-specific transition matrices $\bB^{random}_{i}$ and subject-specific random intercepts $\balpha_{i}$.
The corresponding coefficient tensors are denoted by $\cB^{fixed}$ and $\cB_{i}^{random}$ respectively. 
We also define $\bB_{i} = \bB^{fixed}+\bB_{i}^{random}$ and $\cB_{i} = \cB^{fixed} +\cB_{i}^{random}$.

In (\ref{eq: random-effect}) we subtract $\balpha_{i}$'s from from the $\bx_{i,t}$'s so that the mean of $\by_{i,t}$
takes the form of Equation (\ref{eq:random_eff_mean}), in which $\bnu$ and $\balpha_{i}$ separately estimable. 
{See Section \ref{sec:PTDVAR_derivation} in the supplementary materials for a derivation of this result.} 
\vspace{-5ex}\\
\be
\eE(\by_{i,t}) = (\bI_{K} - \bB_{i})^{-1} \bnu + \balpha_{i}.
\label{eq:random_eff_mean}
\ee
\vspace{-5ex}

For efficient estimation in high-dimensional settings, 
we develop the PTDVAR model 
which can simultaneously estimate both fixed effects $\bB^{fixed}$ and the random effects $\bB^{random}_{i}$ without multiplicatively increasing the number of model parameters. 
We require that the PTDVAR model 
satisfies the the following meaningful criteria. 
(a) The fixed transition is separately identified by the model, and captures the connectivity network common to all subjects in the panel. 
This shared connectivity network includes all the variables in the system, not just a subset of them.
(b) The random transitions are able to capture individual-specific effect size deviations as well as individual-specific connectivity pattern deviations from the fixed effects coefficients.
(c) The model estimated lag is shared between all subjects. 

To satisfy these criteria, the PTDVAR inherits the tensor decomposition structure of the TDVAR defined Equation (\ref{eq: tucker}), but separately estimates $\bbeta_{1}^{fixed}$ to be shared between all subjects, and $\bbeta_{1}^{random,i}$ which are unique to each subject.
At the VAR level, the the fixed and random transition matrices are recovered as
\vspace{-5ex}\\
\be
& \bB^{fixed} = \bbeta_{1}^{fixed} \cG_{(1)} (\bbeta_{3} \otimes \bbeta_{2})^{\trans},~~~~~\bB^{random}_{i} = \bbeta_{1}^{random, i} \cG_{(1)}   (\bbeta_{3} \otimes \bbeta_{2})\trans,
\label{eq: fixed-random-effects}
\ee
\vspace{-5ex}\\
where $\bbeta_{2}, ~ \bbeta_{3}, ~ \cG$ are shared between all subjects.
The shared components $\bbeta_{1}^{fixed}$, $\bbeta_{2}$, $\bbeta_{3}$, $\cG$ are estimated using data from all subjects, whereas $\bbeta_{1}^{i}$ is estimated using only data from the $i\th$ subject. 

It is clear from the definition of the PTDVAR model in (\ref{eq: random-effect}) that it 
satisfies criterion (a). 
While it may seem at first that individual-specific $\bbeta_{2}^{i}$'s would be necessary to satisfy criterion (b), 
we can see from Equation (\ref{eq: fixed-random-effects}) that if $\wt{r} = \textrm{row rank}(\cG_{(1)}   (\bbeta_{3} \otimes \bbeta_{2})\trans) = K$, then $\bbeta_{1}^{random, i}$ can be found for any arbitrary $\bB^{random}_{i}$.
For $\wt{r}<K$, we have $\textrm{row rank}(\bB^{random}_{i}) \leq \wt{r}$, resulting in a low rank random transition matrix that is suitable for modeling individual-level observations with high correlation between variables \citep{alquier2020high},
and is in accordance with the low rank VAR estimation objective presented in section \ref{sec:lowrankVAR}.
Since $\bbeta_{3}$ stores information regarding lag selection, allowing $\bbeta_{3}$ to be shared satisfies criteria (c).
Criteria (c) may appear restrictive 
but allowing each individual to have their own set of lags makes it practically very challenging to interpret the results. 

To gain further insights into model (\ref{eq: tucker}) and its robustness properties, 
note that the $(i_{1},i_{2},i_{3})\th$ element of $\cB_{i} = ((b_{i_{1},i_{2},i_{3}}^{i}))$ can be written as 
\vspace{-5ex}\\
\bse
&\textstyle b_{i_{1},i_{2},i_{3}}^{i} = \sum_{r_{1}}\sum_{r_{2}}\sum_{r_{3}} g_{r_{1},r_{2},r_{3}}\wt\beta_{1,i_{1},r_{1}}^{i}\beta_{2,i_{2},r_{2}}\beta_{3,i_{3},r_{3}} 
= \sum_{r_{1}} \wt\beta_{1,i_{1},r_{1}}^{i} \wt{b}_{i_{2},i_{3},r_{1}},\\
&\textstyle \text{where}~~~ \wt\beta_{1,i_{1},r_{1}}^{i} = (\beta_{1,i_{1},r_{1}}^{fixed} + \beta_{1,i_{1},r_{1}}^{random,i}), 
~~~\text{and}~~~\wt{b}_{i_{2},i_{3},r_{1}} = \sum_{r_{2}}\sum_{r_{3}} g_{r_{1},r_{2},r_{3}}\beta_{2,i_{2},r_{2}}\beta_{3,i_{3},r_{3}}. 
\ese
\vspace{-5ex}\\
The $(i_{1},i_{2},i_{3})\th$ element of the coefficient tensor $\cB_{i}$ can therefore be written as a mixture of unknown `bases' $\wt{b}_{i_{2},i_{3},r_{1}}$ with associated unknown `weights' $\wt\beta_{1,i_{1},r_{1}}^{i}$. 
We have the same set of bases for each pair $(i_{2},i_{3})$ of (column, lag) estimated from all data, 
and the same set of weights for each pair $(i,i_{1})$ of (individual, row) centered around the fixed effects weights estimated from all data
but are also informed by individual variations, 
providing a highly flexible model overall for all tuples $(i,i_{1},i_{2},i_{3})$ of (individual, row, column, lag).

The empirical efficiency and robustness of this carefully designed PTDVAR model 
(under a Bayesian inferential framework developed below)
is demonstrated by extensive numerical experiments for a variety of simulation scenarios summarized in Section \ref{sec: Simulations MP} of the main paper 
and detailed in Section \ref{sec: Simulations SM} in the supplementary materials. 
Notably, the approach provides excellent estimates for generic data generating mechanisms even when they do not exactly conform to our model construction. 
See, e.g., Figure \ref{fig: GC10 circos} in Section \ref{sec: Simulations MP} for a scenario 
where the individual coefficients were simulated by adding random matrices to block diagonal fixed effects.
See also Figure \ref{fig: defaultc_comparison} in Section \ref{sec: Application} for network heterogeneity estimated from real data using our proposed approach.

\subsection{Prior Specification}\label{sec: priors}

As motivated in the previous sections, we introduce priors that can simultaneously shrink our model towards a low rank Tucker decomposition 
while also inducing meaningful sparsity structures in the core tensor and the factor matrices.  
Sparsity in the core tensor decreases the number of interactions between the elements of the factor matrices, thereby reducing the complexity of the model.
Sparsity in the factor matrices can likewise have meaningful implications. 
In particular,
when the $\ell\th$ row in $\bbeta_{3}$ is comprised of all zeros, 
the $\ell\th$ lag gets eliminated from the model (Lemma \ref{lem: lag selection}). 
These desirable sparsity implications motivate us to use priors that can shrink small coefficients towards zero, but leave the large ones intact. 

A well known prior that satisfies these requirements is the horseshoe prior \citep{carvalho2010horseshoe} 
which can be expressed as a global-local scale mixture \citep{polson2012local} as follows 
\vspace{-5ex}\\
\bse
\btheta_{j} &\sim \MVN(\bzero, \btau_{j}^{2} \lambda_{j}^{2}), \quad \tau_{j}^{2} \sim C^{+}(0,1),~~ \lambda_{j}^{2} \sim C^{+}(0,1), 
\ese
\vspace{-5ex}\\
where $(\btheta_{1}, \dots, \btheta_{p})$ are vectors of coefficients, $\lambda_{j}^{2}$ is a global scale for each vector $\btheta_{j}$, and $\btau_{j}$ is a vector of local-scale parameters for each element of $\btheta_{j}$.

In addition to inducing element-wise sparsity in $\bbeta_{j}$ and $\cG$, 
we also wish to shrink each $\bbeta_{j}$ towards a low-rank structure by deleting its unnecessary columns in a data-adaptive manner. 
To this end, we combine the horseshoe prior with the multiplicative gamma process shrinkage (MGPS) prior from \cite{bhattacharya2011sparse} 
to encourage increasing shrinkage of the columns of $\bbeta_{j}$ as the column index increases. 
Our novel combined prior, which applies increasing column-wise shrinkage alongside heavy shrinkage of small coefficients, is implemented using a hierarchical mixture as 
\vspace{-5ex}\\
\be
\label{eq:beta_priors1}
& \beta_{j,k,r}\vert \tau_{j,k,r}, \lambda^{2}_{j},  \psi_{j,r}, \sigma^{2}_{\varepsilon} \sim \Normal\left(0, \frac{ \tau_{j,k,r} \lambda^{2}_{j} \sigma^{2}_{\varepsilon}}{\psi_{j,r}}\right), \label{eq:beta_priors1}\\
& \psi_{j,r} = \prod_{\ell=1}^{r}\delta_{j}^{(\ell)}, ~~ \delta_{1} \sim \Ga(a_{1},1), ~~ \delta_{r} \sim \Ga(a_{2},1), r\geq 2, \label{eq:beta_priors2}\\
& \tau_{j,k,r} \sim C^{+} (0,1),~~~
\lambda^{2}_{j} \sim C^{+} (0,1),~~~
\sigma^{2}_{\varepsilon} \sim \IG(a_{\sigma}, b_{\sigma}),  \label{eq:beta_priors3}
\ee
\vspace{-5ex}\\
where $r=1,\dots R_{j}$ for $j = 1,2,3$.
The global shrinkage parameter for each $\bbeta_{j}$ is $\lambda^{2}_{j}$,
and the local shrinkage parameter for the $k\th$ element in the $r\th$ column of $\bbeta_{j}$ is $\tau_{j,k,r}$. 
The $\delta_{j}^{(\ell)}$ are independent identically distributed gamma variables whose product is 
$\psi_{j,r}$, the shrinkage parameter for the $r\th$ column of $\bbeta_{j}$. 
The variance of the model noise $\sigma^{2}_{\varepsilon}$ is included to prevent multi-modality in the posterior \citep{carvalho2010horseshoe}.

\begin{figure}
 \centering
    \includegraphics[width=\linewidth]{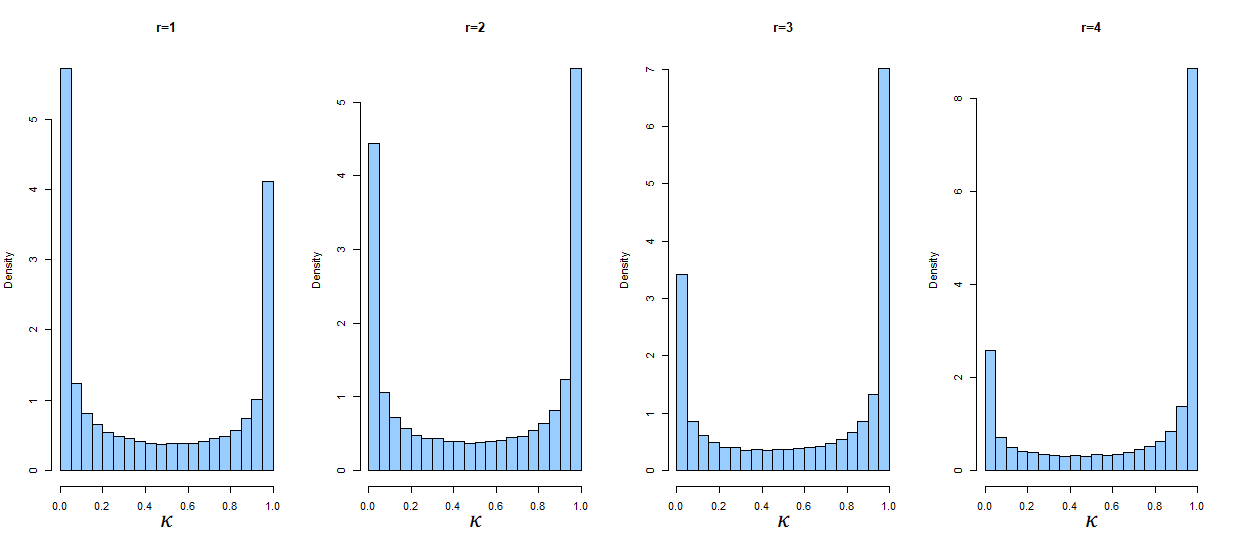}
    \caption{
     Empirical densities of the shrinkage coefficient $\kappa$ for the horseshoe-MGPS prior for increasing values of $r$ based on $100,000$ samples from the prior each.   
   }
    \label{fig:prior_concentration}
\end{figure}

Let $\sigma^{2}_{\beta,j,k,r} = {\tau_{j,k,r} \lambda^{2}_{j} \sigma^{2}_{\varepsilon}}{\psi_{j,r}^{-1}}$ denote the variance of any $\beta_{j,k,r}$, 
and define $\kappa_{r,j,k} = 1/(1+\sigma^{2}_{\beta,r,j,k})$ be the corresponding shrinkage coefficient. 
For a vanilla horseshoe prior, the distribution of $\kappa$ is the horseshoe shaped $\Beta(1/2,1/2)$,  where the left side of the horseshoe, $\kappa \approx 0$, yields virtually no shrinkage and describes signals, and the right side of the horseshoe $\kappa \approx 1$ yields near total shrinkage and describes noise \citep{carvalho2010horseshoe}.
Figure \ref{fig:prior_concentration} shows the distribution of $\kappa$ for our combined horseshoe-MGPS prior is still horseshoe shaped. 
As the column rank $r$ increases, however, the prior shifts more weight towards the right side of the distribution, inducing increasingly heavier shrinkage.

Each Tucker rank $R_{j}$ is upper bounded by its corresponding dimension in $\cB$ \citep{tucker1966some}, 
i.e., $R_{1}\leq K, R_{2} \leq K, R_{3} \leq L$. 
We set the initial ranks based on how much freedom we think the model liberally needs.
We then select the ranks by eliminating the columns of $\bbeta_{j}$ whose whose Frobenius norm falls within a pre-specified small neighbourhood of zero after the MCMC chain settles. 
To account for sampling fluctuations, we use the average of a few successive samples of each $\bbeta_{j}$ to compute the column norms.  
To ensure increased shrinkage for the later columns of $\bbeta_{j}$, we set the hyper-parameters $a_{1}$ and $a_{2}$ 
according to the guidelines set forth in \cite{durante2017note}.

We also allow the dimension of the core tensor $\cG$ to change appropriately with the $\bbeta_{j}$'s, letting the model become increasingly computationally efficient in later iterations.
To make the model more parsimonious, 
we apply the horseshoe prior to the individual elements of $\cG$ as 
\vspace{-5ex}\\
\bse
& g_{r_{1}, r_{2}, r_{3}} \vert \tau_{r_{1}, r_{2}, r_{3}}^{2}, \lambda^{2}_{g}, \sigma^{2}_{\varepsilon} \sim \Normal(0, \tau_{r_{1}, r_{2}, r_{3}}^{2} \lambda^{2}_{g} \sigma^{2}_{\varepsilon}),\\
& \tau_{r_{1}, r_{2}, r_{3}}^{2} \sim C^{+} (0,1), ~~~
\lambda^{2}_{g} \sim C^{+} (0,1), ~~~
\sigma^{2}_{\varepsilon} \sim \IG(a_{\sigma}, b_{\sigma}),
\ese
\vspace{-5ex}\\
as well as to the elements of the intercept term $\nu$ as 
\vspace{-5ex}\\
\bse
& \nu_{k} \vert \tau_{k}^{2}, \lambda^{2}_{\nu}, \sigma^{2}_{\varepsilon} \sim \Normal(0, \tau_{k}^{2} \lambda^{2}_{\nu} \sigma^{2}_{\varepsilon}),\\
& \tau_{k}^{2} \sim C^{+} (0,1), ~~~
\lambda^{2}_{\nu} \sim C^{+} (0,1), ~~~
\sigma^{2}_{\varepsilon} \sim \IG(a_{\sigma}, b_{\sigma}).
\ese
\vspace{-5ex}

In order to sample efficiently from the $C^{+}(0,1)$ distribution, we apply another hierarchical relationship \citep{makalic2015simple} which samples a random variable $x^{2}$ from a $C^{+}(0,A)$ distribution by sampling $a\sim \InvG(1/2, 1/A^{2})$ and $x^2 \vert a \sim \InvG(1/2, 1/a)$. 
Thus, we allow each $\tau^{2}$ to depend on its own unique mixing parameter $\phi$ such that $\tau^{2} \vert \phi \sim \InvG(1/2, 1/\phi)$ and $\phi \sim \InvG(1/2, 1)$. 
Likewise each $\lambda^{2} \vert \xi \sim \InvG(1/2, 1/\xi)$ where $\xi\sim \InvG(1/2, 1)$.

\subsection{Random Effects Distributions}\label{sec: rand effects}
{For the BPTDVAR model introduced in Section \ref{sec: BPTDVAR}, 
we assume $\bbeta_{1}^{fixed}$ and $\bbeta_{1}^{random, i}$ to both be distributed as (\ref{eq:beta_priors1}). 
For the $\bbeta_{1}^{random, i}$'s, the normal distribution in (\ref{eq:beta_priors1}) should now be interpreted as the random effects distributions, 
and (\ref{eq:beta_priors2})-(\ref{eq:beta_priors3}) the associated hyper-priors.
In order to keep rank shrinkage for $\bbeta_{1}$ consistent between all subjects, we let $\bbeta_{1}^{fixed}$ and $\bbeta_{1}^{random, i}$ share the same $\psi_{j,r}$.}
Rank shrinkage for $\bbeta_{1}^{i}$ now occurs if the Frobenius norm of column $r_1$ falls within a neighborhood of zero for \emph{all} $\bbeta_{1}^{i}$'s.

\subsection{Posterior Computation}\label{sec: posterior}
{Posterior inference for all components of the BTDVAR and BPTDVAR models is based on samples drawn from the posterior using a Gibbs sampler. 
The carefully constructed priors in Section \ref{sec: priors} lead to conditionally conjugate posterior distributions for all components of the BTDVAR and BPTDVAR models. 
However, due to the complex relationship between the factor matrices, the core tensor, and the data, computation for the likelihood contribution of $\bbeta_{j}$'s is not straightforward. 
To address this issue, we re-structure the data at each time point to also be tensor-valued, then adapt techniques from \cite{guhaniyogi2017bayesian} in tensor regression settings to compute the corresponding likelihoods.
To keep the length of the main paper manageable, 
we have deferred the details to Section \ref{sec: post inference sm} of the supplementary materials.

\section{Granger Causality Analysis}\label{sec: GCausality}
When using VAR models to perform GCA on fMRI data, including all regions of the brain in the VAR can mitigate the effects of confounding \citep{wang2020large}. 
In a VAR(L) model, the causal variables for $\by_{t}$ are the lagged observations $\{\by_{t-1},\dots,\by_{t-L}\}$. 
There can exist a direct causal relationship between $y_{j,t-\ell}$ and $y_{k,t}$ as well as an indirect effect in which $y_{j,t-\ell}$ influences $y_{i,t-\ell + 1}$ and $y_{i,t-\ell + 1}$ influences $y_{k,t}$.
However, by including lags $1:L$ of all $K$ variables in the VAR(L) model, we eliminate the indirect effects of $y_{j,t-\ell}$ on $y_{k,t}$ by conditioning on the observed value of $y_{i,t-\ell+1}$ \citep{robins2000marginal}. 
Thus, all effects estimated by the coefficients in the VAR can be interpreted as direct effects from $y_{j,t-\ell}$ to $y_{k,t}$.

We represent the causal network of the brain as a directed acyclic graph (DAG) and use the posterior of the VAR coefficients to determine the presence and direction of causal relationships between regions.
Consider a graph $G=(V,E)$, 
where $V$ is the set of nodes and $E\subset V \times V$ is the edge set. 
An edge is called directed if $(j,k)\in E \Rightarrow (k,j) \notin E$. 
In the causal DAG, the nodes of the graph represent random variables $\by = (y_{1},\dots,y_{K})\trans$, 
and the edges represent the causal relationships between them.
The causal effects can be represented using structural equation models \citep{pearl_2009} with each variable $y_{k}$ relating its parents $pa_{k}$ as
\vspace{-5ex}\\
\be
\textstyle y_{k} = \sum_{j\in pa_{k}}\alpha_{jk} y_{j} + \varepsilon_{k}, \quad k=1,\dots,K, 
\label{eq:granger}
\ee
\vspace{-5ex}\\
where $\alpha_{jk}$ is the effect of $y_{j}$ on $y_{k}$, and $j\in pa_{k}$. 
In the special case where $y_{k}$ are Gaussian in \eqref{eq:granger}, it is known that $\alpha_{jk}=0 \Longleftrightarrow j\notin pa_{k}$ \citep{pearl_2009}.
This property has been exploited to determine causality in linear models subjecting the parameters to penalization such as LASSO \citep{shojaie2010discovering}.
We control for false discoveries 
using a decision rule that optimizes a loss function 
{which penalizes both false negatives and false positives \citep{muller2004optimal}}.
See Section \ref{sec:FPR control} in the supplementary materials for a brief review.

To determine whether an element $a_{\ell,j,k} \in \bA_{\ell}$ is zero a-posteriori, we use a decision rule which sets $a_{\ell,j,k} = 0$ when $P(\abs{a_{\ell,j,k}} < \delta/2  \vert \data) > t^{\star}$ for some small $\delta$. 
The threshold probability $t^{\star}$ is chosen such that we minimize the loss function $L(\overline{FP},\overline{FN})= c\overline{FP} + \overline{FN}$, where $\overline{FP}$ and $\overline{FN}$ are the posterior expected count of false positives and false negatives respectively.
Larger values of $c$ will result in decision rules that penalize FPs more harshly while smaller values of $c$ penalize FNs more harshly.
\cite{muller2004optimal} show that under $L(\overline{FP},\overline{FN})$, the optimal decision threshold is $t^{\star} = c/(c+1)$.
See Section \ref{sec:FPR control} in the supplementary materials for additional details.

\section{Simulation Studies} \label{sec: Simulations MP}

We evaluated the performance of our proposed B(P)TDVAR approaches in extensive numerical experiments. 
We present some highlights here, 
but defer most details to Section \ref{sec: Simulations SM} 
in the supplementary materials due to page constraints. 
Though our method scales well with increasing dimensions (see Section \ref{sec: Simulations SM}),
for ease of viewing, we include in this section graphical results from relatively low-dimensional problems.

\begin{figure}[!hb]
    \centering
    \includegraphics[width=0.9\linewidth]{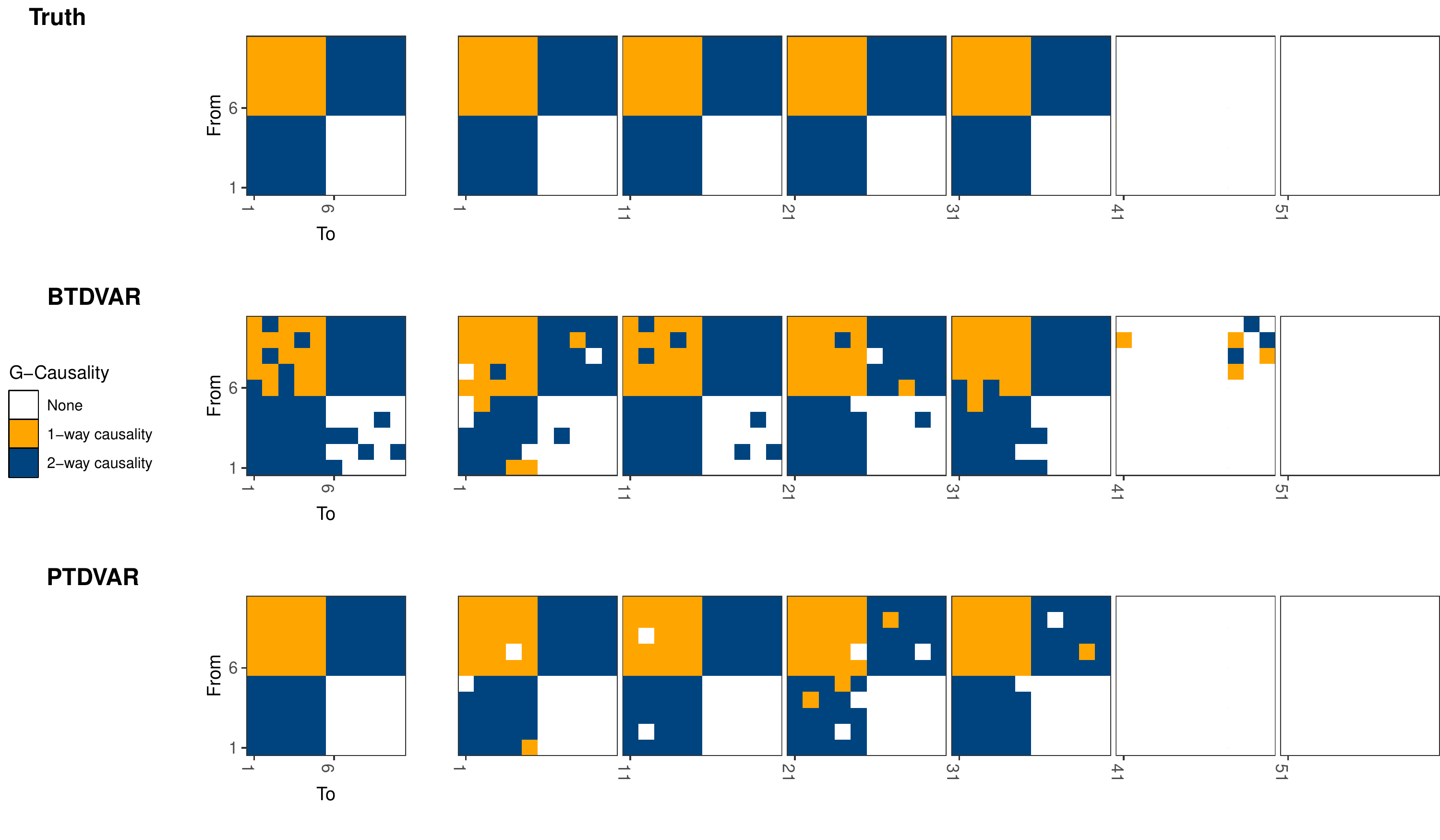}
    \caption{Results from simulated data: 
    GC network results obtained by the B(P)TDVAR models for $N=1 (10)$ subject(s) and $T=150$ observations (for each subject when $N=10$) simulated with $L_{true}=4$ lags and $K=10$ variables. 
    The true network patterns of each $A_{\ell}$ here is block-diagonal. 
    We used $L=6$ initial experimental lags. 
    The left-most matrix in each row shows the composite network summed across all lags, while the 
    6 blocks to the right show the network of each transition matrix $\bA_{\ell}, ~ \ell=1:6$.
    We see that besides very accurately recovering the true Granger network, the B(P)TDVAR very efficiently eliminated the two extra lags (right-most two empty matrices).
    }
    \label{fig: GC10}
\end{figure}

We compared our model estimates with that of ordinary least squares (OLS) and the single-subject Bayesian VAR (SSBVAR) model from \cite{ghosh2018high}. 
We found that the BTDVAR and the BPTDVAR produce robust model fits ($R^2 > 0.7$) for small ($K=10$), medium ($K=50$), and large ($K=200$) simulated VARs (Table \ref{table:sim-accuracy} in the supplementary materials), providing competitive and often better estimates than the alternatives even for small $K$ large $T$ settings. 
The BPTDVAR is generally able to very accurately recover the true underlying GC network by sharing information between individuals (see, e.g., Figure \ref{fig: GC10}).
Importantly, for large $K$ and relatively small $T$ simulations, the B(P)TDVAR produced great model fit, and were able to recover accurate GC networks 
while OLS and SSBVAR \cite{ghosh2018high} were unable to handle the dimensionality issues and could not produce any estimates (see Table \ref{table:sim-accuracy} again).

\begin{figure}[!hb]
\begin{subfigure}{0.33\textwidth}
      \centering
      \includegraphics[width=0.95\linewidth, trim={3cm 3cm 3cm 3cm}, clip=true]{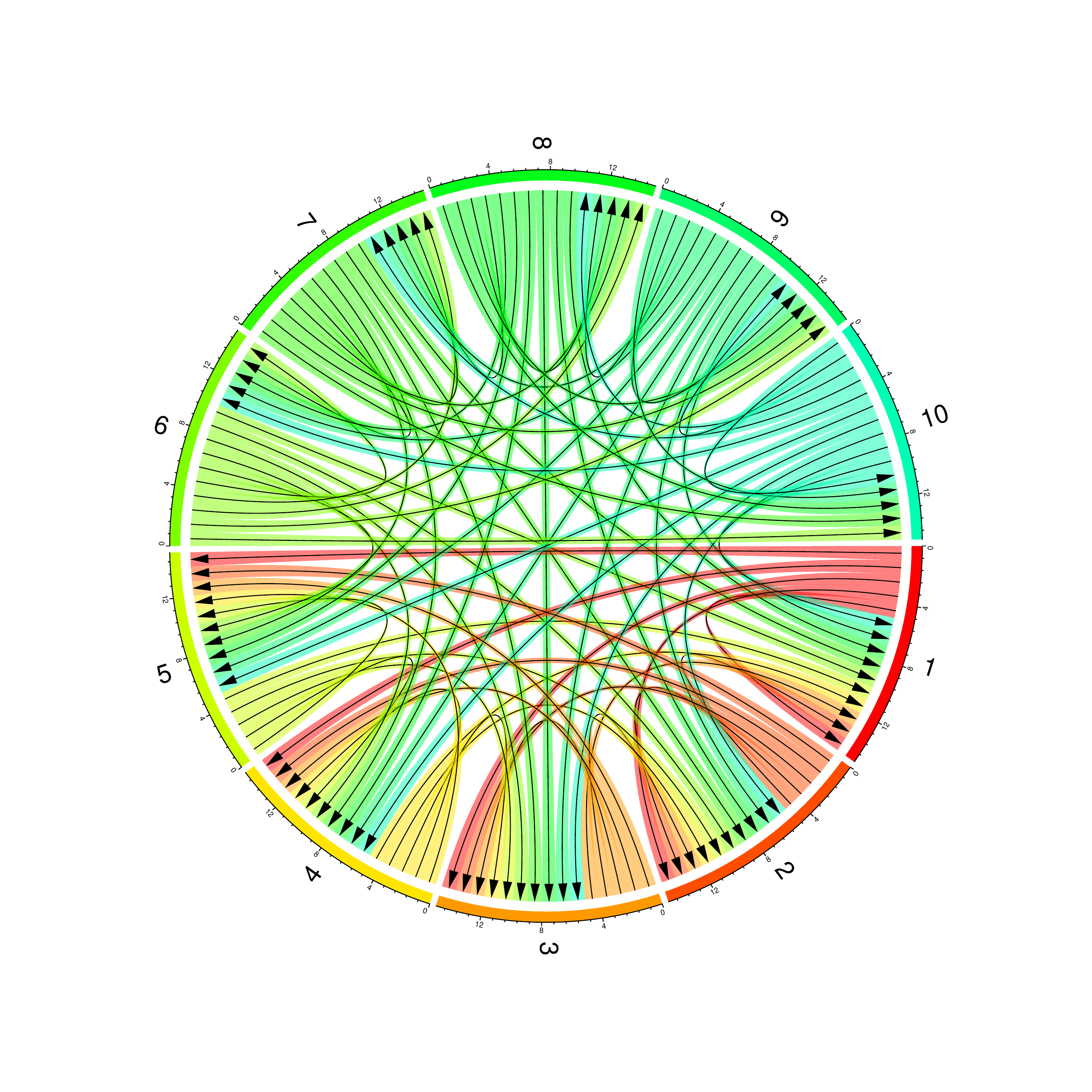}
      \caption{Fixed effects truth.}
    \end{subfigure}%
  \begin{subfigure}{0.33\textwidth}
      \centering
      \includegraphics[width=0.95\linewidth, trim={3cm 3cm 3cm 3cm}, clip=true]{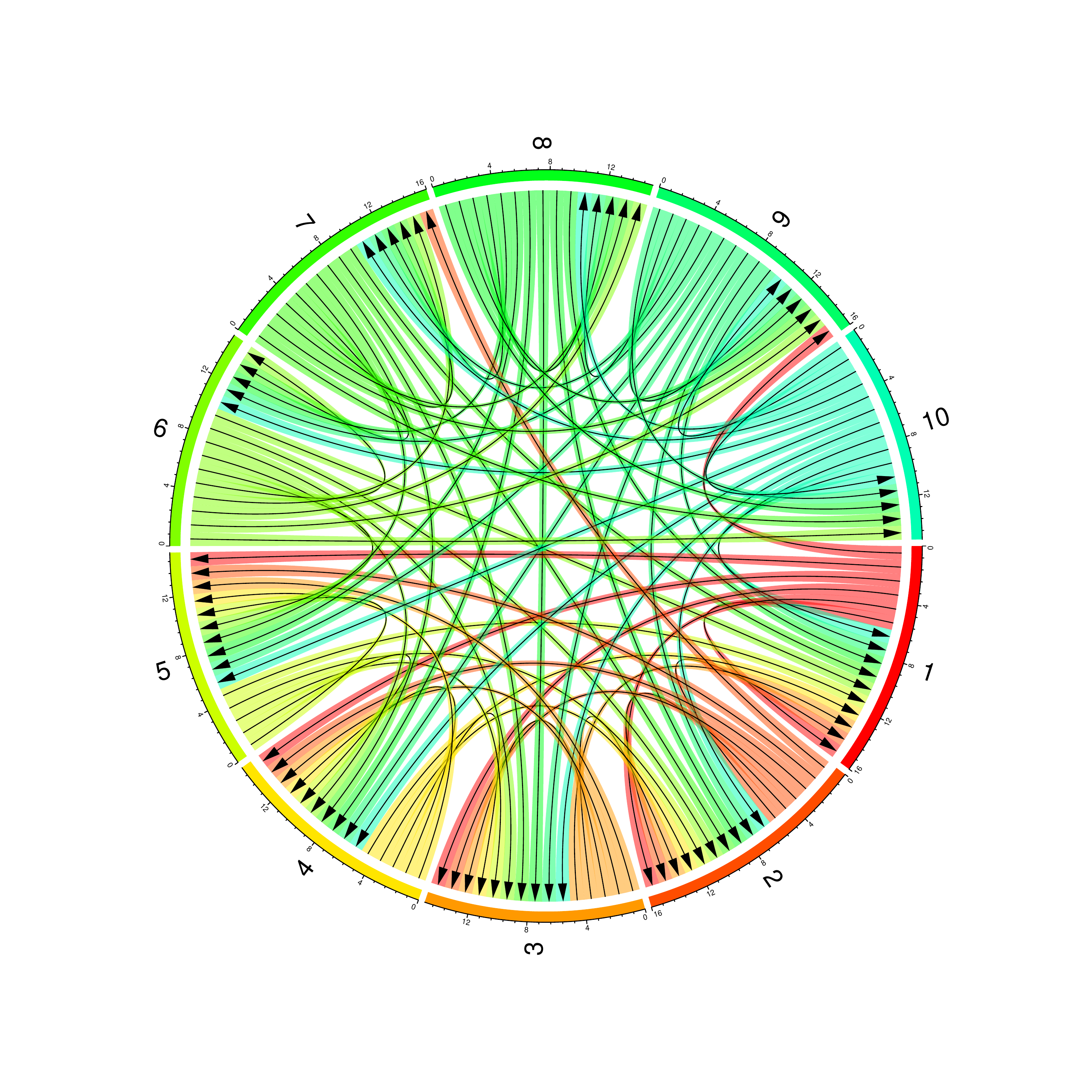}
     \caption{Subject 1 truth.}
  \end{subfigure}%
  \begin{subfigure}{0.33\textwidth}
      \centering
      \includegraphics[width=0.95\linewidth, trim={3cm 3cm 3cm 3cm}, clip=true]{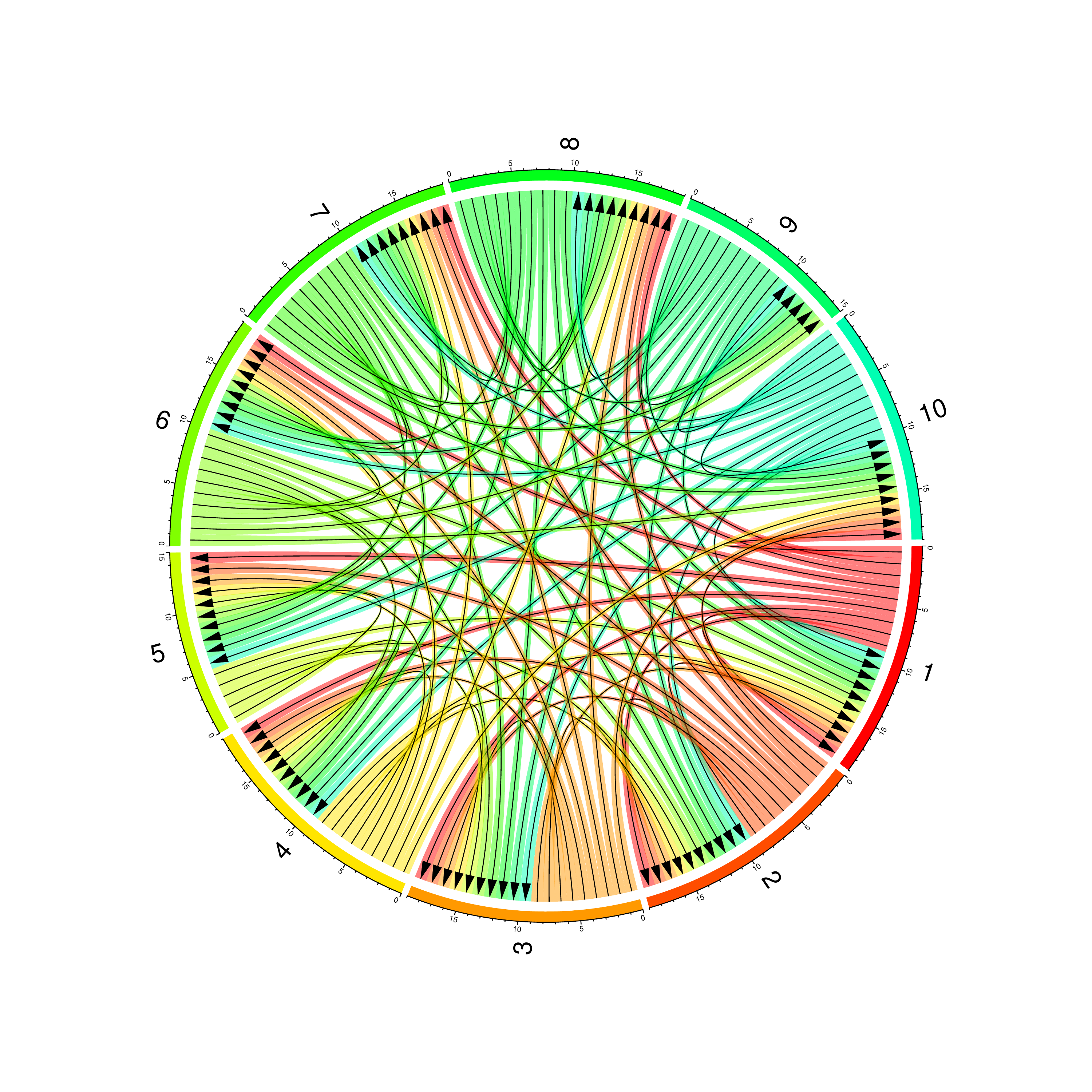}
     \caption{Subject 2 truth.}
    \end{subfigure}\\
 \begin{subfigure}{0.33\textwidth}
      \centering
      \includegraphics[width=0.95\linewidth, trim={3cm 3cm 3cm 3cm}, clip=true]{Figures/blockdiag_sim_fixed.pdf}
      \caption{Fixed effects estimate.}
    \end{subfigure}%
  \begin{subfigure}{0.33\textwidth}
      \centering
      \includegraphics[width=0.95\linewidth, trim={3cm 3cm 3cm 3cm}, clip=true]{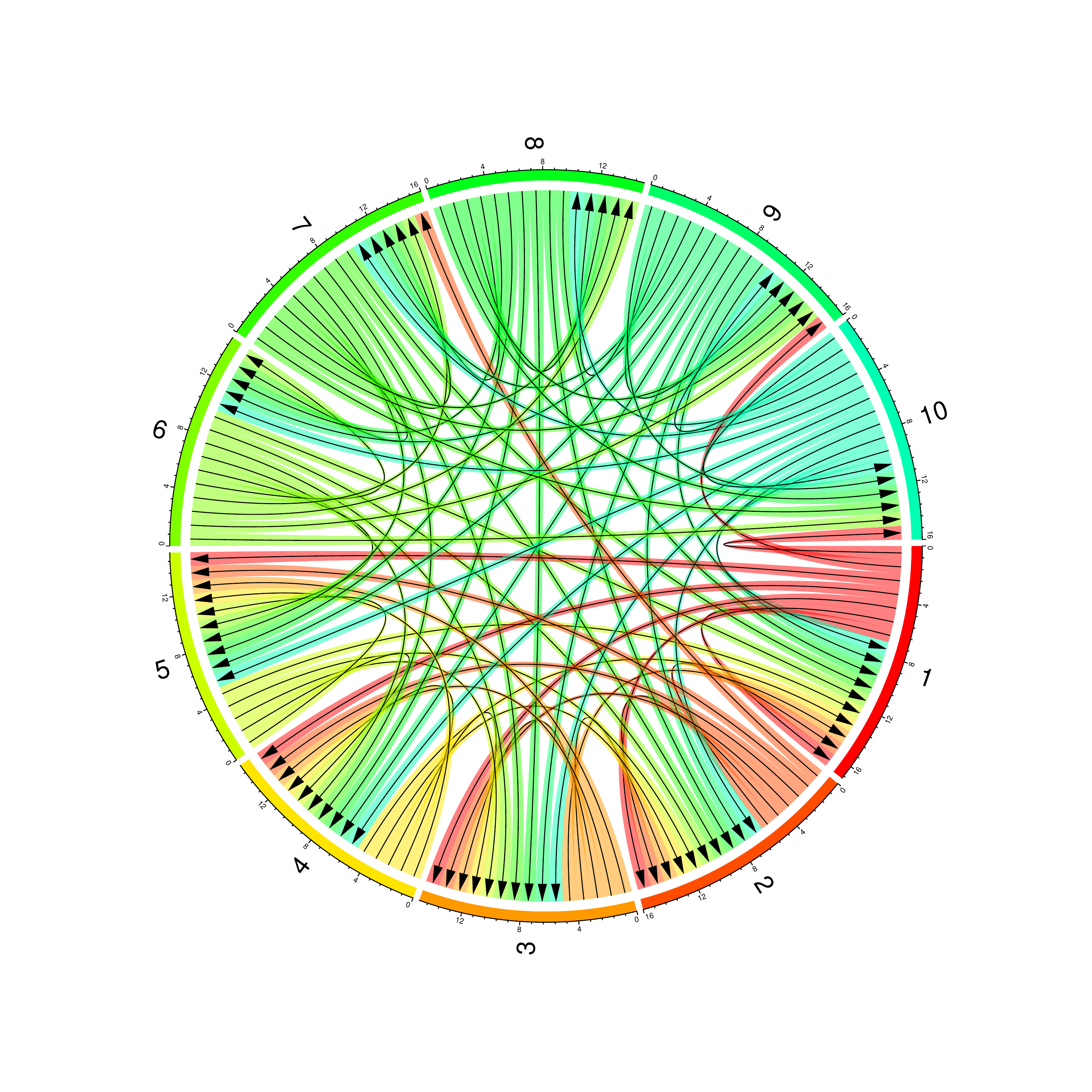}
     \caption{Subject 1 estimate.}
  \end{subfigure}
  \begin{subfigure}{0.33\textwidth}
      \centering
      \includegraphics[width=0.95\linewidth, trim={3cm 3cm 3cm 3cm}, clip=true]{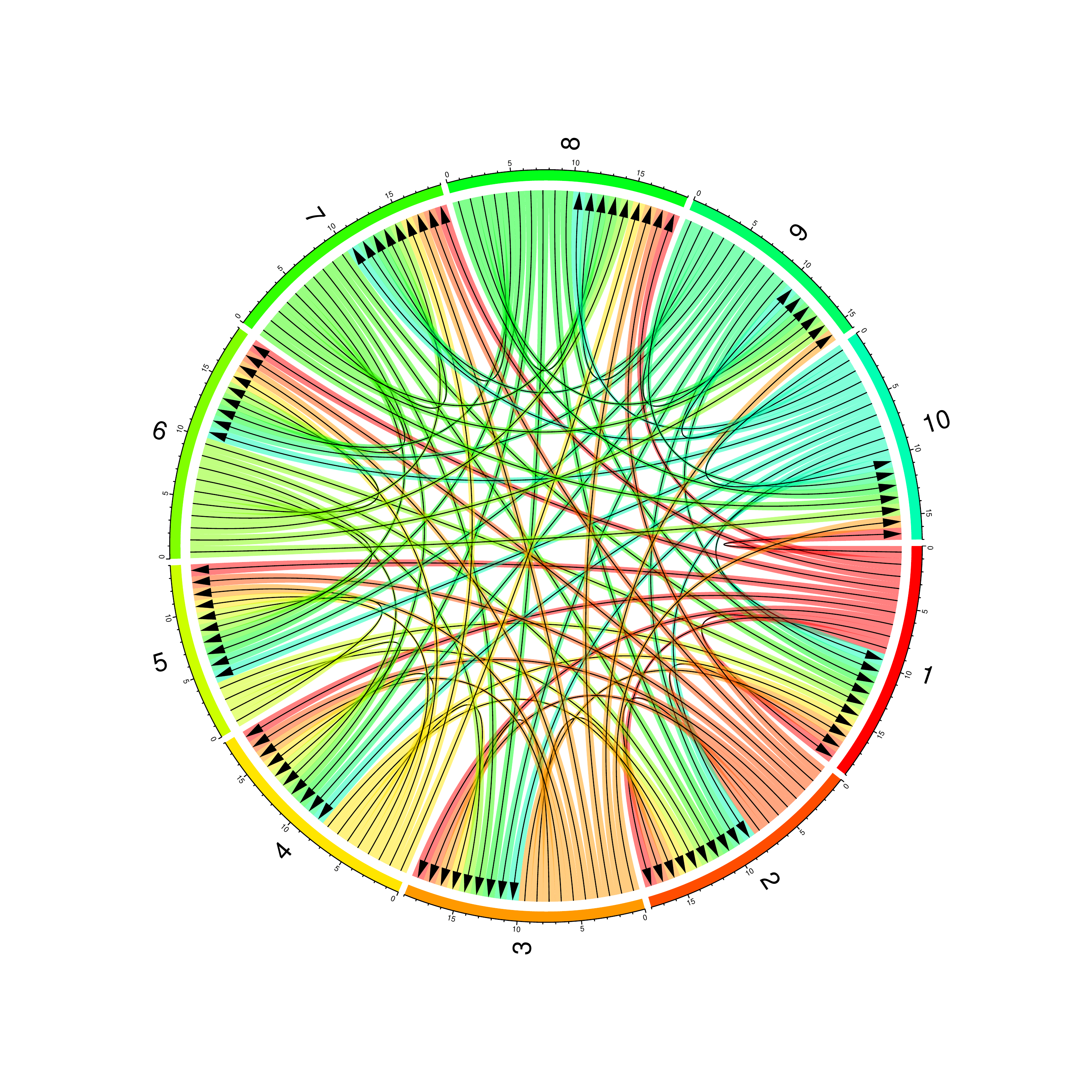}
     \caption{Subject 2 estimate.}
    \end{subfigure}
  \caption{Results from simulated data: 
  GC network results obtained by the BPTDVAR model under the simulation set up shown in Figure \ref{fig: GC10}. 
  Here an arrow going from variable $a$ to $b$ represents a one-way directed GC influence between $a$ and $b$.
  We have perfect recovery for the shared network. 
  The individual-specific estimated networks 
  also show excellent agreement with the corresponding simulation truths. 
  }
  \label{fig: GC10 circos}
\end{figure}
\vskip -0pt

We plot ROC curves in Figure \ref{fig:ROC} to study the interactions of rank compression and GC network threshold probabilities on GC network recovery accuracy. 
We find a substantial increase in the area under the ROC curve when the Tucker ranks are increased from  $R_{1},R_{2}=5$ to $R_{1},R_{2}=10$, 
followed by ever diminishing returns for $R_{1},R_{2}\geq10$.
While overly heavy compression from too small Tucker ranks does not work well, 
the model starts to provide excellent practical performance after the ranks are sufficiently increased, 
increasing the ranks beyond that only marginally improves model performance. 
Figure \ref{fig: GC10} shows the B(P)TDVAR's ability to accurately recover the true GC networks across different lags, while Figure \ref{fig: GC10 circos} shows the the BPTDVAR's ability to recover the heterogeneity between different individuals.
Figure \ref{fig: GC10} also demonstrates that the B(P)TDVAR is able to correctly determine the true number of lags ($L_{true}=4$) from an initial experimental lag of $L=6$.

\section{fMRI Data Analysis Results} \label{sec: Application}
In this section, we discuss the results obtained by our proposed BPTDVAR approach in estimating the GC network across brain regions from the high quality, high resolution 7T-HPC-fMRI data set described in Section \ref{sec: Data}. 
We extracted the mean time series, with $T=500$ seconds, from each of the $K=200$ regions in the Schaefer 200 brain atlas for the resting state fMRI scan of $N=20$ individuals. 
We chose an experimental lag of $L=4$, and Tucker ranks $R_1=17, ~ R_2 = 17, ~R_3 = L$ because of our Yeo 17-Network schema \citep{yeo2011organization}} shown in Figure \ref{fig: Schaefer_atlas}. 
The BPTDVAR picked out the first two lags as significant, the first lag with a much stronger effect than the second. 
We conducted GCA across regions using the estimated fixed effects transition matrices, and group them into the Yeo 17-Network schema.

\begin{figure}[h!]
    \begin{subfigure}{0.5\textwidth}
      \centering
      \includegraphics[width=\linewidth]{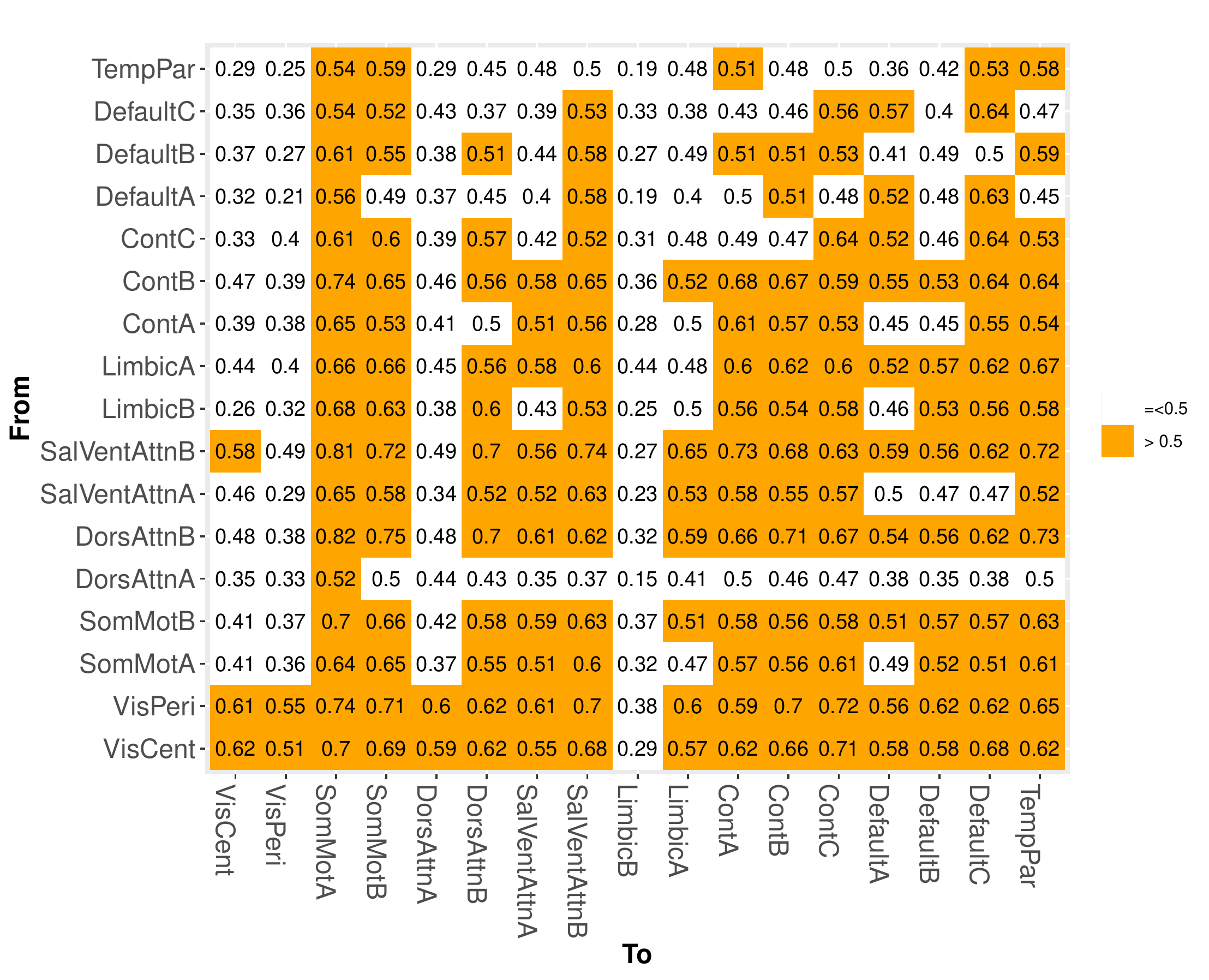}
      \caption{17-network schema.}
    \end{subfigure}%
      \begin{subfigure}{0.5\textwidth}
      \centering
      \includegraphics[width=0.865\linewidth, trim={0cm 0cm 3cm 0cm}, clip=true]{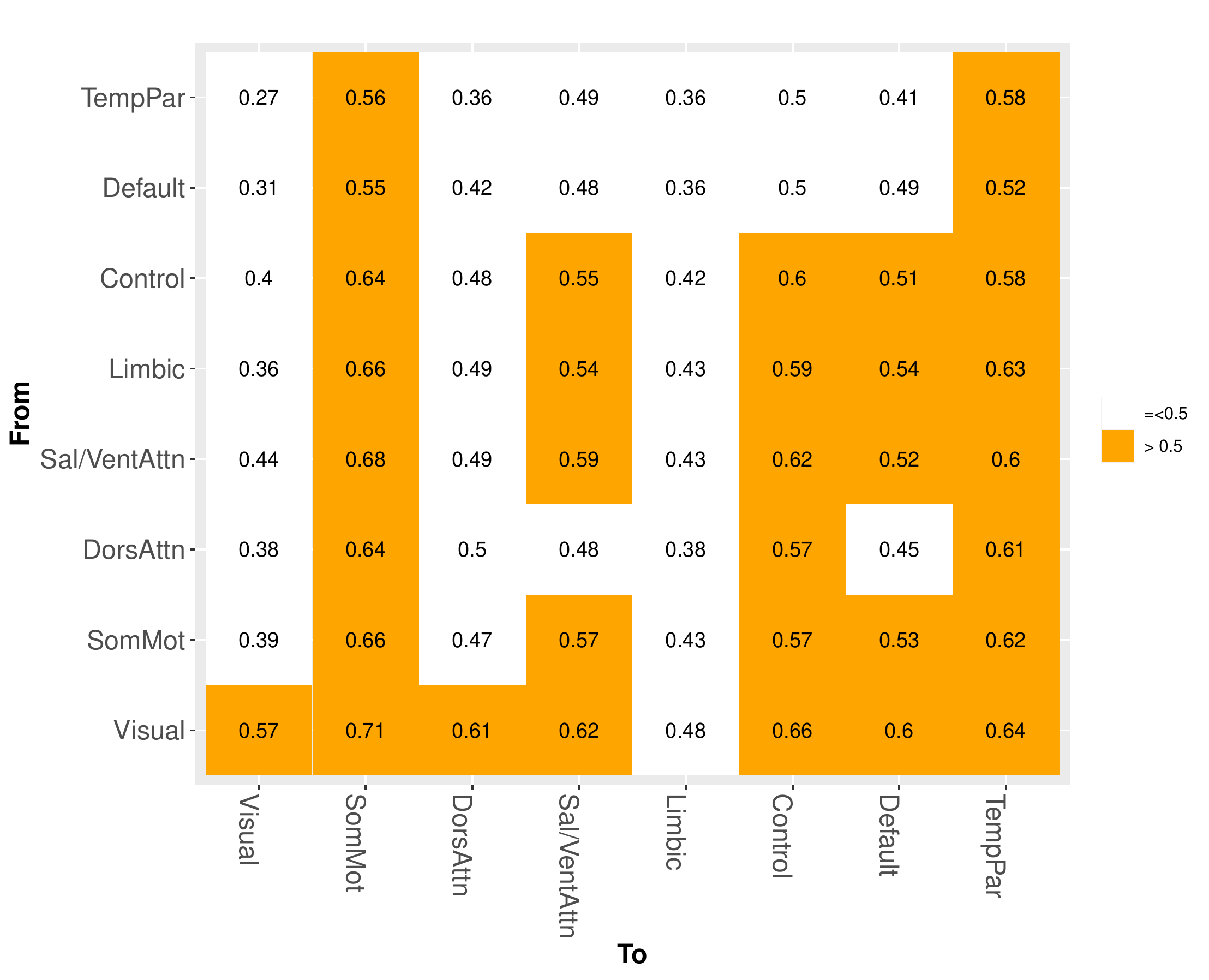}
     \caption{Grouping together similar sub-networks.}
    \end{subfigure}
    \caption{Results from real data: Ratio of connected regions within each network pair.}
    \label{fig: fmri_connected}
\end{figure}

We focus first on the between sub-network connectivity patterns. 
Note that such inference can not be obtained when the 17 sub-networks are examined separately. 
We calculated the proportion of connected edges within each sub-network pair in the shared cortical GC matrix, and designate pairs for which the connectivity ratio is above 0.5 as significantly connected. 
Observing the pattern of significant connections across both the Yeo 17-networks schema and a coarser 8-networks grouped schema (Figure \ref{fig: fmri_connected}) reveals that the Visual networks were a significant driver, with significant connections to every network except the Limbic network (in particular, LimbicB). 
In contrast, the Dorsal Attention network (specifically DorsAttnA) drove the fewest connections. 
The networks that were most driven by other networks were Somatomotor (especially SomMotA) and Temporal Parietal networks.
Meanwhile, the Limbic network (particularly LimbicB) was minimally driven by other networks. 

As the dominant sensory modality in humans, vision is a major input into the central nervous system.
The visual system is organized such that incoming sensory information quickly reaches primary visual cortex, where it is then passed across successively higher order visual processing regions. 
Sensory information from visual cortex is shared across the brain for use in multimodal processing (merging across sensory inputs), attention, decision-making, and many other functions.
Thus, the Visual network's status as a dominant driver across human brain networks in the present investigation converges nicely with decades of research on the principal role of visual sensory input in the human central nervous system.

Meanwhile, our results highlight the Limbic network—particularly the LimbicB sub-network—for its lack of significant connections from any other cortical network.
The human limbic system is involved in emotion processing and regulation and is comprised of a number of subcortical (hypothalamus, amygdala), allocortical (hippocampus, entorhinal cortex), and neocortical (orbitofrontal cortex, temporal pole) structures.
The subcortical and allocortical structures play a major role in driving activity in (neo-)cortical limbic regions \citep{enatsu2015connections}.
As our brain atlas only included neocortical structures, it is thus not surprising that we did not observe significant connections to the Limbic network.

The LimbicB sub-network was particularly underconnected. 
In the Schaefer 200 brain atlas, LimbicB contains the orbitofrontal cortex, while LimbicA  comprises the temporal pole. 
These regions are anatomically proximal but on distinct cortical lobes (frontal vs. temporal lobes) (Figure \ref{fig: limbic_networks} in the supplementary materials). 
Thus, their differing patterns of connectivity with the rest of cortex is not unexpected.

Two other networks stand out for their status as globally driven hubs: Somatomotor (especially SomMotA) and Temporal Parietal networks.
SomMotA contains primary motor cortex (as well as parts of somatosensory cortex), the major output node in human cortex.
Almost all commands effected by the motor system are routed through the motor cortex, which receives input from premotor cortex, supplementary motor area, and a wide swath of cortical and subcortical structures.
The Temporal Parietal network, meanwhile, is a multimodal association network that compiles information from multiple sensory (visual, auditory, and somatosensory) and cognitive regions.
As higher order associative cortex, the Temporal Parietal network is thus expected to receive neural inputs from across the cortex.

\begin{figure}[!ht]
 \begin{subfigure}{\textwidth}
      \centering
      \includegraphics[width=0.7\linewidth, trim={0.75cm 0.75cm 0.75cm 0.75cm}, clip=true]{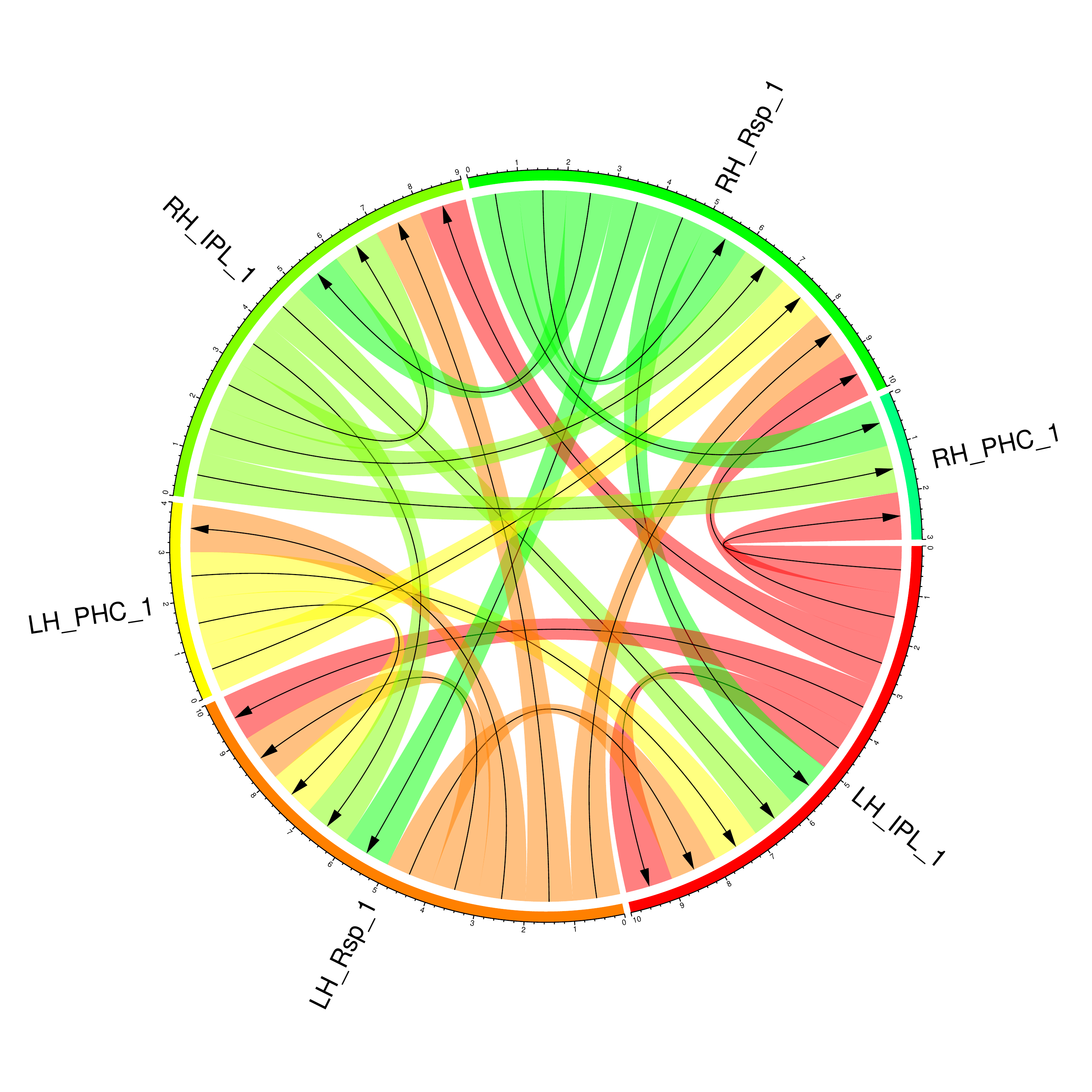}
      \caption{Shared fixed effects network.}
    \end{subfigure}\\
  \begin{subfigure}{0.5\textwidth}
      \centering
      \includegraphics[width=0.94\linewidth]{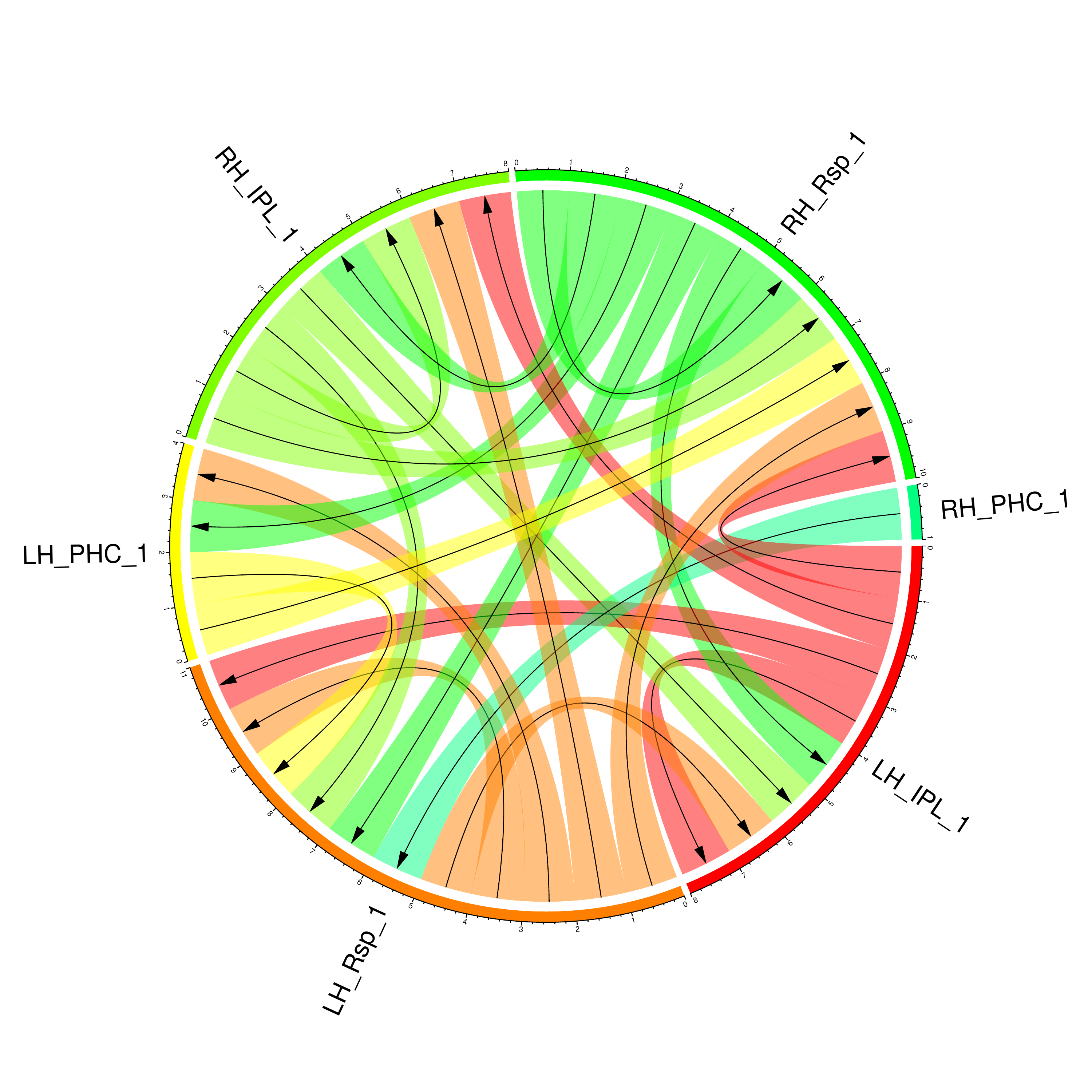}
     \caption{Network for subject 1.}
  \end{subfigure}
  \begin{subfigure}{0.5\textwidth}
      \centering
      \includegraphics[width=0.94\linewidth]{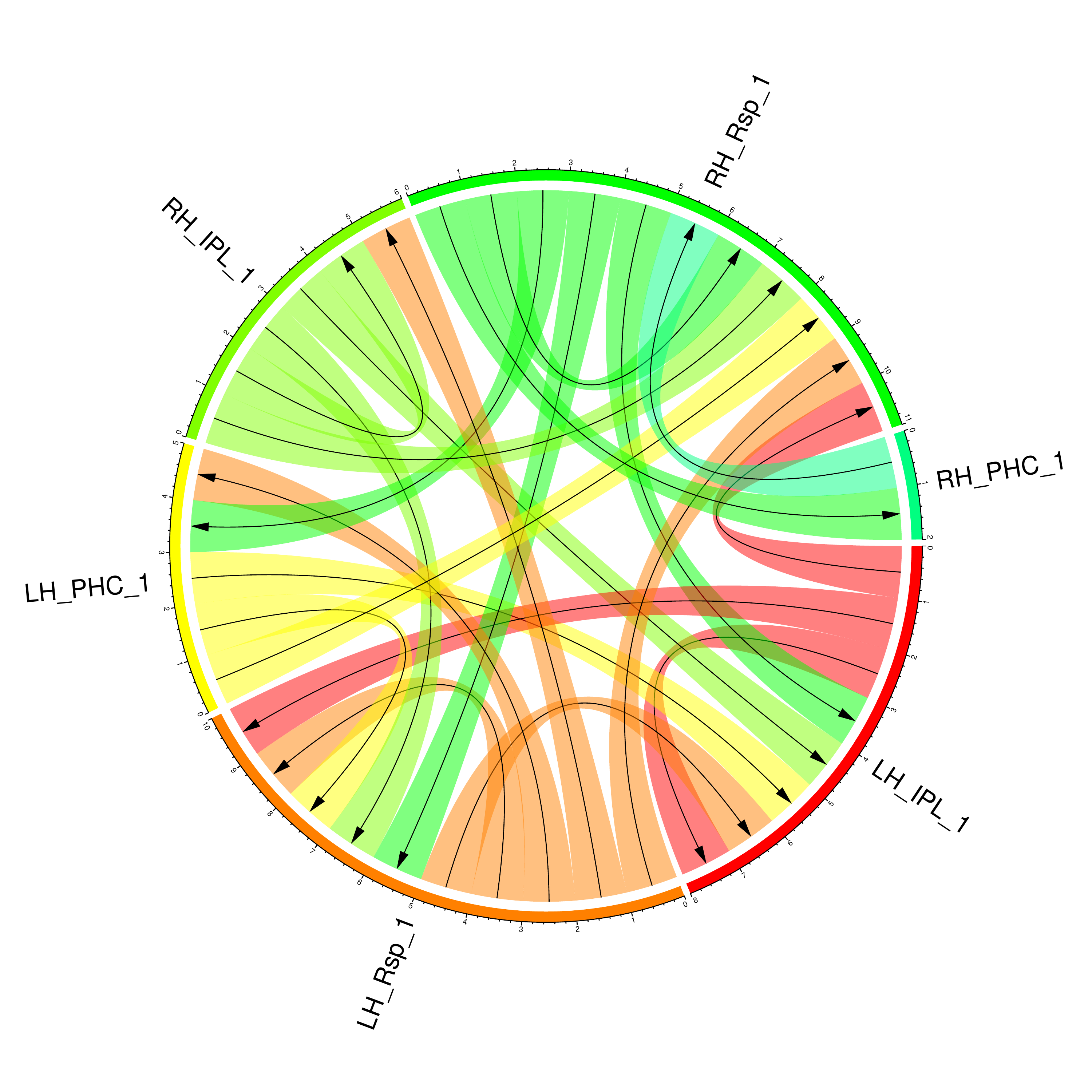}
     \caption{Network for subject 2.}
    \end{subfigure}
  \caption{Results from real data: 
  Shared and subject-specific connectivity patterns for the Default C sub-network. 
  }
  \label{fig: defaultc_comparison}
\end{figure}

Focusing next on the finer within sub-network connectivity patterns and their individual-specific heterogeneity, our results show, in accordance with prior research from \cite{damoiseaux2006consistent}, that between-subject heterogeneity is low. 
As an example, we show the Default C network, which stays activated during rest, and the Control C network, which gets activated during stimulation \citep{uddin2019towards}, to highlight the largely homogeneous connection patterns healthy subjects display. 
Figure \ref{fig: defaultc_comparison} here and Figure \ref{fig: contc_comparison} in the supplementary materials show two examples of within-network heterogeneity for the  Default C and Control C sub-networks respectively. 
Figure \ref{fig: fmri_coefs} in the supplementary materials shows the posterior means and standard deviations of the VAR coefficients for these two sub-regions.

Overall, our results provide 
convergent evidence substantiating prior studies but also uncovers some novel previously unsubstantiated cortical connectivity dynamics.

\section{Discussion} \label{sec: Discussion}

In this article, we introduced the BTDVAR model, which reduces the number of estimated parameters in a VAR(L) model by stacking the VAR transition matrices into a three-way tensor, 
then factorizing it via the Tucker decomposition.
Estimating the components of the Tucker decomposition leads to a highly flexible model while also achieving massive dimension reduction. 
We introduced a horseshoe gamma-gamma shrinkage prior to induce data-adaptive rank shrinkage as well as meaningful sparsity for the tensor components.  
For multi-subject longitudinal data, the model was carefully extended to a structured random effects BPTDVAR model. 
The BPTDVAR model uses information sharing between subjects to estimate a common fixed effect for all subjects, while still allowing for subject-level heterogeneity. 
Granger-causal networks were inferred via posterior analysis.

Using our novel VAR estimation approach, we were able to conduct Granger causality analysis
on our motivating 7T-HCP-fMRI data set, and reveal biologically meaningful connectivity patterns across human cortical networks. 
In particular, the Visual networks were major drivers across most of cortex, only lacking connections with the LimbicB network which did not receive significant connections from any brain networks.
These results provide new insights into our understanding of the human cortical brain networks.

Methodological extensions being pursued as topics of separate research include 
modeling data from multiple modalities, 
data with external covariates, etc.

\baselineskip=14pt

\section*{Supplementary Materials}
Supplementary materials provide 
proofs of theoretical results, 
the choice of hyper-parameters, 
details of the MCMC algorithm to sample from the posterior, 
details of the simulation experiments, 
some additional figures, etc. 
R codes implementing our method are also included as separate files. 

\section*{Funding}
This work was supported in part by the National Science Foundation grant NSF
DMS-1953712 and National Institute on Deafness and Other Communication
Disorders grants R01DC013315 and R01DC015504. 

\section*{Acknowledgments}
Data were provided [in part] by the Human Connectome Project, WU-Minn Consortium (Principal Investigators: David Van Essen and Kamil Ugurbil; 1U54MH091657) funded by the 16 NIH Institutes and Centers that support the NIH Blueprint for Neuroscience Research; and by the McDonnell Center for Systems Neuroscience at Washington University.

\vspace*{-10pt}
\baselineskip=14pt
\bibliographystyle{natbib}
\bibliography{VAR}


\clearpage\pagebreak\newpage
\newgeometry{textheight=9in, textwidth=6.5in}
\pagestyle{fancy}
\fancyhf{}
\rhead{\bfseries\thepage}
\lhead{\bfseries Supplementary Materials}

\baselineskip 25pt
\begin{center}
{\LARGE Supplementary Materials for\\
{\bf Bayesian Tensor Factorized Vector Autoregressive Models 
for Inferring 
Granger Causality Patterns 
from High-Dimensional Multi-subject Panel Neuroimaging Data
}}
\end{center}
\vskip 20pt 
\baselineskip 16pt

\begin{center}
Jingjing Fan$^{a}$ (jfan25@utexas.edu)\\
Kevin Sitek$^{b}$(kevin.sitek@pitt.edu)\\
Bharath Chandrasekaran$^{b}$(b.chandra@pitt.edu)\\
Abhra Sarkar$^{a}$ (abhra.sarkar@utexas.edu)\\
\vskip 7mm
$^{a}$Department of Statistics and Data Sciences, \\
University of Texas at Austin,\\ 2317 Speedway (D9800), Austin, TX 78712-1823, USA
\vskip 8pt 
$^{b}$Department of Communication Science and Disorders,\\ 
University of Pittsburgh,\\
4028 Forbes Tower, Pittsburgh, PA 15260, USA
\end{center}

\setcounter{equation}{0}
\setcounter{page}{1}
\setcounter{table}{1}
\setcounter{figure}{0}
\setcounter{section}{0}
\numberwithin{table}{section}
\renewcommand{\theequation}{S.\arabic{equation}}
\renewcommand{\thesubsection}{S.\arabic{section}.\arabic{subsection}}
\renewcommand{\thesection}{S.\arabic{section}}
\renewcommand{\thepage}{S.\arabic{page}}
\renewcommand{\thetable}{S.\arabic{table}}
\renewcommand{\thefigure}{S.\arabic{figure}}
\baselineskip=15pt

\vskip 10mm
The supplementary materials provide 
proofs of theoretical results, 
the choice of hyper-parameters, 
details of the MCMC algorithm to sample from the posterior, 
details of the simulation experiments, 
some additional figures, etc.

\baselineskip 16pt

\newpage

\section{Proofs and Derivations} \label{sec:derivations}
\subsection{Proof of Lemma 1} \label{sec: proof}

Given a VAR model as defined in (\ref{eq:time series}), mode-3 VAR parameter tensor $\cB \in \mathbb{R}^{K \times K \times L}$, as defined in section \ref{sec: BTDVAR}, and its corresponding Tucker decomposition, 
the $\ell\th$ slice of $\cB$, containing the parameters for the $\ell\th$ lag, has no non-zero elements if and only if the $\ell\th$ row of the factor matrix $\bbeta_{3}$ also contains no non-zero elements. That is, if the $\ell\th$ row of $\bbeta_{3}$ contains all zeros, then $A_{\ell}$ is entirely zero, and $\by_{t-\ell}$ has no effects on $\by_{t}$. 

\textbf{Proof:} We begin with the following two assumptions:

\begin{enumerate}
    \item $\cB$ contains at least one non-zero element not in the $\ell\th$ frontal slice. That is, at least one element $b_{i_{1}, i_{2}, i_{3} \neq \ell} \neq 0$.
    \item Each column of $\bbeta_{1}, \bbeta_{2}, \bbeta_{3}$ must make a nonzero contribution to $\cB$. 
\end{enumerate}

\noindent The second assumption is to prevent the formulation of a Tucker decomposition in which each factor matrix is artificially inflated with zero-valued columns which do not contribute to the reconstructed parameters. 
Given the model for the parameter tensor $\cB$ in (\ref{eq: tucker}, 
we have that the matricization $\cB_{(1)} = [\bA_{1},\dots, \bA_{L}]$ can be constructed using,
\be
\cB_{(1)} = \bbeta_{1} \cG_{(1)}(\bbeta_{3} \otimes \bbeta_{2})\trans,
\label{eq:tucker-reparam}
\ee
where $\otimes$ denotes the Kronecker product. 
Using this notation, it is obvious that if the $\ell\th$ row of $\bbeta_{3}$ consists of all zeros, then the $\ell k\th$ to $(\ell + 1) k\th$ columns of $\cB_{(1)}$ will also be zero. 
This block of zeros corresponds directly to the matrix of parameters for the $\ell\th$ lag, $\bA_{\ell}$.
Thus, a row of zeros in the $\ell\th$ row of $\bbeta_{3}$ loading matrix implies that the $\ell\th$ lag is not relevant to the response. 

Now let us assume that the $\ell\th$ row of $\bbeta_{3}$ contains a nonzero element in column $r' \in \{1, 2, \dots, R_{3}\}$. Set $r_{3}=r'$.
By the definition of the tucker decomposition,
\bse
\cB = \sum_{r_{1}=1}^{R_{1}} \sum_{r_{2}=1}^{R_{2}}\sum_{r_{3}=1}^{R_{3}} g_{r_{1}r_{2}r_{3}} \bbeta_{1,\cdot,r_{1}} \circ  \bbeta_{2,\cdot,r_{2}} \circ  \bbeta_{3,\cdot,r_{3}},
\ese
 the contribution of $\bbeta_{3,\cdot,r'}$ is $\sum_{r_{1}=1}^{R_{1}} \sum_{r_{2}=1}^{R_{2}} g_{r_{1}r_{2}r'} \bbeta_{1,\cdot,r_{1}} \circ  \bbeta_{2,\cdot,r_{2}} \circ  \bbeta_{3,\cdot,r'}$. Given that the $\ell\th$ element of $\bbeta_{3,\cdot,r'}$ is not zero, $\cB{i_{1}, i_{2}, i_{3}=\ell} = 0$ for all $(i_{1}, i_{2})$ only if:

\begin{enumerate}
    \item $\bbeta_{1,\cdot,r_{1}} \circ \bbeta_{2,\cdot,r_{2}}$ equals the zero matrix for all ranks $r_{1}$ and $r_{2}$. In this case, however, $\cB$ would consist entirely of zeros, contradicting assumption 1.
    \item $g_{r_{1} r_{2} r'} =0$ for all $r_{1}$ and $r_{2}$. In this scenario, however, each product $g_{r_{1}r_{2}r'} \bbeta_{1,\cdot,r_{1}} \circ  \bbeta_{2,\cdot,r_{2}} \circ  \bbeta_{3,\cdot,r'}$ results in a zero tensor, contradicting assumption 2.
\end{enumerate}

Thus, if the $\ell\th$ frontal slice of $\cB$ contains only zeros, then the $\ell\th$ row of $\bbeta_{3}$ must also contain all zeros.

\subsection{Expectation of BPTDVAR} \label{sec:PTDVAR_derivation}
In this section, we derive (\ref{eq:random_eff_mean}), the unconditional mean of the random effects model defined in (\ref{eq: random-effect}).
Consider a VAR(1) process with $K$ variables and mean-zero errors defined as $\by_{t} = \bnu + \balpha + \bA_{1} (\by_{t-1}-\balpha) + \bu_{t}.$
If the generation mechanism starts at some time $t=1$, we get the following equivalence after some algebraic simplification 
\bse
& \by_{t} = \bnu + \balpha + \bA_{1} (\by_{t-1}-\balpha) + \bu_{t}\\
& = (\Ind_{K} + \bA_{1} + \dots + \bA_{1}^{t-1})\bnu + (\Ind_{K} + \bA_{1}^{t})\balpha + \bA_{1}^{t}\by_{0} + \sum_{i=0}^{t-1} \bA_{1}^{i} u_{t-i}.
\ese

Since all the eigenvalues of $\bA_1$ have modulus less than 1, $\bA_{1}^{t}$ converges to zero quickly as $t\rightarrow \infty$ so $\bA_{1}^{t}\by_{0}$ can be ignored in the limit. Thus, 
\be
\eE(\by_{t}) = (\Ind_{K} + \bA_{1} + \dots + \bA_{1}^{t})\bnu + (\Ind_{K} + \bA_{1}^{t})\balpha \underset{t\rightarrow\infty}{\longrightarrow} (\Ind_{K} - \bA_{1})^{-1} \bnu + \balpha.
\label{eq:ar1_random_mean}
\ee

A VAR$(L)$ process can be written as a VAR$(1)$ process using notation from section \ref{sec:VAR}. 
Define a $K \times L$ vector $\bmu = [\balpha, \dots, \balpha]^{T}$. 
The VAR$(L)$ process $\bY_{t} = \bV + \bmu + \bA(\bX_{t} - \bmu)$ has expectation $\eE(\bY_{t}) = (\Ind_{KL} - \bA)^{-1} \bV + \bmu$.
Since $\by_{t} = \bJ \bY_{t}$ where $\bJ = [\bI_{K}:\bzero:\dots:\bzero]$, 
the mean of $\by_{t}$ is 
$$\eE(\by_{t}) = (\Ind_{K} - \bB)^{-1} \bnu + \balpha,$$
where $\bB = \bJ\bA$.

\newpage
\section{FPR Control for GCA with the TDVAR Model} \label{sec:FPR control}
In this section, we briefly review the loss function introduced in \cite{muller2004optimal}, 
which is minimized using the decision rule in \ref{sec: GCausality}. 
Let us assume that for each element of $a_{\ell,j,k}\in \bA_{\ell}$, 
there exists a ground truth indicator variable $z_{\ell,j,k}$ where $z_{\ell,j,k} = 1$ if $a_{\ell,j,k} \neq 0$ and $z_{\ell,j,k} = 0$ if $a_{\ell,j,k} = 0$.
Let the discovery of $a_{\ell,j,k} \neq 0$ be represented by $d_{\ell,j,k} = 1$, and a negative $a_{\ell,j,k} = 0$ be represented by $d_{\ell,j,k} = 0$.
Finally, let $v_{\ell,j,k} = P(z_{\ell,j,k}=1|\by_{1,\cdots,T})$ denote the posterior marginal probability for $a_{\ell,j,k} \neq 0$.
For our model, $v_{\ell,j,k}$ is computed by approximating $P(\abs{a_{\ell,j,k}} < \delta/2  \vert \by_{1,\cdots,T}), ~\delta = 0.01$ using the posterior samples of $\bB$.

The true number of false discoveries and false negatives can be computed as 
\bse
FP(d,z) = \sum d_{\ell,j,k}(1-z_{\ell,j,k}),\\
FN(d,z) = \sum (1-d_{\ell,j,k})z_{\ell,j,k}.
\ese
The posterior expected count of false discoveries and false negatives is then  
\bse
\overline{FP}(d,z) = \sum d_{\ell,j,k}(1-v_{\ell,j,k}),\\
\overline{FN}(d,z) =  \sum (1-d_{\ell,j,k})v_{\ell,j,k}.
\ese

To combine the goals of minimizing false discoveries and false negatives, define a loss function $L_{R} = c\overline{FP} + \overline{FN}$. Using this loss function, the optimal decision rule takes the form $d_{\ell,j,k} = I(v_{\ell,j,k} \geq t^{\star})$, where the optimal $t^{\star}$ is $t^{\star} = {c}/{(c+1)}$.

\newpage
\section{Posterior Computation for B(P)TDVAR Model} \label{sec: post inference sm}

\subsection{BTDVAR}
Posterior inference for the VAR model given in (\ref{eq:likelihood}) in the main paper, 
is based on samples drawn from the posterior using a message passing MCMC algorithm. 
In the following, $\btheta$ includes all the variables not explicitly mentioned in the conditioning. 
Also, $P_{j}$ denotes the length of each $\bbeta_{j,\cdot,r_{j}}$. We have $P_{1}=P_{2}=K, P_{3}=L$.
The sampler cycles through the following steps of updating each parameter from its posterior full conditional. 

\begin{enumerate}
    \item Update the variance $\sigma_{\varepsilon}^{2}$: 
    \bse
        & (\sigma_{\varepsilon}^{2} \mid \bY, \bX,\bbeta_{j}, \cG, \bnu, \btheta)  \sim \IG(a^{\star}, b^{\star}),\\
        & \text{where} ~~~a^{\star} = a+\frac{K(T-L)}{2} + \frac{R_{1}K}{2} + \frac{R_{1}K}{2} + \frac{R_{3}L}{2} + \frac{R_{1}R_{2}R_{3}}{2} + \frac{K}{2},\\
        & b^{\star} = b + \frac{1}{2}\sum_{t=L}^{T} (\by_{t} - \bmu_{t})\trans(\by_{t} - \bmu_{t}) + \frac{1}{2}\sum_{j=1}^{3}\sum_{r=1}^{R_{j}}\frac{\psi_{j,r}}{\lambda_{j}^{2}} \bbeta_{j,\cdot,r}\trans\bD_{\tau_{\beta_{j}}}^{-1}\bbeta_{j,\cdot,r}, \\
        & + \frac{1}{2} \sum_{p=1}^{R_1}\sum_{q=1}^{R_2}\sum_{r=1}^{R_3} \frac{g_{r_{1}, r_{2}, r_{3}}^{2}}{\lambda_{g}^{2} \tau_{g,r_{1}, r_{2}, r_{3}}^{2}} + \frac{1}{2\lambda_{\nu}^{2}}\bnu^{T}\bD_{\nu}\bnu,\\
        & \text{and} ~~~ \bmu_{t} = \bnu + \bB\bx_{t}.
    \ese

    \item Update the component specific local shrinkage parameters $\btau_{\beta_{j}}, ~ \btau_{g}, ~ \btau_{\nu}$:
    \bse
        & (\tau_{j,k,r}^{2} \mid \lambda_{j}^{2}, \sigma_{\varepsilon}^{2},  \beta_{j,k,r}, \phi_{j,k,r}) \sim \IG\left(1, \frac{1}{\phi_{j,k,r}} + \frac{\beta_{j,k,r}^{2}}{2\lambda_{j}^2 \sigma_{\varepsilon}^2}\right),\\
        & (\tau_{g,r_{1}, r_{2}, r_{3}}^{2} \mid \lambda_{g}^{2}, \sigma_{\varepsilon}^{2},  \cG, \phi_{g, r_{1}, r_{2}, r_{3}}) \sim \IG\left(1, \frac{1}{\phi_{g,r_{1}, r_{2}, r_{3}}} + \frac{g_{r_{1}, r_{2}, r_{3}}^{2}}{2\lambda_{g}^2 \sigma_{\varepsilon}^2}\right),\\
       & (\tau_{\nu,k}^{2} \mid \lambda_{\nu}^{2}, \sigma_{\varepsilon}^{2},  \cG, \phi_{\nu, k}) \sim \IG\left(1, \frac{1}{\phi_{\nu,k}} + \frac{\nu_{k}^{2}}{2\lambda_{\nu}^2 \sigma_{\varepsilon}^2}\right).
    \ese
    
     \item Update the component specific regularization parameters $\lambda^{2}$:
    \bse
        & (\lambda_{j}^{2} \mid \btau_{\beta_{j}}^{2}, \bbeta_{j,\cdot,r}, \btheta) \sim \IG\left(\frac{1+\abs{\bbeta_{j}}}{2}, \frac{1}{\xi} + \frac{1}{2\sigma_{\varepsilon}^{2}}\sum_{r=1}^{R_{j}} \sum_{k} \frac{\psi_{j,r}\beta_{j,k,r}^{2}}{\tau_{j,k,r}^{2}}  \right),\\
        & (\lambda_{g}^{2} \mid \btau_{g}^{2}, \cG, \btheta) \sim \IG \left(\frac{1+\abs{\cG}}{2}, \frac{1}{\xi} + \frac{1}{2\sigma_{\varepsilon}^{2}}\sum_{r_{1}}\sum_{r_{2}}\sum_{r_{3}} \frac{g_{r_{1}, r_{2}, r_{3}}^{2}}{\tau_{g,r_{1}, r_{2}, r_{3}}^{2}}\right),\\
        & (\lambda_{\nu}^{2} \mid \btau_{\nu}^{2}, \bnu, \btheta) \sim \IG \left(\frac{1+K}{2}, \frac{1}{\xi}+\frac{1}{2\sigma_{\varepsilon}^{2}} \sum_{k} \frac{\nu_{k}^{2}}{\tau_{\nu}^{2}}\right),
    \ese
    where $\abs{.}$ denotes the cardinality.
    
    \item Update the multiplicative gamma process components $\delta_{j}^{(\ell)}$:
    \bse
        & (\delta_{j}^{(1)} \mid \bbeta_{j}, \btau_{\beta_{j}}^{2}, \lambda_{j}^{2},\btheta) \sim \Ga \left(a_{1} + \frac{P_{j}R_{j}}{2}, 1+\frac{1}{2}\sum_{r=1}^{R_{j}}\frac{\prod_{\ell=2}^{r}\delta_{j}^{(\ell)}}{\lambda_{j}^{2} \sigma_{\varepsilon}^{2}}\bbeta_{j,\cdot,r}^{\rm T} \bD_{\beta_j}^{-1}\bbeta_{j,\cdot,r}\right),\\
        & (\delta_{j}^{(\ell)}, \ell \geq 2 \mid \bbeta_{j}, \btau_{\beta_{j}}^{2}, \lambda_{j}^{2},\btheta) \sim \Ga \left(a_{2} + \frac{P_{j}R_{j}-\ell+1}{2}, 1+\frac{1}{2}\sum_{r=\ell}^{R_{j}}\frac{\prod_{m=1, m\neq \ell}^{r}\delta_{j}^{(m)}}{\lambda_{j}^{2} \sigma_{\varepsilon}^{2}}\bbeta_{j,\cdot,r}^{\rm T} \bD_{\beta_j}^{-1}\bbeta_{j,\cdot,r}\right).
    \ese
    \item Update $\bpsi_{j}$:
    \bse
        \psi_{j,r} = \prod_{\ell=1}^{r}\delta_{j}^{(\ell)}.
    \ese
    \item Update the factor matrices $\bbeta_{j}$: We update each $\bbeta_{j,\cdot,r_{j}}$, the $r_{j}\th$ column of $\bbeta_{j}$, as a block. 
    To keep track of how the likelihood relates to each $\bbeta_{j,\cdot,r_{j}}$, we create a covariate tensor $\cX_{t}$ for each $t$ such that each frontal slice consists of $(\by_{t-1}, \by_{t-2},\dots, \by_{t-L})$, 
    and $K$ identical slices are then stacked to create a $K \times L \times K$ tensor. 
    We rotate the parameter tensor $\cB$ so that $y_{t,k}$, the $k\th$ element of $\by_{t}$, can be found by summing the element-wise product of each frontal slice of $\cB$ with the corresponding frontal slice of $\cX$.  
    In this new orientation, $\cB$ has covariates as rows, lags as columns, and variables as depth. 
    Each element $x\in \cX_{t}$ be indexed by $(t,d_{2}, d_{3}, d_{1})$ where $d_{j}$ corresponds to the direction controlled by $\bbeta_{j}$. 
    Note the switching of the order of the indices to match the orientation of $\X_{t}$. 
    Suppose we are updating $\bbeta_{j=1, \cdot, r_1 = r}$.  
    For $k=1,\dots,K$ and $p=1,\dots,P_{j}$, define 
    \bse
         h^{(r_{1}=r)}_{t,j=1,k, p} = \sum_{d_{1}=1,d_{2} = 1, d_{3}=1}^{K,L,K} \left\{\Ind(d_{1}=k) \Ind(d_{j} = p) x_{t, d_{2},d_
         {3},d_{1}} \sum_{r_{2}, r_{3}}^{R_{2},R_{3}} g_{r, r_{2},r_{3}}\bbeta_{2, r_{2}} \circ \bbeta_{3, r_{2}}\right\}
    \ese
    For each $\bbeta_{j=1,\cdot,r_{1}=r}$, we can then create a $K \times P_{j}$ matrix  
    \bse
    \bH^{(r)}_{t,j=1} =
    \begin{bmatrix}
        h^{(r)}_{t,j,1,1} & \hdots & h^{(r)}_{t,j,1,P_{j}}\\
        \vdots & \ddots & \vdots\\
        h^{(r)}_{t,j,K,1} &\hdots & h^{(r)}_{t,j,K,P_{j}}
    \end{bmatrix}.
    \ese
    Additionally, define $\wt{\by}_{t} = \by_{t} - \bnu - \bB^{\star} \bx_{t}$, where $\bB^{\star}$ is the mode-3 matricization of $\cB^{\star} = \sum_{r_1' \neq r, r_2, r_3} g_{r_1',r_2, r_3 }\bbeta_{1,\cdot,r_{1}'} \circ  \bbeta_{2,\cdot,r_{2}} \circ  \bbeta_{3,\cdot,r_{3}}$, and $\bx_{t}^{KL \times 1} = (\by_{t-1}\trans,\dots,\by_{t-L}\trans)\trans$. 
    Then the full conditional of $\bbeta_{j=1,\cdot,r_{1}=r}$ is
    \bse
        & (\bbeta_{j=1,\cdot,r_{1}=r} \mid \bnu, \sigma^{2}_{\varepsilon}, \btau_{\beta_{j,r}}, \btheta) \sim \MVN_{P_{j}}(\bmu_{j,r}, \bSigma_{j,r}),\\
        &\text{where}~~~\bSigma_{j,r} = \left(\frac{\sum_{t=1}^{T}\bH_{t,j}^{(r)T}\bH_{t,j}^{(r)} +  \psi_{j,r}\lambda_{j}^{-2}\bD_{\tau_{\beta_{j,r}}}^{-1}}{\sigma_{\varepsilon}^{2}}\right)^{-1} 
        ~~~\text{and}~~~\bmu_{j,r} = \bSigma_{j,r}\left(\frac{\sum_{t=1}^{T} \bH^{(r)T}_{t,j} \wt\by_{t}}{\sigma_{\varepsilon}^{2}}\right).
    \ese
    The vectors $\bbeta_{2,\cdot,r_{2}}$ and $\bbeta_{3,\cdot,r_{3}}$ can be updated analogously.

    \item Update the core tensor $\cG$: 
    \bse
        & (g_{r_{1}, r_{2}, r_{3}} \mid \bbeta_{j}, \bnu, \bY, \bX, \btheta) \sim \Normal(\mu_{g}, \sigma_{g}^{2}), \\
        & \text{where}~~~\mu_g = \left(\frac{1}{\sigma_{\varepsilon}^{2}} \sum_{t = L}^{T}  \bomega_{t}\trans \wt{\by_{t}}\right) \sigma_{g}^{2},
        ~~~\sigma_{g}^{2} = \frac{1}{\sigma_{\varepsilon}^{2}}\left(\sum_{t} \bomega_{t}\trans \bomega_{t} + \frac{1}{\lambda_{g}^{2} \tau_{r_{1}, r_{2}, r_{3}}^{2}}\right)^{-1},\\
        & \bomega_{t} = (\bbeta_{1,\cdot,r_{1}} \circ \bbeta_{2,\cdot,r_{2}} \circ \bbeta_{3,\cdot,r_{3}})_{(1)} \bx_{t},\\
        & \text{and}~~~\wt\by_{t} = \by_{t} - (\sum_{r_1', r_2', r_3' \neq r_1, r_2, r_3} g_{r_{1}' r_{2}' r_{3}'} \bbeta_{1,\cdot,r_{1}'} \circ  \bbeta_{2,\cdot,r_{2}'} \circ  \bbeta_{3,\cdot,r_{3}')})_{(1)} \bx_{t}.
    \ese
    
     \item Update the intercept $\bnu$:
    \bse
        & (\bnu \mid \bB, \sigma_{\varepsilon}^{2}, \btau_{\nu}, \btheta) \sim \MVN\left(\bmu_{\nu}, \bSigma_{\nu}\right),\\
        & \text{where}~~~\bSigma_{\nu} = \left(\frac{T}{\sigma_{\varepsilon}^{2}}\bI_{K} + \frac{\bD_{\tau_{\nu}}^{-1}}{\lambda^2\sigma_{\varepsilon}^2}\right)^{-1}
        ~~~\text{and}~~~\bmu_{\nu} = \bSigma_{\nu}\left\{\frac{1}{\sigma_{\varepsilon}^{2}}\sum_{t=1}^{T}(\bB \bx_{t} - \by_{t})\right\}.
    \ese    
    
    \item Update $\bphi$ (see Section 4.3 in the main paper):
    \bse
        & (\phi_{j,k,r}\mid \tau_{j,k,r}^{2}) \sim \IG\left(1,1+\frac{1}{\tau_{j,k,r}^{2}}\right),\\
        & (\phi_{g, r_{1}, r_{2}, r_{3}}\mid \tau_{g, r_{1}, r_{2}, r_{3}}^{2}) \sim \IG\left(1,1+\frac{1}{\tau_{g, r_{1}, r_{2}, r_{3}}^{2}}\right),\\
        & (\phi_{\nu,k}\mid \btau_{\nu}^{2}) \sim \IG\left(1,1+\frac{1}{\tau_{\nu,k}^{2}}\right).
    \ese
    \item Update $\xi$ (see Section 4.3 in the main paper):
    \bse
        (\xi \mid \lambda_{j}^{2}, \lambda_{g}^2, \lambda_{\nu}^{2}) \sim \IG\left(\frac{1+3+1+1}{2},1+\sum_{j=1}^{3}\frac{1}{\lambda_{j}^2} + \frac{1}{\lambda_{g}^{2}} + \frac{1}{\lambda_{\nu}^{2}}\right).
    \ese
    
\end{enumerate}

\subsection{BPTDVAR}

Posterior calculations for the BPTDVAR model follows similarly as the BTDVAR model.
We assign separate local shrinkage parameters $\btau_{\bbeta_1}^{fixed}$ and $\btau_{\bbeta_1}^{i}$ to $\bbeta_{1}^{fixed}$ and $\bbeta_{1}^{i}$, respectively. 
To update the $\bbeta_{1}^{fixed}$, we redefine the $\wt{\by}_{t}$ from step 6 of the BTDVAR posterior computation as $\wt{\by}_{i,t} = \by_{i,t} - \bnu - \balpha_{i}- \bB^{\star}_{i} \bx_{t}$ and sum the likelihood contribution across all $i$. 
To update $\bbeta_{1}^{i}$, define $\wt{\by}_{i,t} = \by_{i,t} - \bnu - \balpha_{i}- \bB^{\star}_{fixed} \bx_{t}$ and use the likelihood contribution of each $\wt{\by}_{i,t}$ to update each $\bbeta_{1}^{i}$. 
We use $\bbeta_{1}^{random,i} = \bbeta_{1}^{fixed} + \bbeta_{1}^{i}$ to compute $\bB_i$ for updating the other parameters, and sum across $i$ to account for the input from all subjects.

\newpage
\section{Simulation Studies} \label{sec: Simulations SM}
In this section, we use two simulation scenarios to demonstrate the BTDVAR and BPTDVAR's ability to recover VAR parameters and accurately estimate the associated GC networks. 
For comparison, we report the performance of the ordinary least squares (OLS) model, and the single-subject Bayesian VAR (SSBVAR) developed by \cite{ghosh2018high}.

\vskip 3mm
{\bf Scenario 1:}
In the first experiment, we test our models' ability to estimate small $(K=10)$, medium $(K=50)$, and large $(K=200)$ dimensional VARs using $T=150$ and $T=500$ observations, 
where the bottom right quadrant of each of $L_{true}=4$ transition matrices is set to zero, 
leading to a block diagonal GC network shown in Figure \ref{fig: GC10} in the main paper.
Since we are mainly interested in panel fMRI data where prediction is almost never a goal, 
we do not require the data to be strictly stationary. 
For numerical convenience, 
we still ensure that all simulations are stable by requiring that the eigenvalues of the companion matrix of $\bB$ all lie within the unit circle.
The number of initial lags used is $L=6$.
We set $R_{1} = R_{2} = 10, ~ R_{3} = L$ for small, medium, and large VAR simulations 
\ul{to demonstrate the robustness of our reduced-rank approach for varying dimensions}.  

For the BTDVAR, the OLS and the SSBVAR, we simulate single-subject data with $N=1$.
For the BPTDVAR model, we simulate multi-subject data with $N=10$.
For the multi-subject simulations, we use the same parameters from the single-subject simulations as the shared effects, 
and add mean-zero random effects to give each subject unique random effects. 
In addition, the intercept term for each subject has a shared intercept and a random component as specified by equation (\ref{eq: random-effect}).
\ul{In this way, we kept the true data generating process for the multi-subject simulations very general to evaluate the practical performance of the B(P)TDVAR model in settings which do not conform exactly to our model construction.} 
The results of simulation scenario 1 are summarized in \ref{table:sim-accuracy}.

\vskip 3mm
{\bf Scenario 2:}
In the second simulation scenario, \ul{we showcase our models' ability to recover complex realistic GC networks 
by simulating data using VAR parameters whose sparsity patterns mimic the GC networks recovered from our real fMRI data set} (Figure \ref{fig: SM GC50}).   
We report the performance for a medium $(K=50)$ size VAR with $T=150$ and $R_{1} = R_{2} = 10, ~ R_{3} = L$ for easy comparison with the block-diagonal simulation.
We also include the results for a large $K=200$ VAR with $T=500$, $R_{1}=17, R_{2}=10, R_{3}=L$ 
\ul{which are comparable to the dimensions of our real data application}. 
For these experiments in simulation scenario 2, we choose $L_{true}=2$ and and initial maximal lag  $L=4$.
Again, we simulate single-subject data with $N=1$ for the BTDVAR model, and multi-subject data with $N=10$ for the BPTDVAR model.
We report the results of simulation scenario 2 in Table \ref{table:sim-accuracy2}.
 
\vskip 3mm
{\bf Performance Metrics:}
To measure the goodness of model fits, 
we calculate $R^2$ values for in-sample and one-step-ahead out-of-sample fits, averaged across variables, for each model.
The $R^2$ value is calculated as
\vspace{-4ex}\\
\bse
R^{2} = 1- \frac{\sum_{k=1}^{K}\sum_{t=1}^{T} (y_{k,t} - \wh{y}_{k,t})^2}{\sum_{k=1}^{K}\sum_{t=1}^{T} (y_{k,t}-\bar{y}_{k})^2},
\ese
\vspace{-3ex}\\
where $\wh{y}_{k,t}$ are the fitted values and $\bar{y}_k$ is the mean of the series $k$.
$R^2$ has a domain of $(-\infty,1]$ where values closer to 1 are desirable, and negative values indicate that the model fit is worse than using just the mean of the series. 
While prediction is almost never an objective in fMRI data applications, 
we still evaluate the out-of-sample $R^2$ as an important indicator of model performance. 
\ul{For multi-subject simulations, we report the $R^2$ averaged across all subjects.}

\begin{figure}[h!]
    \centering
    \includegraphics[width=0.5\linewidth]{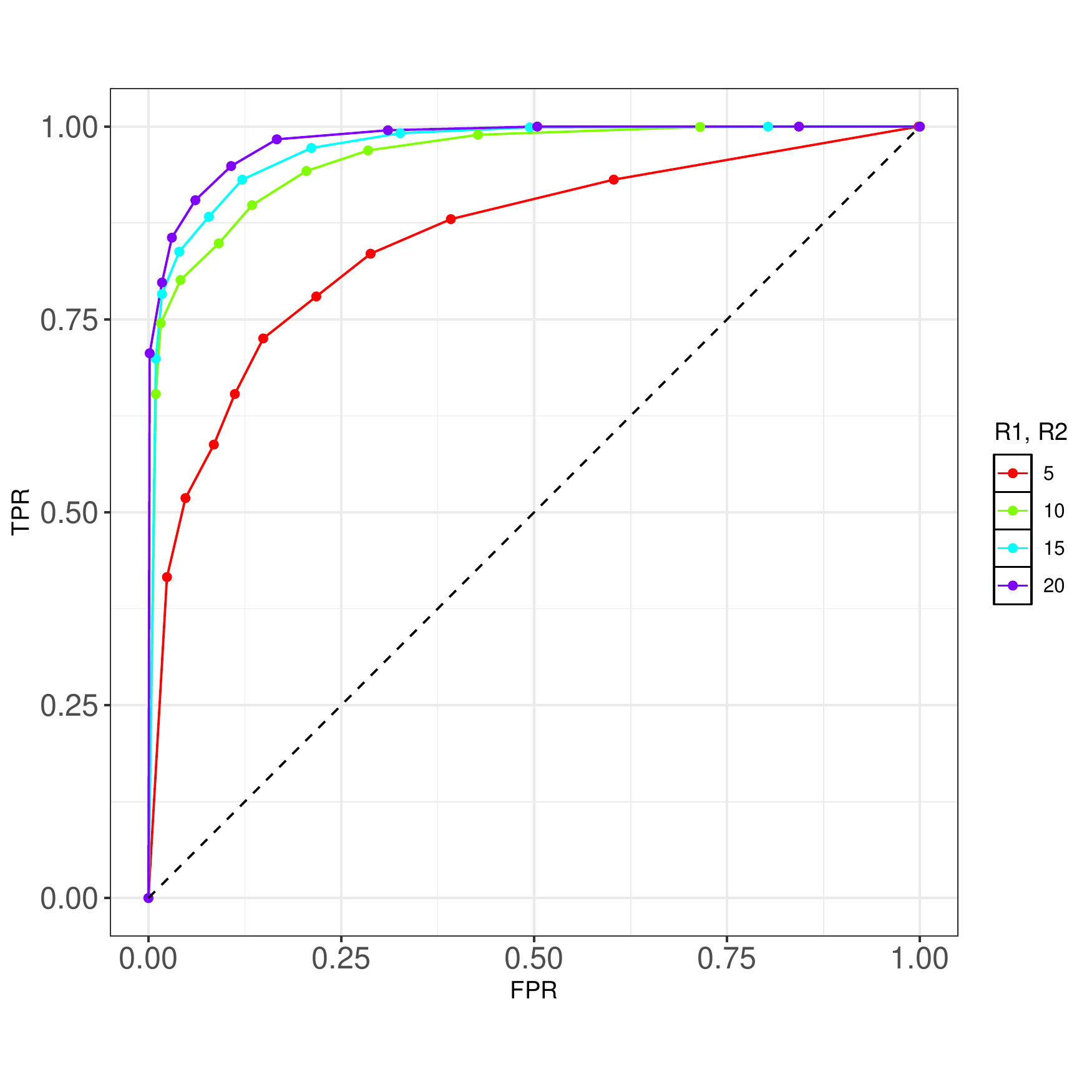}
    \caption{Results from simulated data: ROC curves for the BTDVAR model $K=50, ~T=150$. 
    We test Tucker ranks  $R_{1} = R_{2} \in \{5, 10,  15, 20\}$ to evaluate the impact of rank compression on GC network recovery accuracy. 
    Each ROC curve is generated by allowing the FPR control threshold $t^{\star}$ to vary from [0,1] with a step-size of 0.1.} 
    \label{fig:ROC}
\end{figure}

GC networks are determined using the method described in Section \ref{sec: GCausality}, 
with $c = 1 \Rightarrow t^{\star}= 0.5$, 
which in Figure \ref{fig:ROC} represents that inversion point of each receiving operating characteristic curve.
For the BPTDVAR, the reported shared GC networks are derived from the fixed effects, $\bB^{fixed}$, which is shared between all subjects. 
To evaluate the accuracy of the recovered GC networks, 
we report the true positive rate (TPR), true negative rate (TNR), false positive rate (FPR), and the false negative rate (FNR), defined as 
\vspace{-4ex}\\
\bse
& TPR = \frac{\textrm{no. of discovered true positives}}{\textrm{total no. of true positives}}\times 100,\\
& TNR = \frac{\textrm{no. of discovered true negatives}}{\textrm{total no. of true negatives}}\times 100,\\
& FPR = (1-TNR)\times 100,\\
& FNR = (1-TPR)\times 100.
\ese
\vspace{-4ex}\\
For the TPR and TNR, values closer to 100 are desirable; for the FPR and FNR, values closer to 0 are desirable.

For the SSBVAR method, we use the posterior mean of transition matrix estimates to calculate the model fit, 
and we apply the posterior FDR control method described in Section \ref{sec: GCausality} of the main paper 
to determine the GC network. 
The SSBVAR, as originally proposed, does not include any intercept in the model. 
For fair comparisons, we thus detrend our simulated data before applying the SSBVAR. 
We do not calculate FPR, FNR, etc. for the OLS model since the OLS does not provide estimates of the GC network. 
Simulations for which the OLS and the SSBVAR are unable to produce an estimate altogether due to insufficient number of time points are marked as `NC' in Table \ref{table:sim-accuracy}.

\begin{table}[h!]
\centering
\resizebox{\textwidth}{!}
{\begin{tabular}{  c  c  c  c | c | c  c  c  | c | c  c  c | c  }
\cline{2-13}
\multirow{4}{*}{}& \multicolumn{12}{c}{T=150} \\
\cline{2-13}
 & \multicolumn{4}{c|}{K = 10} & \multicolumn{4}{|c|}{K=50} & \multicolumn{4}{|c}{K=200} \\
\cline{2-13}
& \multicolumn{3}{c|}{N = 1} & N=10 & \multicolumn{3}{|c|}{N = 1} & N=10 & \multicolumn{3}{|c|}{N = 1} & N=10\\
\cline{2-13}
 & SSBVAR & OLS & BTDVAR & BPTDVAR & SSBVAR & OLS & BTDVAR & BPTDVAR  & SSBVAR & OLS & BTDVAR & BPTDVAR\\
\hline 
\multicolumn{1}{c|}{In-sample $R^{2}$}   & 0.405 & 0.939 & 0.900& 0.754  & NC & NC & 0.892 & 0.774  & NC& NC & 0.753 & 0.704\\
\multicolumn{1}{c|}{Out-of-sample $R^{2}$}  & 0.348 & 0.859 & 0.884& 0.623  & NC & NC & 0.860 & 0.709 & NC& NC & 0.702 & 0.631\\
\multicolumn{1}{c|}{TPR}  & 100.0\% & - & 100.0\% & 100.0\%  & NC & - & 88.1\% & 88.6\% & NC & - & 64.6\% & 70.2\%\\
\multicolumn{1}{c|}{TNR}  & 0.0\% & - & 76.0\% & 100.0\%  & NC & - & 88.8\% & 89.2\% & NC & - & 95.4\%& 97.3\%\\
\multicolumn{1}{c|}{FPR}  & 100.0\% & - & 24.0\% & 0.0\%  & NC & - & 11.2\% & 10.8\%& NC& -& 4.6\% &2.7\%\\
\multicolumn{1}{c|}{FNR}  & 0.0\% & - & 0.0\% & 0.0\%   & NC & - & 11.9\% & 11.4\%&NC & - & 34.4\% & 29.8\% \\
\hline
&\multicolumn{12}{ c }{T=500} \tabularnewline
\hline 
\multicolumn{1}{c|}{In-sample $R^{2}$}  & 0.754 & 0.954 & 0.950  & 0.800  & -0.600 & 0.936 & 0.941 & 0.823 &NC& NC & 0.802 & 0.716\\
\multicolumn{1}{c|}{Out-of-sample $R^{2}$}  & 0.738 &  0.917 & 0.928 & 0.682  & -0.907 & 0.837 & 0.935 & 0.817  & NC & NC & 0.751 & 0.686\\
\multicolumn{1}{c|}{TPR}  & 100.0\% & - & 100.0\% & 100.0\%  & 100.0\% & - & 89.9\% & 90\% & NC & -& 68.8\% & 73.5\%\\
\multicolumn{1}{c|}{TNR}  & 4.0\% & - & 76.0\% & 100.0\% & 0.0\% & - & 86.4\% & 89.4\% & NC & - & 99.9\%& 100.0\%\\
\multicolumn{1}{c|}{FPR}  & 96.0\% & - & 24.0\% & 0.0\% & 100.0\% & - & 13.6\% & 10.6\% & NC & - & 0.1\%& 0.0\%\\
\multicolumn{1}{c|}{FNR}  & 0.0\% & - & 0.0\% & 0.0\% & 0.0\% & - & 10.1\% & 10.0\% & NC & - & 31.2\%& 26.5\% \\
\hline\end{tabular}}%
\caption{Results of simulation scenario 1: The GC network of the true parameters used in these simulations is block-diagonal. 
Scenarios for which the OLS and the SSBVAR are unable to produce an estimate are marked as `NC'.}

\label{table:sim-accuracy}
\end{table}

\begin{table}[h!]
\small
\centering
\begin{tabular}{ c  c  c  |  c  c }
\cline{2-5}
 & \multicolumn{2}{c|}{K=50} & \multicolumn{2}{|c}{K=200}\\
\cline{2-5}
& BTDVAR & BPTDVAR & BTDVAR & BPTDVAR \\
\hline 
\multicolumn{1}{c|}{In-sample $R^{2}$} & 0.913 & 0.882 & 0.905 & 0.856\\
\multicolumn{1}{c|}{Out-of-sample $R^{2}$} & 0.889 & 0.876 & 0.900 & 0.841\\
\multicolumn{1}{c|}{TPR} & 81.3\%  & 89.5\% & 82.7\%& 85.6\%\\
\multicolumn{1}{c|}{TNR} & 83.7\%  & 95.8\% & 90.4\% & 93.8\%\\
\multicolumn{1}{c|}{FPR} &  16.3\% & 4.2\% & 9.6\% & 6.2\%\\
\multicolumn{1}{c|}{FNR} & 19.7\% & 10.5\% & 17.3\% & 14.4\%\\
\hline
\end{tabular}%
\caption{Results of simulation scenario 2: The GC network of the true parameters used in these simulations mimics that of the GC network recovered from our real fMRI data set.}
\label{table:sim-accuracy2}
\end{table}

\vskip 3mm
{\bf Results and Findings:}
Under scenario 1, the BTDVAR model produces model fits with high in-sample and out-of-sample $R^2$ for all $K=10,50,200$. 
Importantly, even in settings with $T>>K$, the in-sample $R^{2}$ values obtained by the BTDVAR are very competitive to the OLS, 
while the out-of-sample $R^{2}$ values are in fact significantly better than the OLS. 
Increasing the number of time points from $T=150$ to $T=500$ results in marginal improvements for BTDVAR $R^2$ across all simulation settings, which affirms the BTDVAR's ability to produce excellent estimates in small $T$ settings. 
This result is especially important since neither the SSBVAR nor the OLS were able to produce estimates for $K=50,200$ and $T=150$. 
Though $T=150>K=50$, using a large experiment lag $L=6$ makes $(\bx_{t} \trans \bx_{t})$ singular.
An invertible $\bx_{t} \trans \bx_{t}$ is, however, required for both the OLS and the SSBVAR.

For the medium and large VAR simulations under scenario 1, the BTDVAR and the BPTDVAR models were able to produce good in-sample and out-of-sample model fits ($R^2 > 0.7$) despite $R_{1} = R_{2} = 10$ being much smaller than $K= 50, 200$. 
More evidence to support the efficacy of low rank B(P)TDVAR estimation can be seen in Figure \ref{fig:ROC}, 
where the ROC curves generated by the BTDVAR for values of  $R_{1} = R_{2} \in \{5, 10,  15, 20\}$ for the $K=50$ case are compared.
We can see that the area under the ROC curve is already near optimal for $R_{1},R_{2}\geq10$. 
Furthermore, the substantial jump in accuracy between $R_{1}=R_{2}=5$ and $R_{1}=R_{2}=10$ followed by ever diminishing returns for $R_{1},R_{2}\geq10$ implies that while overly-heavy compression from too-low Tucker ranks can negatively impact the BTDVAR's ability to recover the true positives, 
sufficiently large values of the Tucker ranks that are still much smaller than the original VAR dimensions can vastly improve model performance to near optimality. 
The drastic improvement in the FNR values for $K=200$ simulations between scenario 1, where $R_1 = R_2  = 10$, and scenario 2, where $R_{1}= R_{2} = 17$, also demonstrates this phenomenon. 
These results validate the excellent practical effectiveness of our low rank B(P)TDVAR approaches.

In all simulations, the SSBVAR estimate was unable to recover the underlying sparsity patterns in the true parameters, as seen by the extremely low TNR in Table \ref{table:sim-accuracy}.

\begin{figure}[h!]
    \begin{subfigure}{0.5\textwidth}
      \centering
      \includegraphics[width=\linewidth]{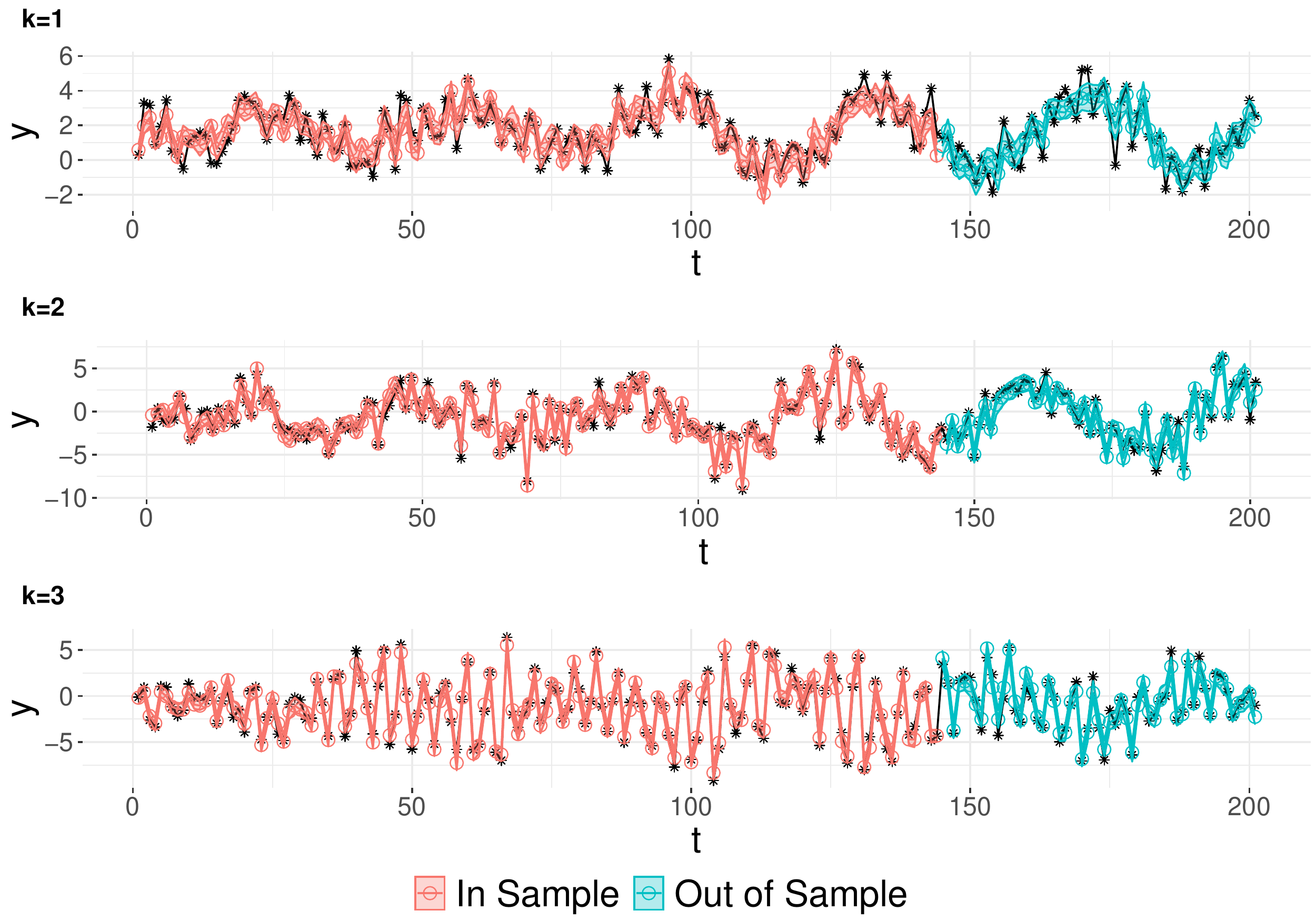}
      \caption{K=10}
    \end{subfigure}
    \bigskip
    \begin{subfigure}{0.5\textwidth}
      \centering
      \includegraphics[width=\linewidth]{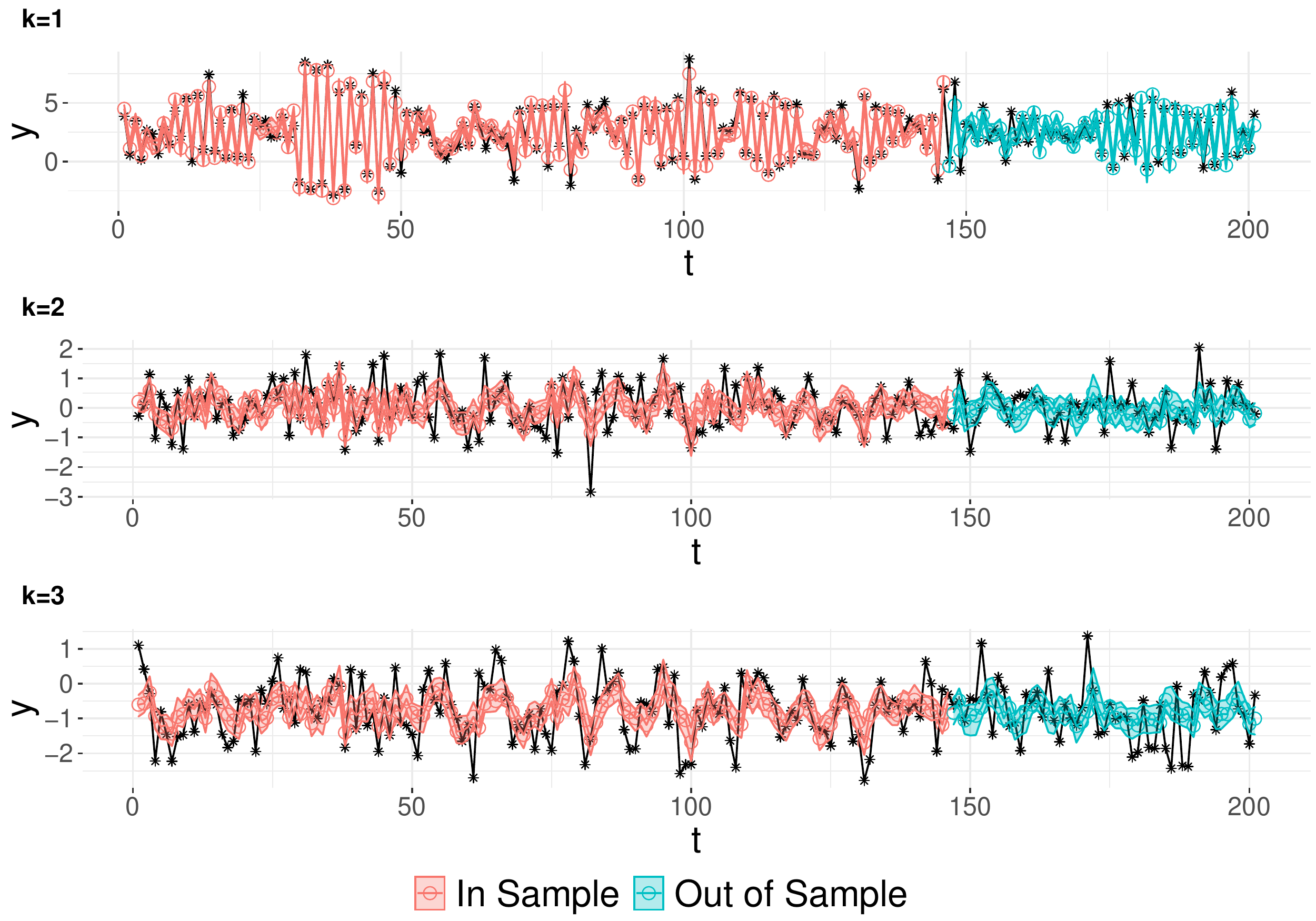}
      \caption{K=50}
    \end{subfigure}
    \caption{Results from simulated data: 
    The true network patterns of each $A_{\ell}$ here is block-diagonal. 
    In-sample (red) and one-step-ahead out-of-sample (teal) BTDVAR fits with 95\% credible intervals for the first three series for the $K=10$ and $50$ cases respectively with $T=150$ initial observations each. }
    \label{fig:BTDVAR_fit}
\end{figure}

\begin{figure}[h!]
    \begin{subfigure}{0.5\textwidth}
      \centering
      \includegraphics[width=0.95\linewidth]{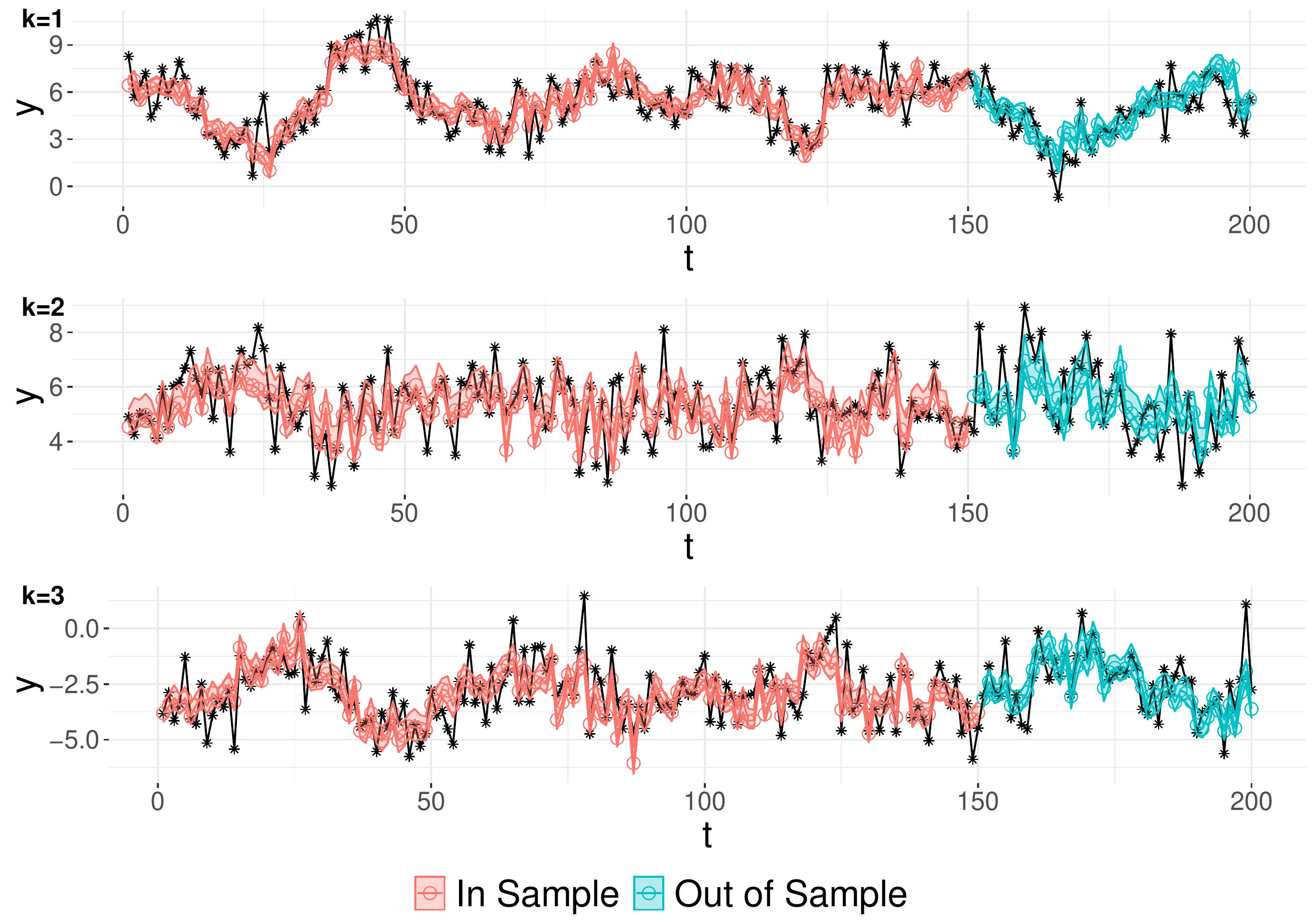}
      \caption{K=10}
    \end{subfigure}%
      \begin{subfigure}{0.5\textwidth}
      \centering
      \includegraphics[width=0.95\linewidth]{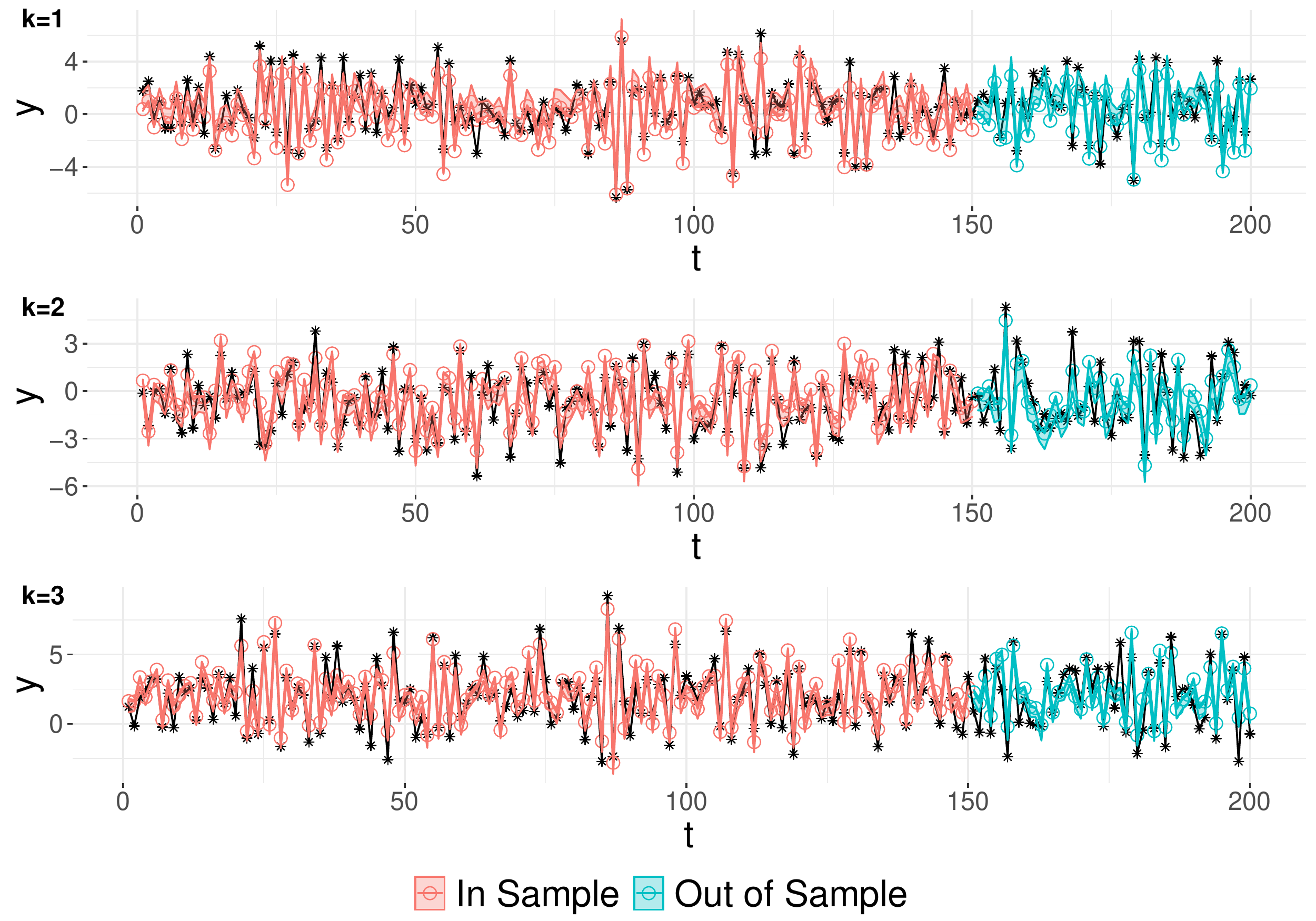}
     \caption{K=50}
    \end{subfigure}
    \caption{Results from simulated data: 
    The true network patterns of each $A_{\ell}$ here is block-diagonal. 
    In-sample (red) and one-step-ahead out-of-sample (blue) BPTDVAR fits with 95\% credible intervals for the first three series for the $K=10$ and $50$ cases respectively with $N=10$ subjects and $T=150$ initial observations each.}
    \label{fig:PTDVAR_fit}
\end{figure}

\begin{figure}
    \centering
    \includegraphics[width=\linewidth]{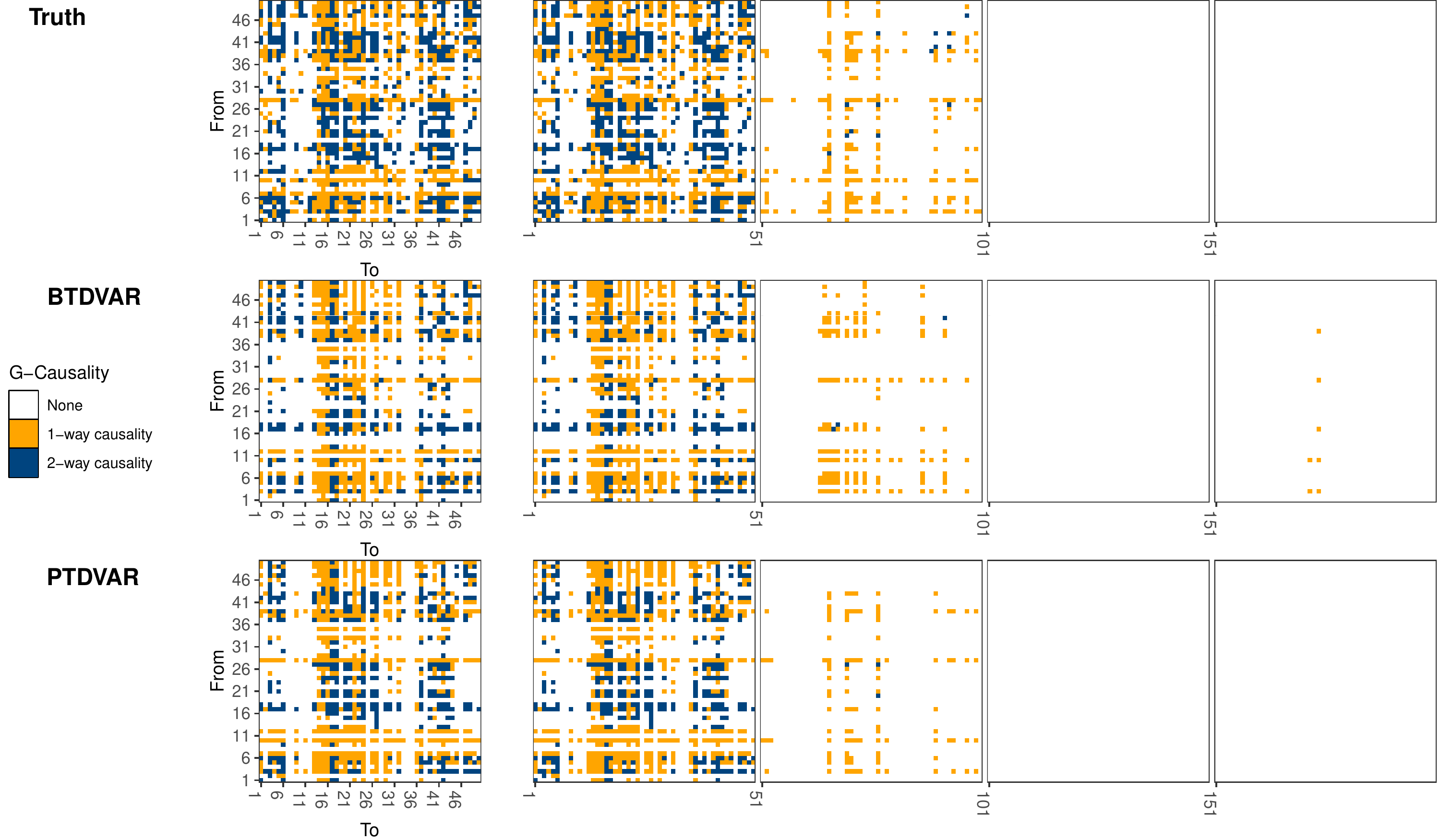}
    \caption{
    Results from simulated data: 
    GC network results obtained by the B(P)TDVAR models for $N=1 (10)$ subject(s) and $T=150$ observations (for each subject when $N=10$) simulated with $L_{true}=4$ lags and $K=50$ variables. 
    The true network patterns of each $A_{\ell}$ here mimic the patterns recovered from real data experiments.
    We used $L=4$ initial experimental lags. 
    The left-most matrix in each row shows the composite network summed across all lags, while the 
    4 blocks to the right show the network of each transition matrix $\bA_{\ell}, ~ \ell=1:4$.
    We see that the B(P)TDVAR have very efficiently eliminated the two extra lags (right-most two empty matrices).
    }
    \label{fig: SM GC50}
\end{figure}

\begin{figure}[!ht]
 \begin{subfigure}{\textwidth}
      \centering
      \includegraphics[width=\linewidth]{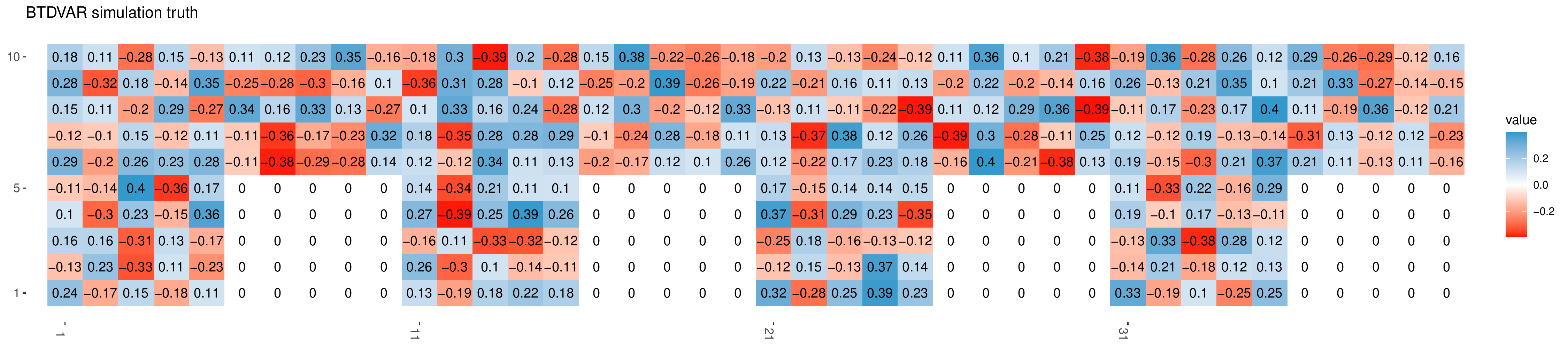}
    \end{subfigure}\\
  \begin{subfigure}{\textwidth}
      \centering
      \includegraphics[width=\linewidth]{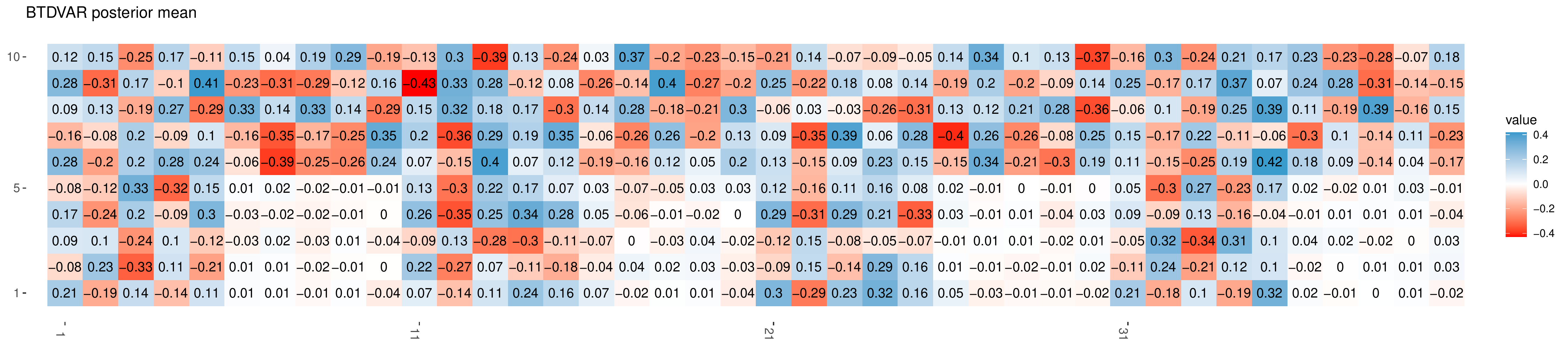}
  \end{subfigure}\\
  \begin{subfigure}{\textwidth}
      \centering
      \includegraphics[width=\linewidth]{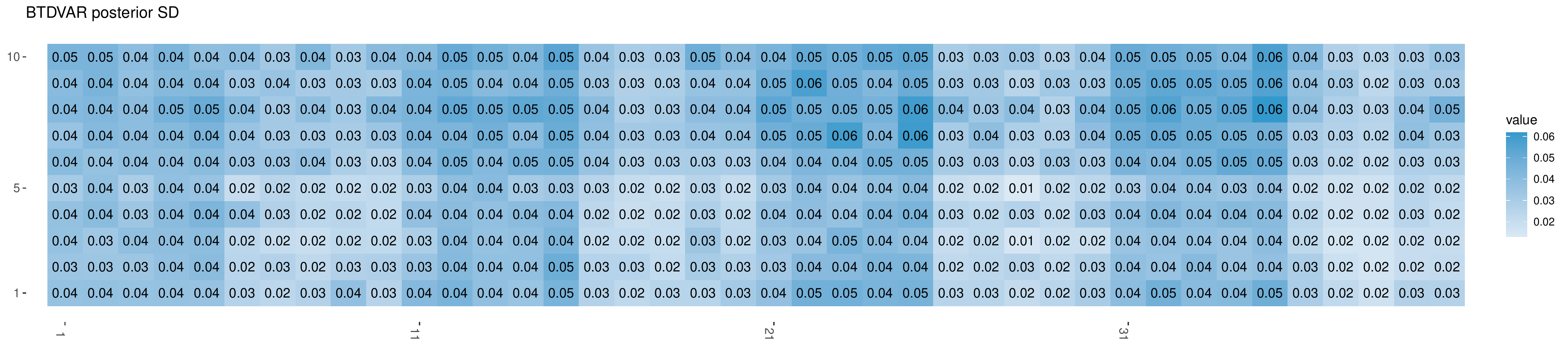}
    \end{subfigure}
  \caption{Results from simulated data: The true data generating VAR parameters $\bB_{true}$ (top), the posterior mean of the VAR coefficients estimated by the BTDVAR model (middle), the posterior standard deviation of the VAR coefficients estimated by the BTDVAR model (bottom).  
  }
  \label{fig: BTDVAR_coeff_comparison}
\end{figure}

\begin{figure}[!ht]
 \begin{subfigure}{\textwidth}
      \centering
      \includegraphics[width=\linewidth]{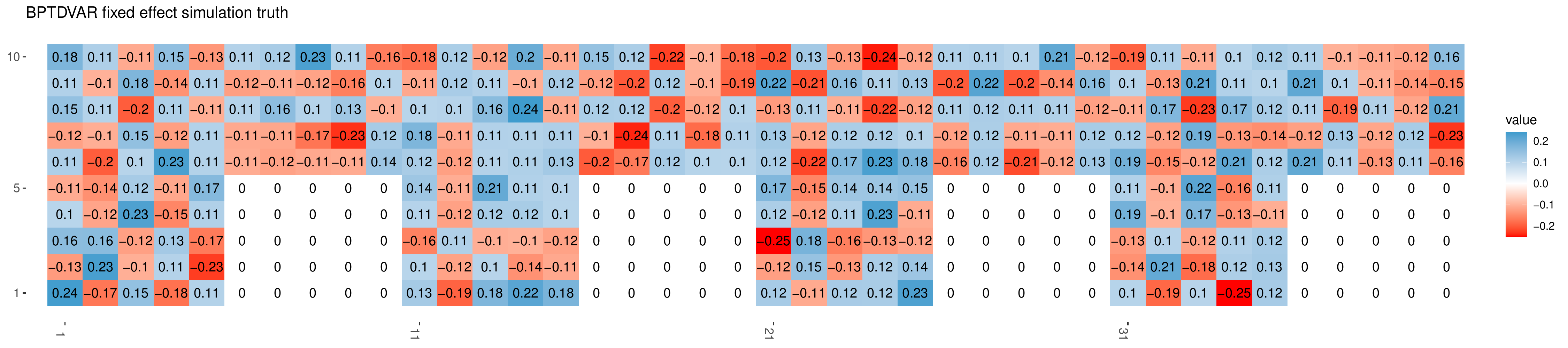}
    \end{subfigure}\\
  \begin{subfigure}{\textwidth}
      \centering
      \includegraphics[width=\linewidth]{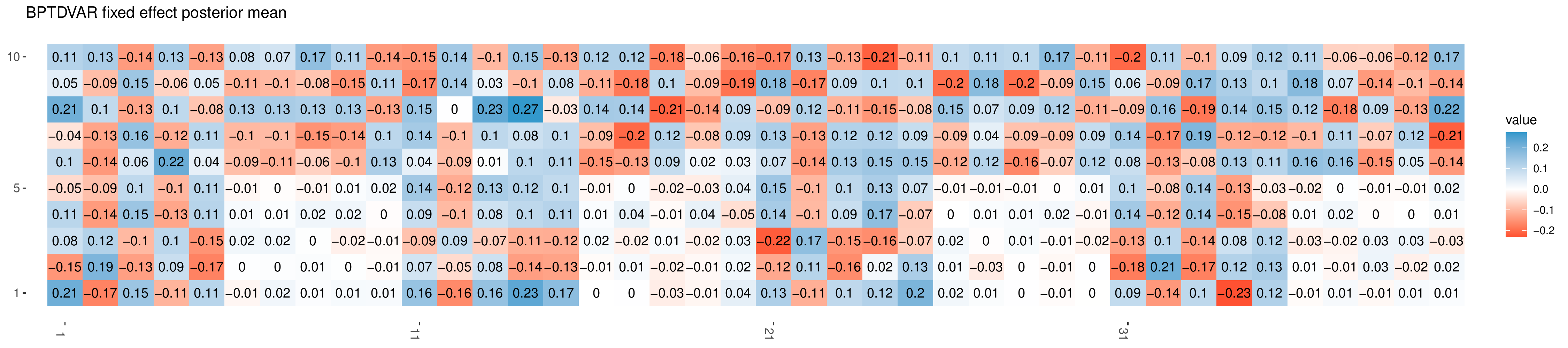}
  \end{subfigure}\\
  \begin{subfigure}{\textwidth}
      \centering
      \includegraphics[width=\linewidth]{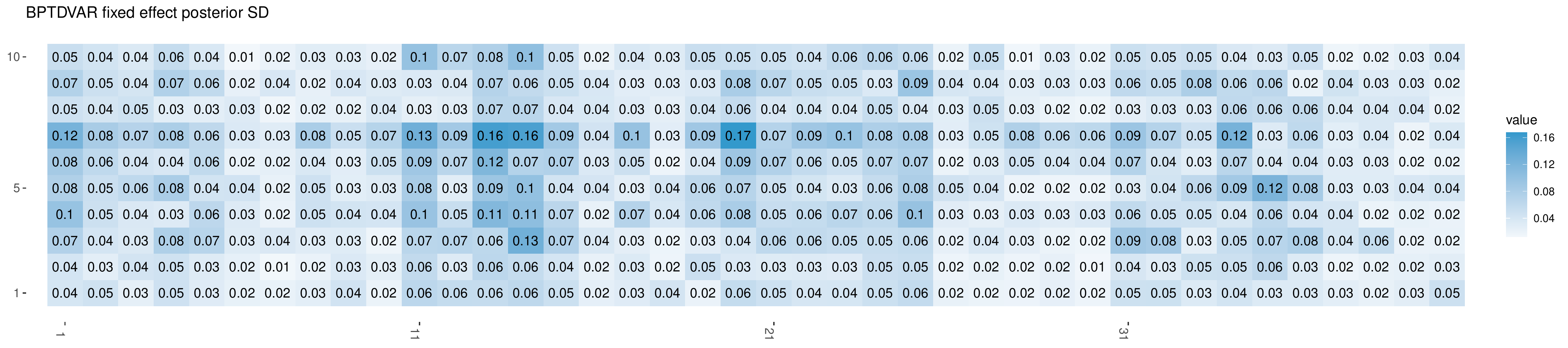}
    \end{subfigure}
  \caption{Results from simulated data: The true data generating fixed-effect VAR parameters $\bB^{fixed}_{true}$ (top), the posterior mean of the fixed-effect VAR coefficients estimated by the BPTDVAR model (middle), the posterior standard deviation of the fixed-effect VAR coefficients estimated by the BPTDVAR model (bottom).  
  }
  \label{fig: PBTDVAR_coeff_comparison}
\end{figure}

\begin{figure}[!ht]
 \begin{subfigure}{0.5\textwidth}
      \centering
      \includegraphics[width=\linewidth]{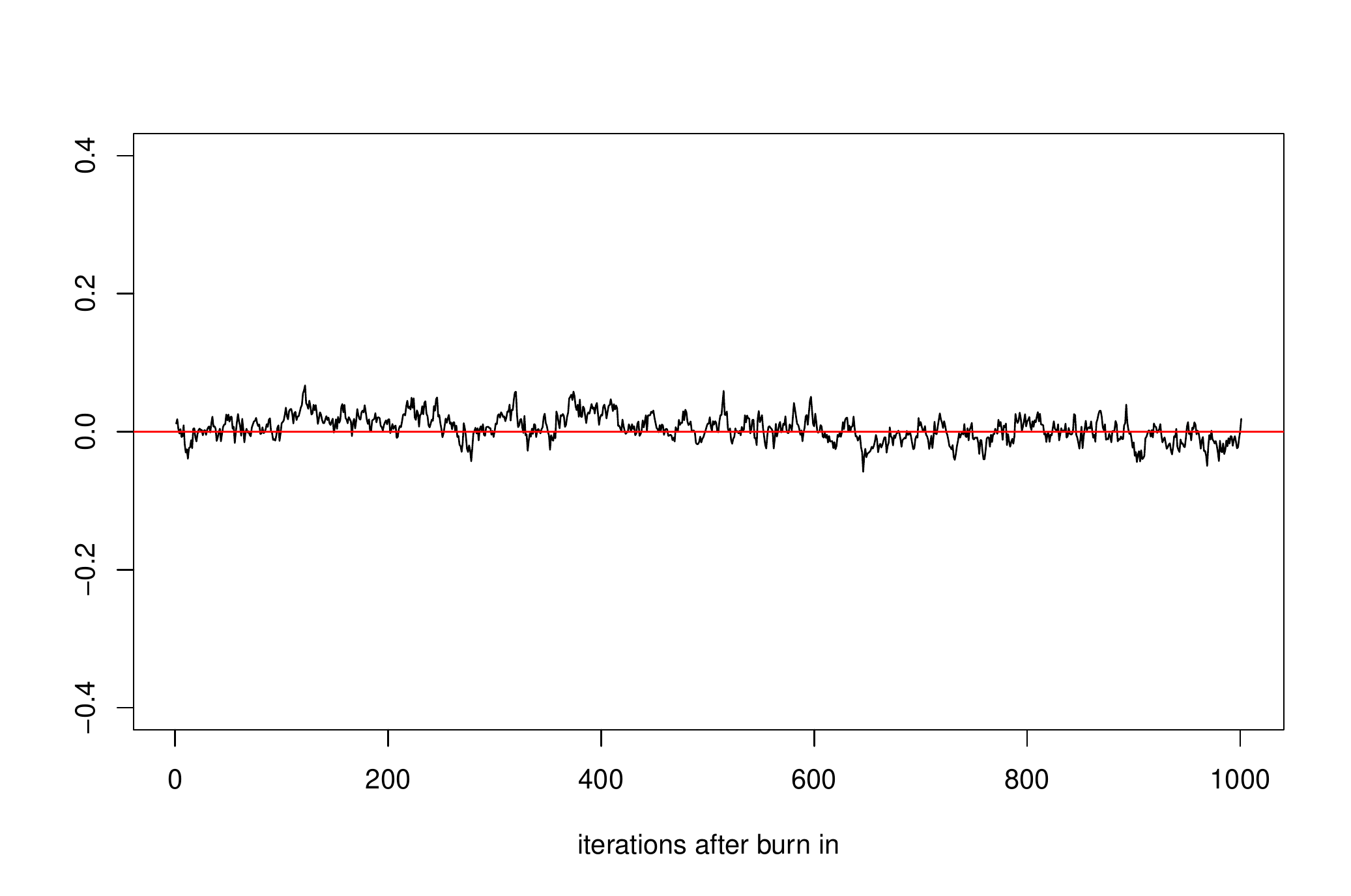}
      \caption{BTDVAR}
    \end{subfigure}
  \begin{subfigure}{0.5\textwidth}
      \centering
      \includegraphics[width=\linewidth]{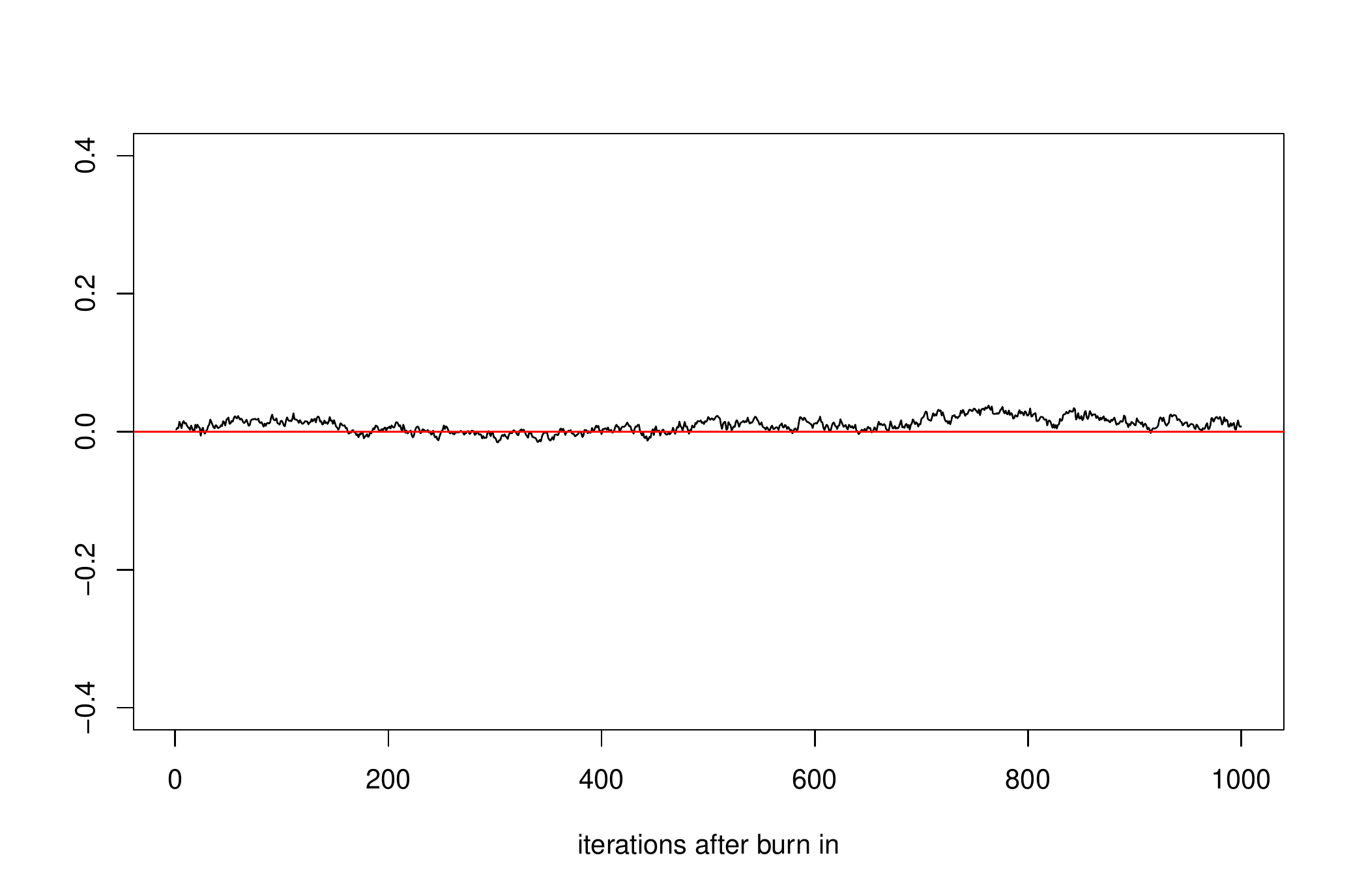}
      \caption{BPTDVAR}
  \end{subfigure}\\
  \begin{subfigure}{0.5\textwidth}
      \centering
      \includegraphics[width=\linewidth]{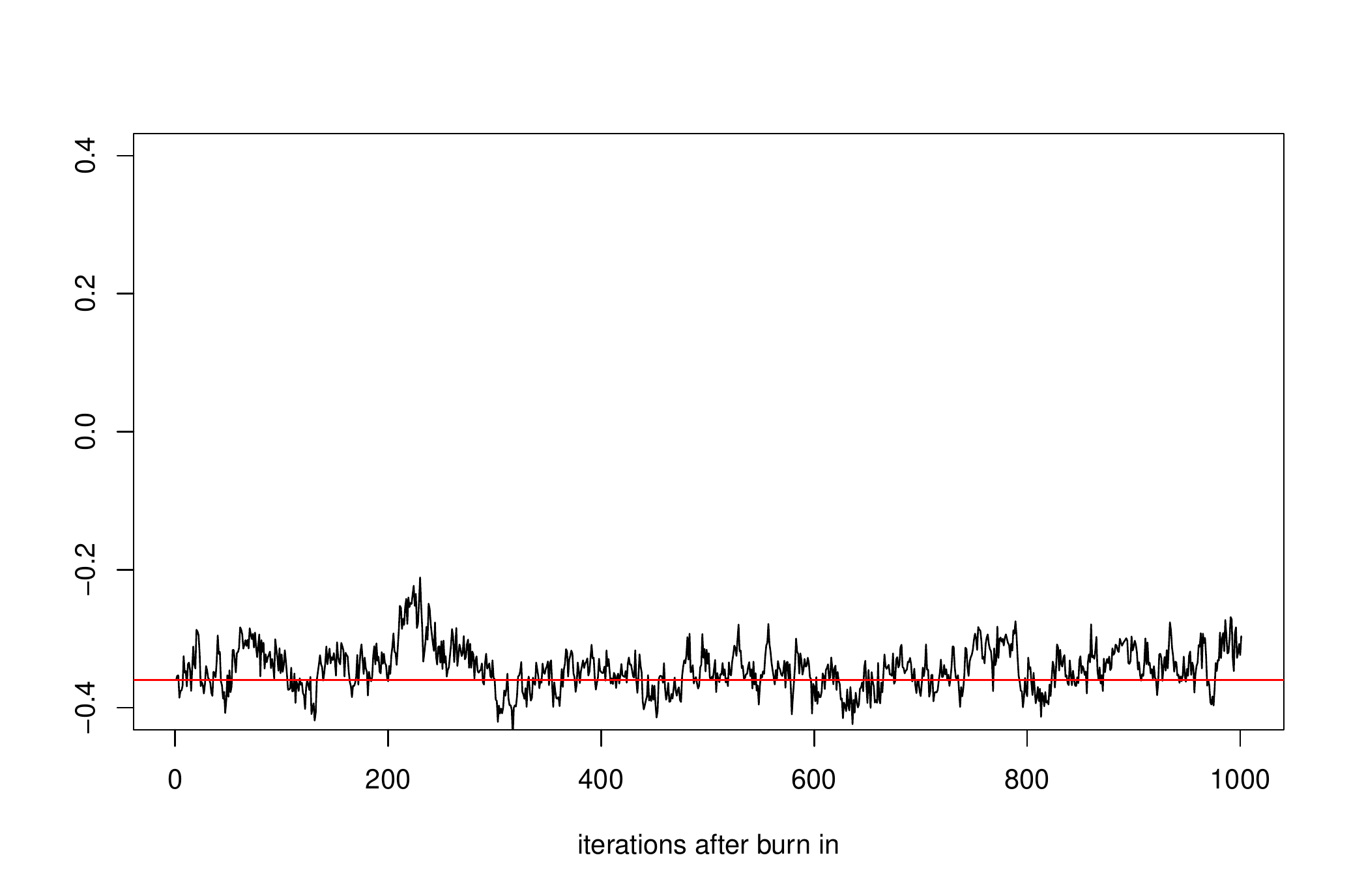}
      \caption{BTDVAR}
    \end{subfigure}
  \begin{subfigure}{0.5\textwidth}
      \centering
      \includegraphics[width=\linewidth]{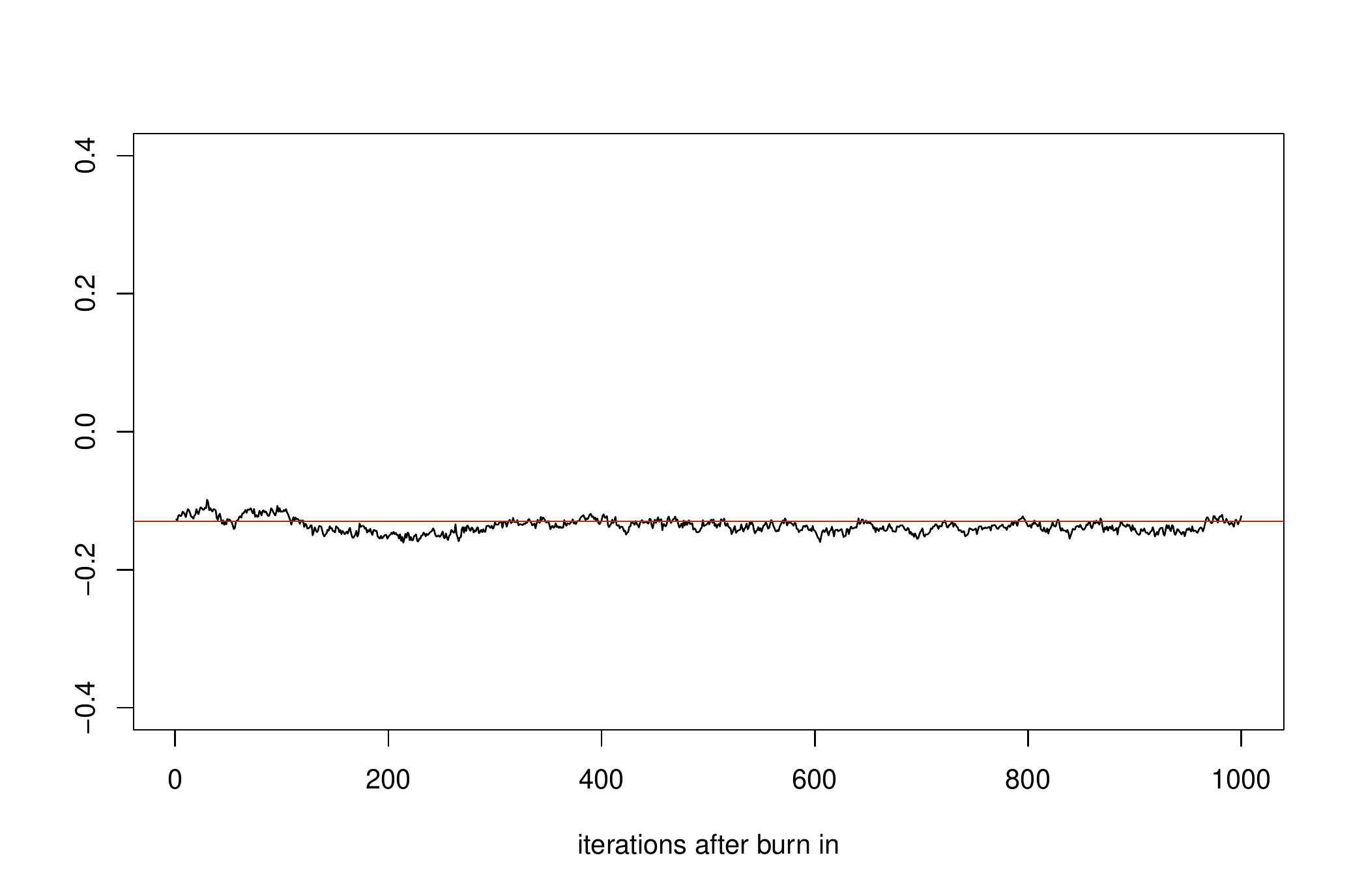}
      \caption{BPTDVAR}
  \end{subfigure}\\
  \begin{subfigure}{0.5\textwidth}
      \centering
      \includegraphics[width=\linewidth]{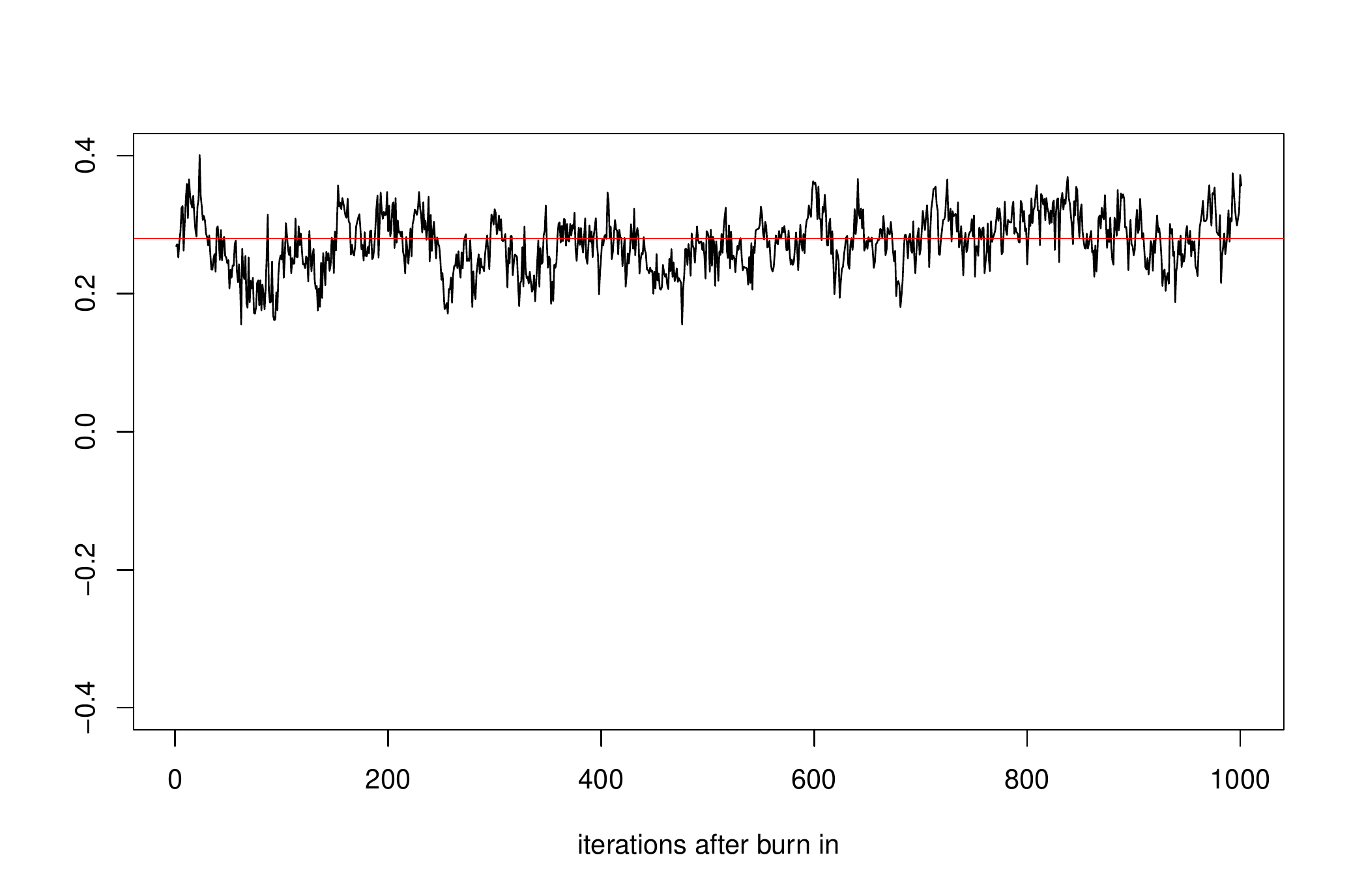}
      \caption{BTDVAR}
    \end{subfigure}
  \begin{subfigure}{0.5\textwidth}
      \centering
      \includegraphics[width=\linewidth]{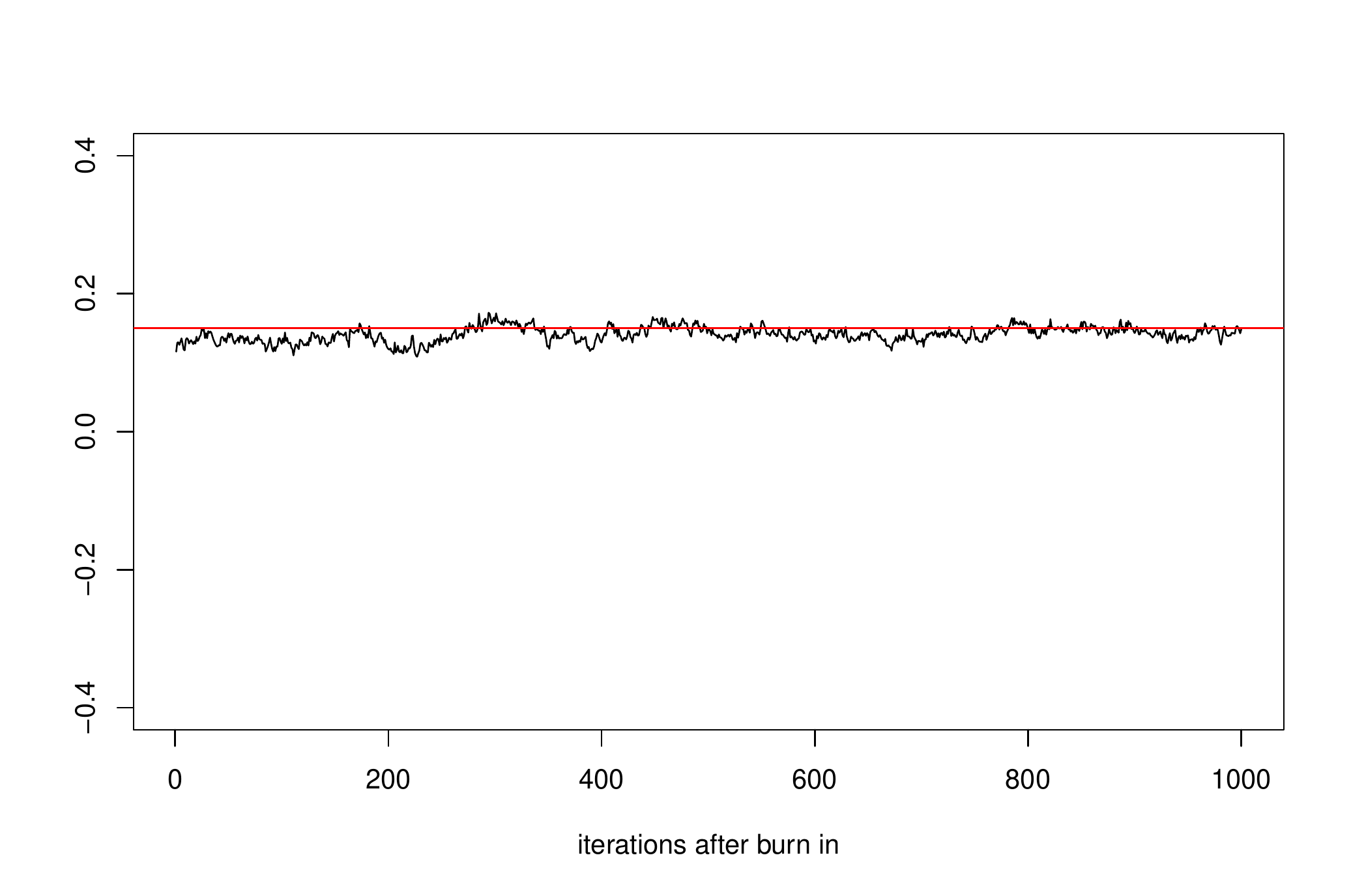}
      \caption{BPTDVAR}
  \end{subfigure}\\
  \caption{Results from simulated data: Trace plots for the BTDVAR (left) and BPTDVAR (right) for three VAR coefficients 
  with the corresponding truths (one zero, one positive, and one negative) shown in red.
  }
  \label{fig: trace_comparison}
\end{figure}

At a first glance, the single-subject BTDVAR may appear to produce better $R^{2}$ values than the BPTDVAR. 
Recall, however, that for the multi-subject BPTDVAR, the reported $R^{2}$ values are obtained by averaging across the $R^{2}$ values for all individuals. 
By sharing information between the individuals, 
the BPTDVAR is indeed able to more accurately recover the underlying shared fixed effects parameters, 
as evidenced by the performance metrics evaluating the recovery of the true shared GC networks -- 
we can see in Tables \ref{table:sim-accuracy} and \ref{table:sim-accuracy2} that the BPTDVAR produces higher TPR and TNR, and lower FPR and FNR as compared to the BTDVAR. 
Figures \ref{fig: GC10} and \ref{fig: GC10 circos} from the main paper 
visually confirm that the GC networks recovered by the BPTDVAR are more accurate than those recovered by the BTDVAR. 
Figures \ref{fig:BTDVAR_fit} and \ref{fig:PTDVAR_fit} provide a visual assessment of the model fits. 
Figures \ref{fig: BTDVAR_coeff_comparison} and \ref{fig: PBTDVAR_coeff_comparison} show the true VAR coefficients and the corresponding posterior means and the associated posterior standard deviations obtained by the B(P)TDVAR models. 
Figure \ref{fig: trace_comparison} shows the trace plots of three VAR coefficients for the B(P)TDVAR models 
with the corresponding truths (one zero, one positive, and one) shown in red.

Figures \ref{fig: GC10} and \ref{fig: GC10 circos} from the main paper also show, 
respectively for the two simulation scenarios, 
the B(P)TDVAR models' ability to recover the true lag-specific GC networks 
as well as their ability to correctly select the true number of lags. 
For both experiments, our initial experimental lag is $L = L_{true} + 2$. 
The lag specific GC matrix estimates show that the BTDVAR is able to shrink most of the parameters in the extra lags to zero, 
while the BPTDVAR is able to shrink all of the parameters in the extra lags to zero.
Figure \ref{fig: GC10 circos} further illustrates the BPTVAR's ability to capture the heterogeneity in subject-specific composite GC networks. 
The agreement between the estimated GC networks and the corresponding simulation truths is clearly excellent.
These results validate the BPTDVAR's ability to very accurately recover the underlying shared parameters 
as well as the heterogeneity across individuals.

\clearpage\newpage
\section{Additional Figures}

 \begin{figure}[h!]
    \centering
    \includegraphics[width=0.5\linewidth, trim={7cm 14cm 7cm 14cm},clip=true]{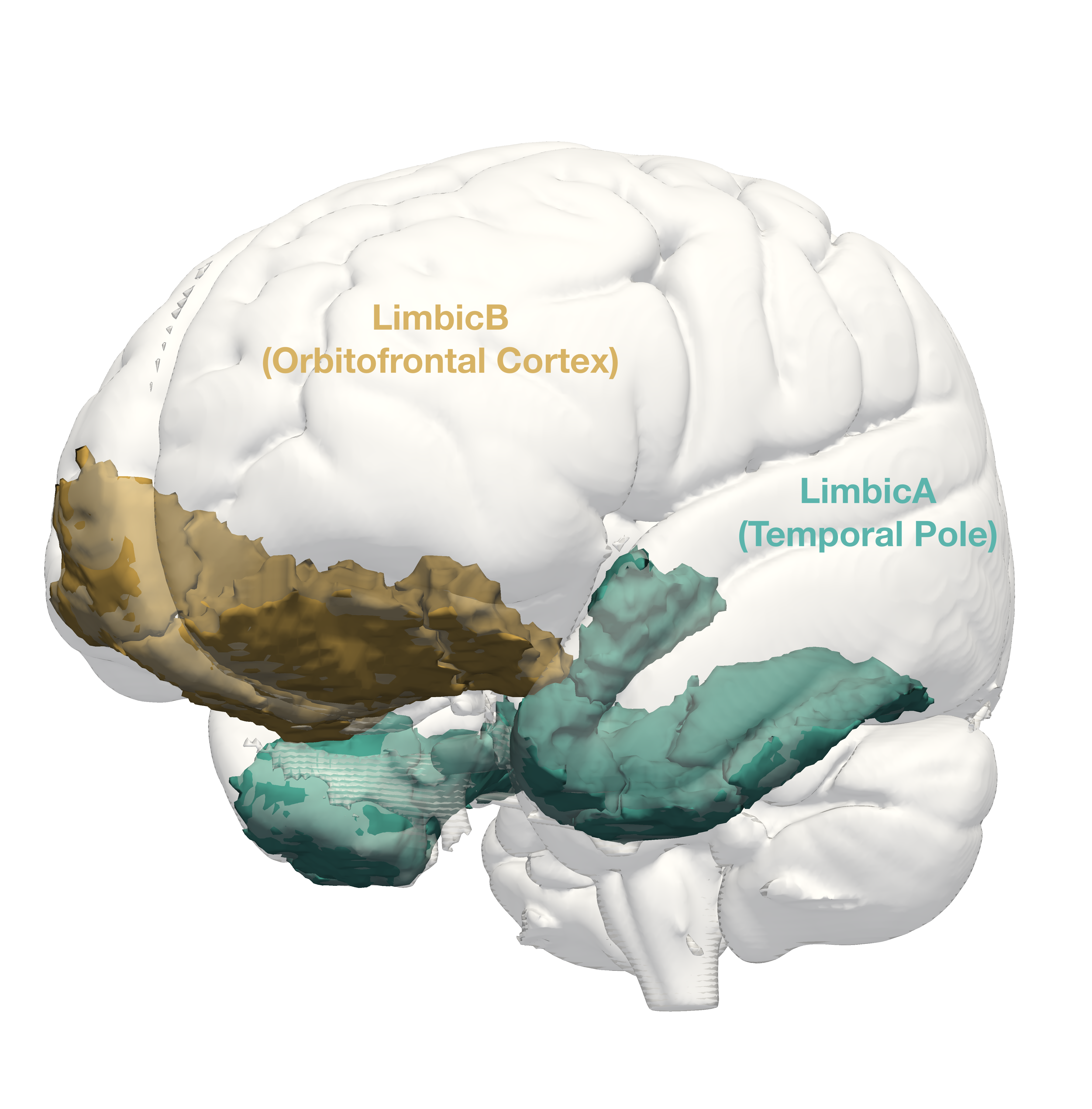}
    \caption{The LimbicA and LimbicB are physically distinct areas of the brain.}
    \label{fig: limbic_networks}
\end{figure}

\begin{figure}[!ht]
 \begin{subfigure}{\textwidth}
      \centering
      \includegraphics[width=0.75\linewidth,trim={0.75cm 0.75cm 0.75cm 0.75cm}, clip=true]{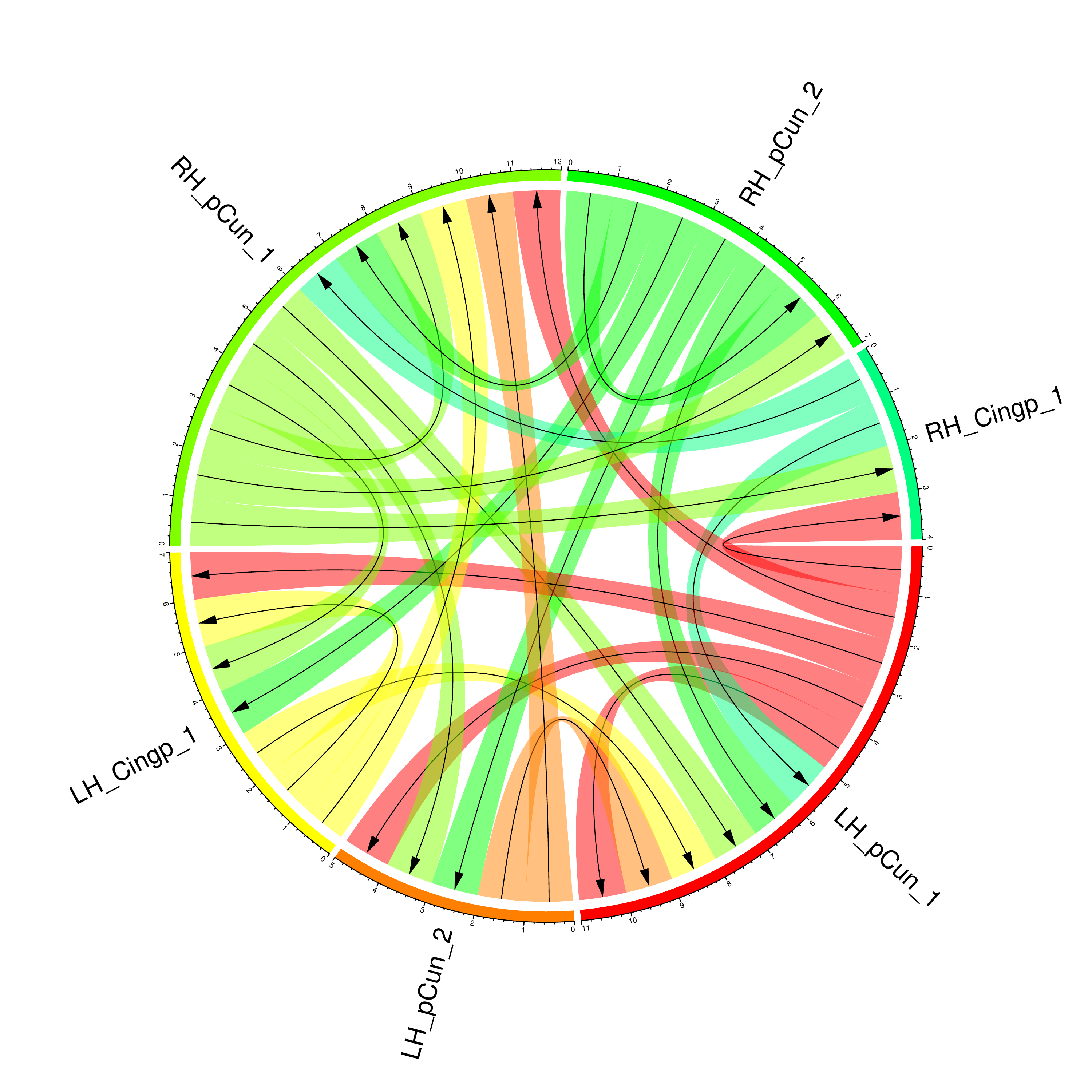}
      \caption{Shared fixed effects network.}
    \end{subfigure}\\
  \begin{subfigure}{0.5\textwidth}
      \centering
      \includegraphics[width=\linewidth]{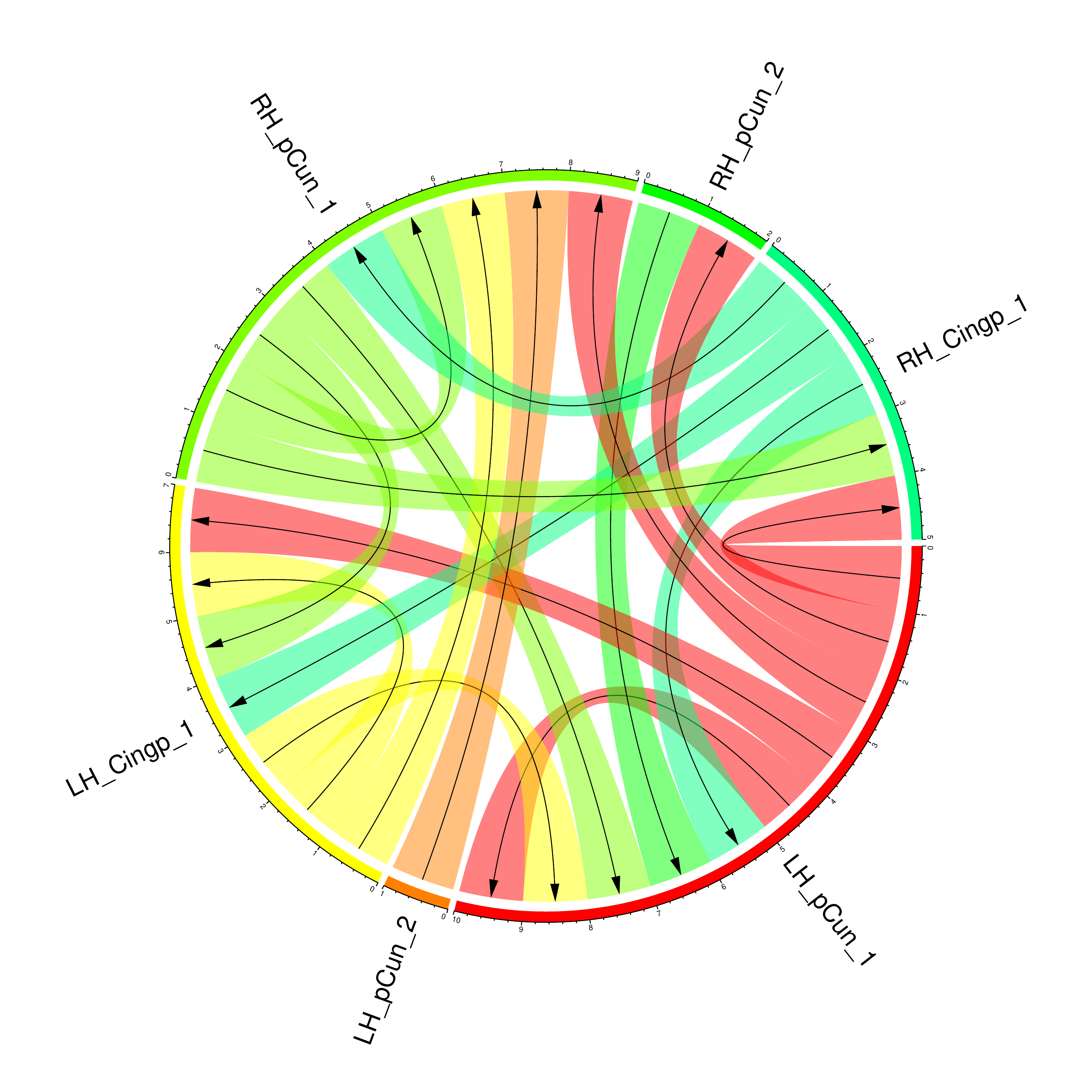}
     \caption{Network for subject 1.}
  \end{subfigure}
  \begin{subfigure}{0.5\textwidth}
      \centering
      \includegraphics[width=\linewidth]{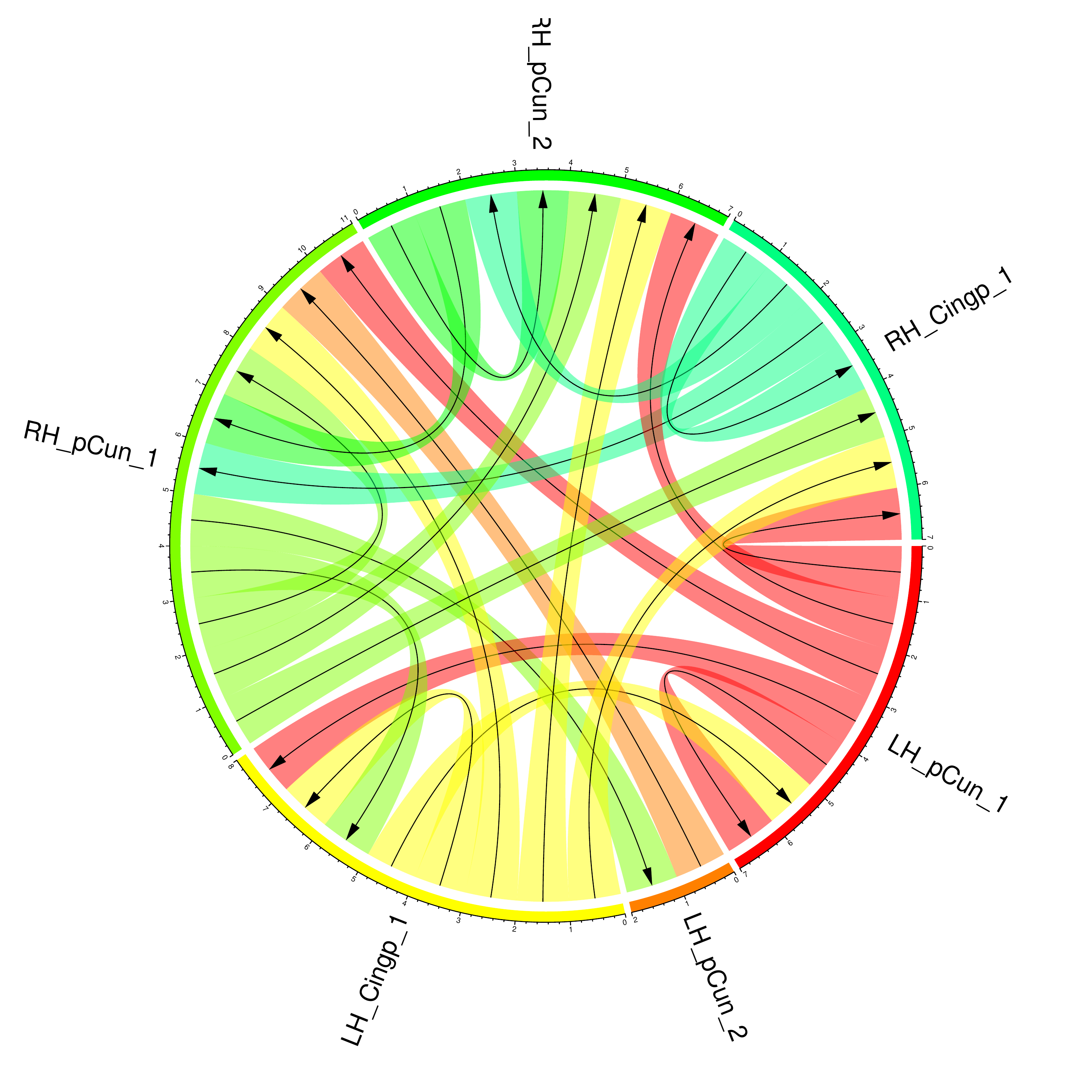}
     \caption{Network for subject 2.}
    \end{subfigure}
  \caption{Results from real data: 
  Shared and subject-specific connectivity patterns for the Control C sub-network. 
  }
  \label{fig: contc_comparison}
\end{figure}

\begin{figure}[!ht]
 \begin{subfigure}{0.5\textwidth}
      \centering
      \includegraphics[width=\linewidth]{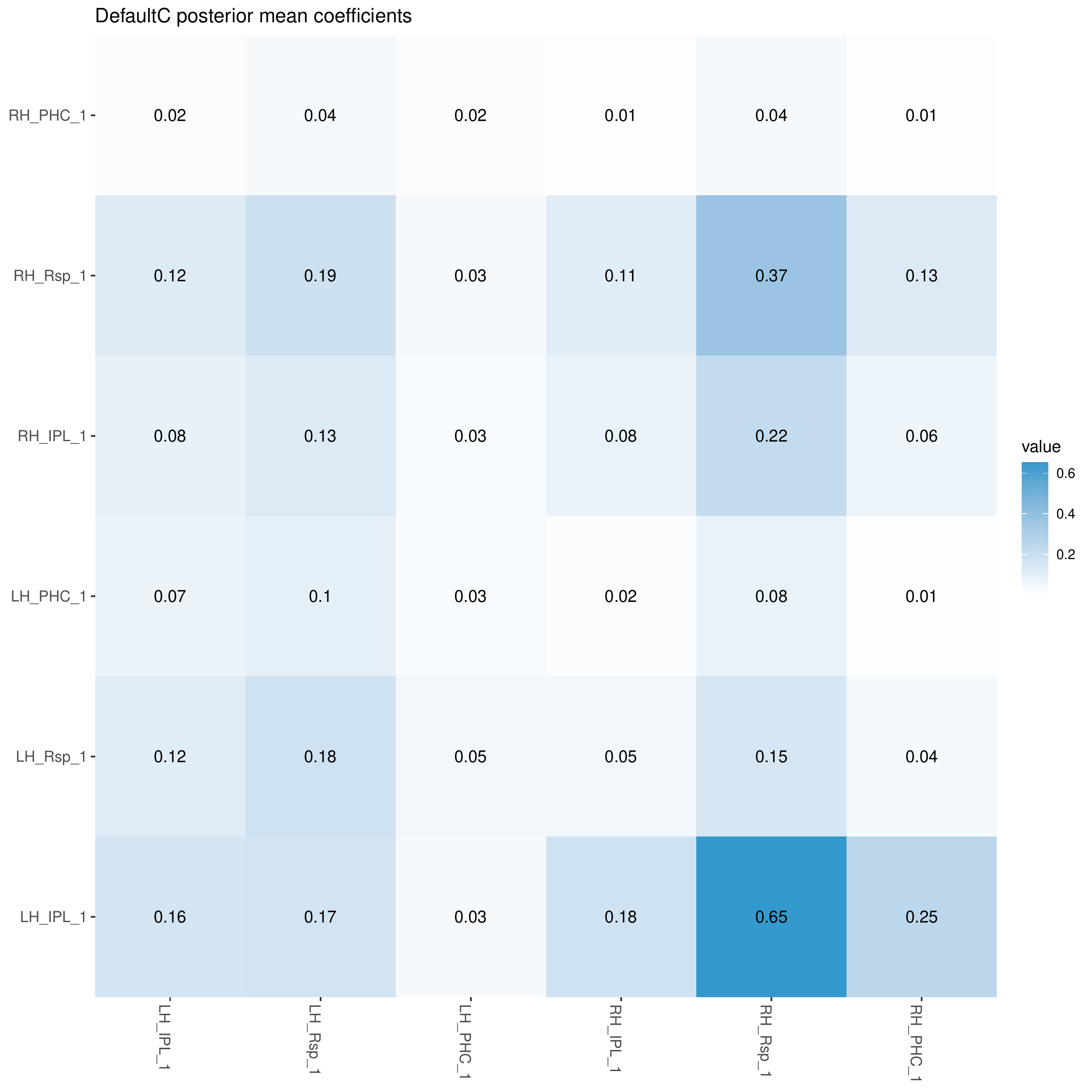}
    \end{subfigure}
  \begin{subfigure}{0.5\textwidth}
      \centering
      \includegraphics[width=\linewidth]{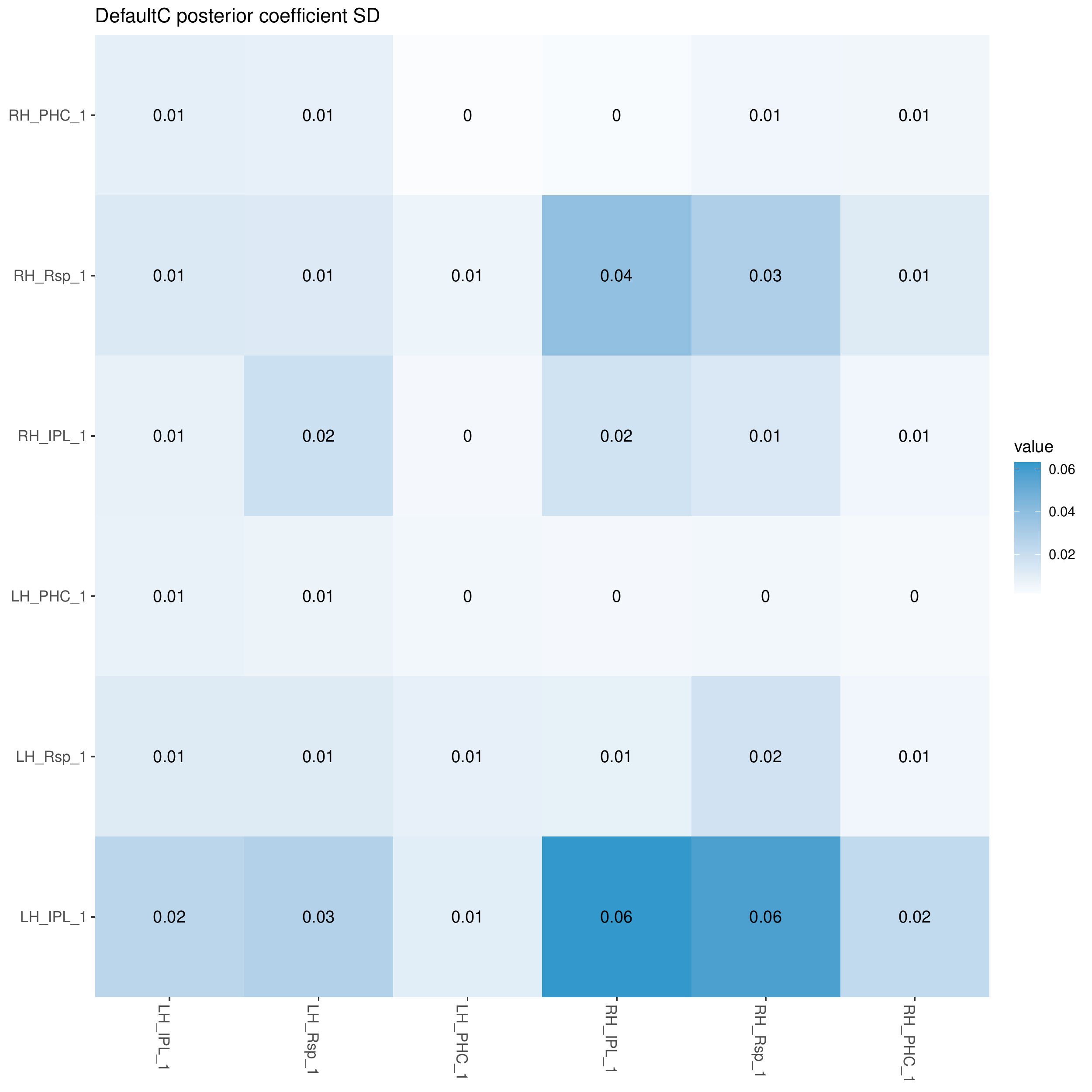}
  \end{subfigure}\\
  \begin{subfigure}{0.5\textwidth}
      \centering
      \includegraphics[width=\linewidth]{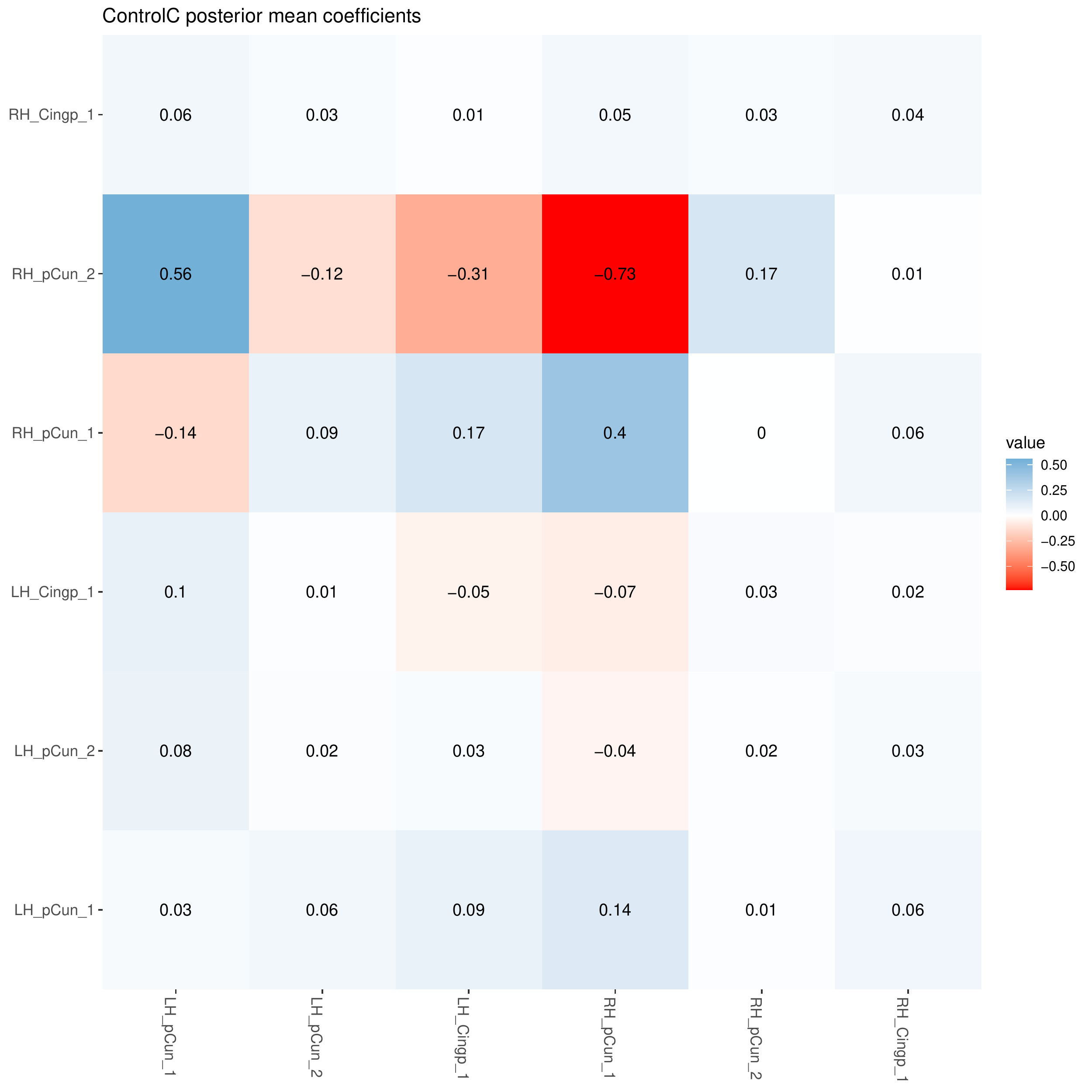}
    \end{subfigure}
  \begin{subfigure}{0.5\textwidth}
      \centering
      \includegraphics[width=\linewidth]{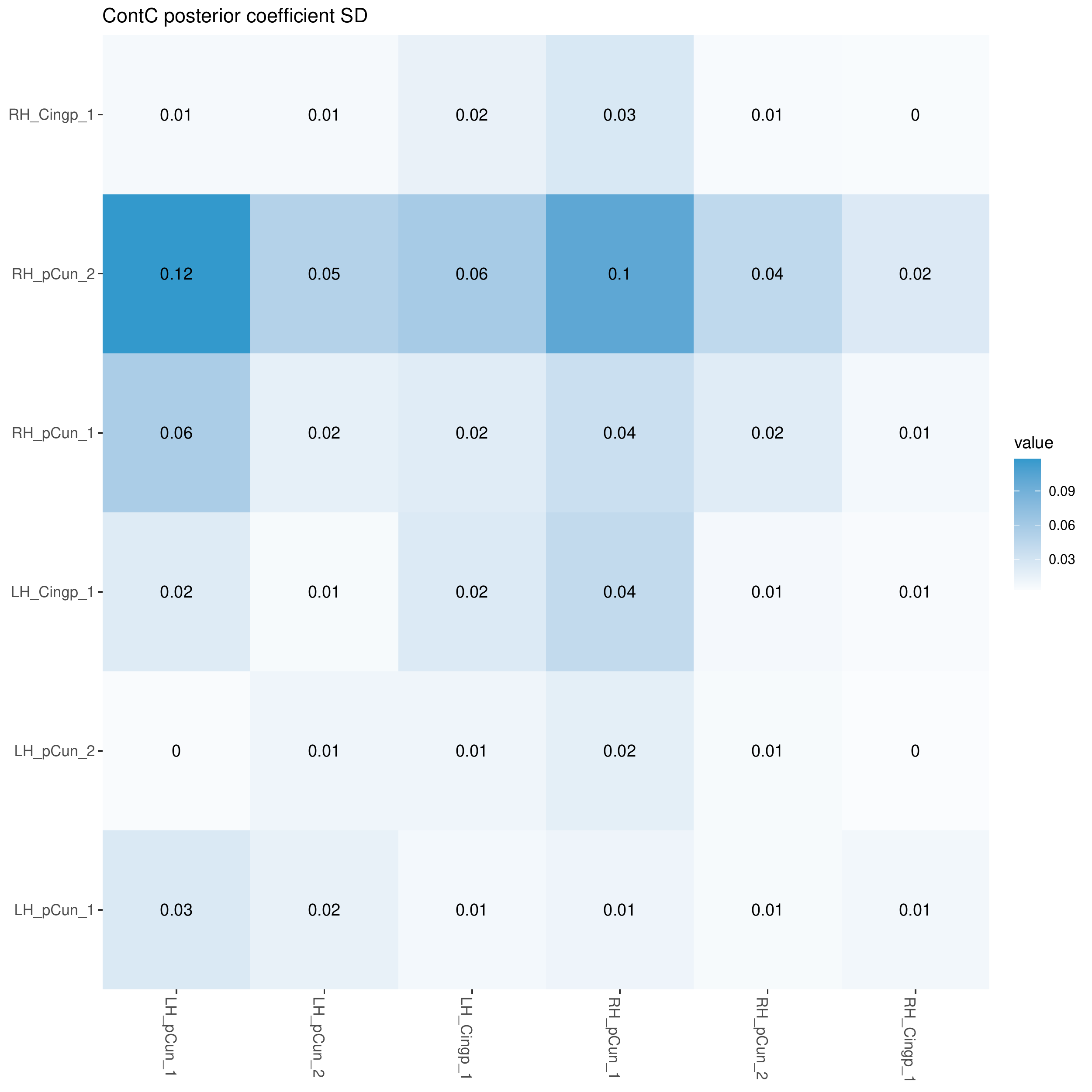}
  \end{subfigure}\\
  \caption{Results from real data: Posterior mean and standard deviations for the lag-1 fixed-effect VAR coefficients corresponding to the Control C and Default C brain regions. 
  }
  \label{fig: fmri_coefs}
\end{figure}

\end{document}